\documentstyle[12pt,fleqn]{article}
\input{psfig}

\def\nsection#1{\setcounter{equation}{0}\section{#1}}
\newenvironment{eq}{%
\setlength{\mathindent}{0 cm}\begin{equation}}{\end{equation}}
\newenvironment{diseq}{\setlength{\mathindent}{0 cm}\[}{\]}

\newcommand{\Integer}{\:\mbox{\sf Z} \hspace{-0.82em} \mbox{\sf Z}\,}

\def\m{\!-\!}
\def\p{\!+\!}
\def\e{\mbox{e}}
\def\ib{\,\mbox{i}\,}
\def\is{\,\mbox{\scriptsize i}\,}
\def\la{\lambda}
\def\te{\vartheta_1}
\def\tt{\vartheta_2}
\def\td{\vartheta_3}
\def\tv{\vartheta_4}
\def\case#1#2{{\textstyle{#1\over #2}}}
\def\W#1#2#3#4#5{W #1 \! \left(\hspace{-1mm}
         \begin{array}{cc}#5 & #4 \\ #2 & #3 \end{array}
         \hspace{-1mm}\right)}
\def\Wt#1#2#3#4#5{\widetilde{W} #1 \! \left(\hspace{-1mm}
         \begin{array}{cc}#5 & #4 \\ #2 & #3 \end{array}
         \hspace{-1mm}\right)}
\def\Mult#1#2#3{\left[#1 \atop #2,#3 \right]}
\def\Mults#1#2#3{\left[{\textstyle {#1 \atop #2,#3} } \right]}
\def\half{\frac{1}{2}}

\begin{document}

%  #] Preamble:

%  #[ Title and author:
\title{Order Parameters of the Dilute A Models}

\author{
S. Ole Warnaar\thanks{
Instituut voor Theoretische Fysica,
Universiteit van Amsterdam,
Valckenierstraat~65,
1018~XE Amsterdam,
The Netherlands; email: warnaar@phys.uva.nl and
nienhuis@phys.uva.nl},
Paul A. Pearce\thanks{
Mathematics Department,
University of Melbourne,
Parkville,
Victoria 3052,
Australia; email: pap@mundoe.maths.mu.oz.au and
kseaton@mundoe.maths.mu.oz.au} , \\
Katherine A. Seaton$^{*\dagger}$
and Bernard Nienhuis$^*$}
\date{ITFA 93-15}
\maketitle
%  #] Title and author:

%  #[ Abstract:
\begin{abstract}
The free energy and local height probabilities of the dilute A models with
broken $\Integer_2$ symmetry are calculated analytically using
inversion and corner transfer matrix methods.
These models possess four critical branches. The first two branches
provide new realisations of the unitary minimal series and the other two
branches give a direct product of this series with an Ising model.
We identify the integrable perturbations which move the dilute A models
away from the critical limit.
Generalised order parameters are defined and their critical
exponents extracted. The associated conformal weights are found to occur on
the diagonal of the relevant Kac table. In an appropriate regime the
dilute A$_3$ model lies in the universality class of the Ising model in a
magnetic field. In this case we obtain the magnetic exponent $\delta=15$
directly, without the use of scaling relations.
\end{abstract}
\thispagestyle{empty}
\newpage
%  #] Abstract:

%  #[ Introduction:
\newlength{\mathin}
\setlength{\mathin}{\mathindent}
\setcounter{page}{1}
\nsection{Introduction}
In the last decade many infinite hierarchies of exactly solvable models
have been found.
Of foremost importance among these models
are the restricted solid-on-solid (RSOS) models
of Andrews, Baxter and Forrester (ABF) \cite{ABF}.
In these models each site of the lattice carries a height variable,
restricted to the values $1,\ldots,h-1$ with $h=4,5,\ldots$, subject to
the rule that heights on neighbouring lattice sites differ by $\pm 1$.
If the allowed heights are represented by the nodes in the following diagram,
\[
\psfig{file=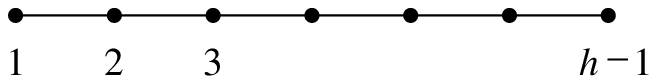}
\]
\noindent
the adjacency rule requires that neighbouring lattice sites take
values that are adjacent on the diagram.

Andrews, Baxter and Forrester considered four different regimes,
labelled I--IV.
It was pointed out by Huse \cite{Huse} that the critical line
separating regimes III and IV
realises the complete unitary minimal series
of conformal field theory. This series is described by a central charge
\begin{equation}
c = 1-\frac{6}{h(h-1)} \qquad h=4,5,\ldots
\label{eq:A.ccharge}
\end{equation}
and a set of conformal weights, given by the Kac formula
\begin{equation}
\Delta_{r,s}^{(h)}=
\frac{[hr-(h-1)s]^2 - 1}{4h(h-1)} \qquad 1\leq r\leq h-2,\quad
1\leq s \leq h-1.
\label{eq:A.Kactable}
\end{equation}
The corresponding modular invariant partition function is
\cite{cardy} \begin{equation}
Z=\half\sum_{r=1}^{h-2}\sum_{s=1}^{h-1}|\chi_{r,s}^{(h)}(q)|^2
\label{diagmipf}
\end{equation}
where $q$ is the modular parameter
and the $\chi_{r,s}^{(h)}$ are the Virasoro characters given by
\begin{equation}
\chi_{r,s}^{(h)} (q) = \frac{q^{\Delta_{r,s}^{(h)}-c/24}}{Q(q)}
\sum_{j=-\infty}^{\infty}\left\{
q^{h(h-1)j^2+[hr-(h-1)s]j}
-q^{h(h-1)j^2+[hr+(h-1)s]j+rs}\right\}
\label{eq:A.Virasoro}
\end{equation}
with $Q(q)=\prod_{n=1}^\infty (1-q^n)$.

\begin{figure}[hbt]
\centerline{\psfig{file=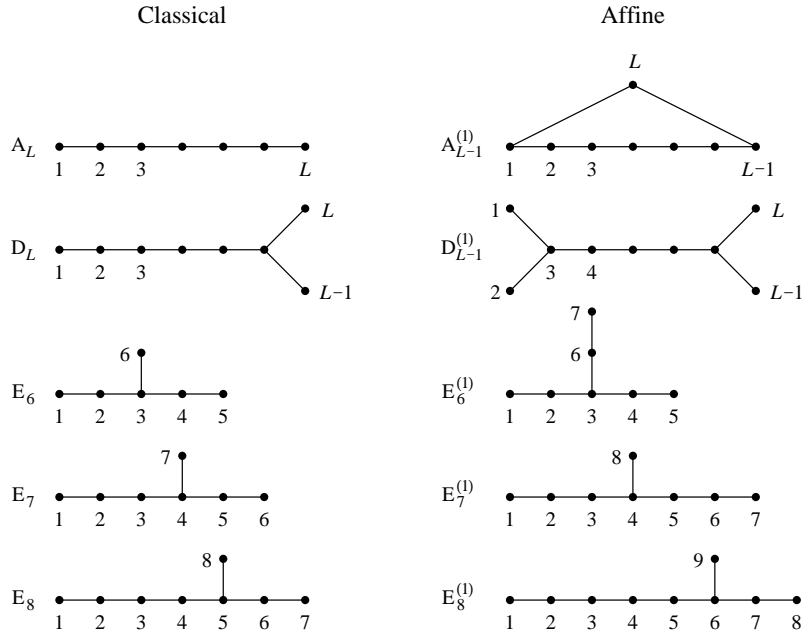,height=8.5cm}}
\caption{Dynkin diagrams of the simply-laced Lie algebras.}
\label{fig:A.ADEdiagrams}
\end{figure}

By giving a loop or polygon interpretation to the critical
ABF models, Pasquier \cite{Pasquier1,Pasquier2} extended
these models to arbitrary adjacency graphs.
Demanding that these new models be critical
restricts the graphs to the Dynkin diagrams
of the classical and affine simply-laced Lie algebras shown in
Fig.~\ref{fig:A.ADEdiagrams}.

Recently a new construction of solvable RSOS
models was found \cite{WNS1,WNS2,WN}.
Basically, the method is an extension of the work of Pasquier, and
related work of Owczarek and Baxter \cite{Owczarek},
to more general loop models.
Application to the O$(n)$ model \cite{Nienhuis},
which is closely related to the Izergin-Korepin model \cite{IK},
has led to a new family of critical RSOS models labelled by Dynkin
diagrams.
The same models were found independently by Roche \cite{Roche}.

In the approach of Pasquier, the polygons, which are interpreted as
domain walls separating regions of different height,
densely cover the edges of the dual lattice.
As a consequence, heights on adjacent sites are always different.
In the new RSOS models, two neighbouring sites of the lattice either
have the same or different height, so that the domain walls occupy
some but not all edges of the dual lattice.
Therefore  it is natural, following \cite{Roche},
to term these new models {\em dilute} A-D-E models.

Each member of the dilute A$_L$ hierarchy possesses
four distinct critical branches.
The central charge is given by
\begin{equation}
c = \left\{\begin{array}{ll}
1-\frac{\displaystyle 6}
{\displaystyle h(h-1)} & \qquad \mbox{branch 1 and 2} \\
%& \\
\frac{3}{2}-\frac{\displaystyle 6}
{\displaystyle h(h-1)} & \qquad \mbox{branch 3 and 4}
\end{array}\right.
\end{equation}
where
\begin{equation}
h=\left\{\begin{array}{ll}
L+2 & \qquad \mbox{branch 1 and 3} \\
L+1 & \qquad \mbox{branch 2 and 4.}
\end{array}\right.
\label{eq:A.Ltoh}
\end{equation}
The first two branches give new realisations of the unitary minimal
series with the modular invariant partition functions (\ref{diagmipf}).
The other two branches appear to be a
direct product of this same series and an Ising model, with
modular invariant partition functions
\begin{equation}
Z={1\over4}\sum_{r'=1}^{2}\sum_{s'=1}^{3}
\sum_{r=1}^{h-2}\sum_{s=1}^{h-1}
|\chi_{r',s'}^{(4)}(q)\;\chi_{r,s}^{(h)}(q)|^2\label{mipf}.
\end{equation}

As reported in \cite{WNS1,WNS2}, the models related to the A$_L$
Dynkin diagrams admit an off-critical extension.
A remarkable feature of these off-critical models is that,
for odd values of $L$, they break
the $\Integer_2$ symmetry of the underlying Dynkin diagram.
The simplest of these symmetry breaking models belongs to the
universality class of the Ising model.
This allows the calculation of the magnetic exponent $\delta=15$
without the use of scaling relations.

This paper is devoted to the investigation of the
models of the dilute A$_L$ hierarchy.
First we briefly describe the whole family of dilute A-D-E models.
Then, in section~\ref{sec:A.offc}, we
define the off-critical A$_L$ model and in
section~\ref{sec:A.free} we calculate its
free energy. From this we extract the critical exponent $\alpha$
when $L$ is even and $\delta$ when $L$ is odd.
The main body of the paper is concerned with the
calculation of the
order parameters of the dilute A models for odd values of $L$.
In section~\ref{sec:A.lhp} we compute the local height probabilities
and in the subsequent section we use these results to evaluate
generalised order parameters. We also extract the set of associated
critical exponents $\delta_k$ and derive the corresponding conformal
weights. In section~\ref{sec:A.phdiag} we discuss the phase diagram,
concentrating on $L=3$, and in section~\ref{sec:Isfi} we collect
results concerning the Ising model in a field.
Finally, we summarise and discuss our main results.

The results for the order parameters when $L$ is even will be
presented in a future publication.
Likewise, results for the critical models related
to the other adjacency diagrams,
among which is a solvable tricritical Potts model
\cite{TriPotts}, will be reported elsewhere.
%  #] Introduction:

%  #[ A-D-E models:
\nsection{The dilute A-D-E models}
In this section we define the family of dilute A-D-E models.
Although we restrict the description to the square lattice, they can
be defined on any planar lattice.

Consider an arbitrary connected graph $\cal G$ consisting of
$L$ nodes and a number of bonds connecting distinct nodes.
Label the nodes by an integer {\em height} $a\in\{1,\ldots,L\}$.
Nodes $a$ and $b$ are called adjacent on $\cal G$
if they are connected via a single bond.
Such a graph is conveniently represented by an {\em adjacency
matrix} $A$
with elements
\begin{equation}
A_{a,b}=\left\{
\begin{array}{cl} 1 & \quad \mbox{if $a$ and $b$ adjacent} \\
                    0 & \quad \mbox{otherwise}.
\end{array}\right.
\end{equation}
Let $n$ denote the largest eigenvalue of $A$ and $S$ the
Perron-Frobenius vector, {\em i.e.,} $AS=nS$.

With these ingredients we define an RSOS model on the
square lattice $\cal L$ as follows.
Each site of $\cal L$ can take one of $L$ different heights.
The Boltzmann weight of a configuration
is non-zero only if all pairs of
neighbouring sites carry heights which are
either equal or adjacent on $\cal G$.
The weight of an elementary face
of the RSOS model is given by
\begin{eqnarray}
\W{}{a}{b}{c}{d}
&=&
\rho_1 \delta_{a,b,c,d} +
\rho_2 \delta_{a,b,c} A_{a,d} +
\rho_3 \delta_{a,c,d} A_{a,b} +
\left(\frac{S_a}{S_b}\right)^{1/2}\rho_4
\delta_{b,c,d} A_{a,b} \nonumber \\
&+&
\left(\frac{S_c}{S_a}\right)^{1/2}\rho_5
\delta_{a,b,d} A_{a,c} +
\rho_6 \delta_{a,b} \delta_{c,d} A_{a,c} +
\rho_7 \delta_{a,d} \delta_{b,c} A_{a,b}
\label{eq:dA.RSOS} \\
&+&
\rho_8 \delta_{a,c} A_{a,b} A_{a,d} +
\left(\frac{S_a S_c}{S_b S_d}\right)^{1/2}
\rho_9 \delta_{b,d} A_{a,b} A_{b,c} \nonumber
\end{eqnarray}
where $S_a$ is the {\em a}-th entry of $S$ and
$a,b,c$ and $d$ can take any of the $L$ heights
of the graph $\cal G$.
The generalised Kronecker $\delta$ is defined as
$\delta_{i_1,\ldots,i_m} \equiv
\prod_{j=2}^m \delta_{i_1,i_j}$.
If we parametrise $n$ by
\begin{equation}
n = -2 \cos 4\la  \label{eq:A.parn}
\end{equation}
then $\rho_1,\ldots,\rho_9$ are given by\footnote{We note that we have
changed the variable $\lambda$ of reference \cite{WNS1}
to $\case{1}{2}\pi-\lambda$.}  \addtocounter{footnote}{-1}
\begin{eqnarray}
\rho_1 &=& (\sin 2\la \sin 3\la + \sin u \sin(3\la-u))
/(\sin 2\la \sin 3\la) \nonumber \\
\rho_2 = \rho_3 &=& \sin(3\la-u)/\sin 3\la \nonumber \\
\rho_4 = \rho_5 &=& \sin u/\sin 3\la \nonumber \\
\rho_6 = \rho_7 &=& \sin u \sin(3\la-u)
/(\sin 2\la \sin 3\la) \label{eq:A.rhos} \\
\rho_8 &=& \sin(2\la-u)\sin(3\la-u)/(\sin 2\la \sin 3\la)
\nonumber \\
\rho_9 &=& -\sin u \sin(\la-u)/(\sin 2\la \sin 3\la) .
\nonumber
\end{eqnarray}
We note that the weights $\rho_4=\rho_5$ and the weights
$\rho_6=\rho_7$ are determined only up to a sign.
For any graph $\cal G$, the RSOS model defined by
(\ref{eq:dA.RSOS})-(\ref{eq:A.rhos})
satisfies the Yang-Baxter equation \cite{BaxterBook}. In fact, in
\cite{WNS1,WNS2,WN} it was shown that all models defined above have the
same partition function as the O$(n)$ model \cite{Nienhuis}.

{}From equations (\ref{eq:A.parn}) and (\ref{eq:A.rhos})
it follows that there
are four different branches that yield the same values of $n$.
Using the periodicity of the weights, we can restrict $\la$ to
the interval $\case{1}{6}\pi \leq \la <\case{1}{3}\pi$:
\begin{equation}
\begin{array}{lcl}
\mbox{branch 1} & \qquad 0<u<3\la &
\qquad \case{1}{6}\pi \leq \la \leq \case{1}{4}\pi
\vphantom{\displaystyle \frac{1}{2}}\\
\mbox{branch 2} & \qquad 0<u<3\la &
\qquad \case{1}{4}\pi \leq \la < \case{1}{3}\pi \\
\mbox{branch 3} & \qquad -\pi+3\la<u<0 &
\qquad \case{1}{4}\pi \leq \la < \case{1}{3}\pi
\vphantom{\displaystyle \frac{1}{2}}\\
\mbox{branch 4} & \qquad -\pi+3\la<u<0 &
\qquad \case{1}{6}\pi \leq \la \leq \frac{1}{4}\pi. \\
\end{array}          \label{eq:A.branch}
\end{equation}
These four branches correspond to (part of) the four branches defined
in \cite{Blote} for the O$(n)$ model.
For $n>2$ the value of $\la-\case{1}{4}\pi$ must be chosen imaginary,
and the weights become complex, unlike the ordinary A-D-E models.
As was pointed out in \cite{Pasquier1,Pasquier2},
the only adjacency graphs that have
largest eigenvalue $n\leq 2$ are the Dynkin diagrams of the classical
$(n<2)$ and affine $(n=2)$
simply-laced Lie algebras shown in Fig.~\ref{fig:A.ADEdiagrams}.
For the classical case the respective values of $n$ are listed in
Table~\ref{tab:A.thetable}.
The corresponding Perron-Frobenius vector can be found in
\cite{Pasquier1}.

{}From the equivalence with the O$(n)$ model, the central
charge of the dilute
A-D-E models is known \cite{WBN}. The values on the four branches are
listed in  Table~\ref{tab:A.thetable}
for the classical algebras. For the affine algebras we have
$\la=\case{1}{4}\pi$ and $c=1$ or $\frac{3}{2}$.
%  #[ table:
\begin{table}[hbt]
\centering
$$
\begin{array}{|c|l|l|l|c|}
\hline
\mbox{algebra} & \qquad \!n & \; \; \; \qquad \la &
\; \; \qquad \qquad c & \mbox{branch} \\
\hline
\mbox{A}_L & 2\cos \frac{\pi}{L+1} &
\frac{\pi}{4} \left( 1 \mp \frac{1}{L+1}\right) &
\left\{\begin{array}{l} 1-\frac{6}{(L+1)(L+1 \pm 1)} \\
                \frac{3}{2}-\frac{6}{(L+1)(L+1 \mp 1)}
\end{array}  \right. &
\begin{array}{c}1,2 \\ 4,3 \end{array}\\
&&&& \\
\mbox{D}_L & 2\cos \frac{\pi}{2L-2} &
\frac{\pi}{4} \left(1 \mp \frac{1}{2L-2}\right) &
\left\{\begin{array}{l} 1-\frac{6}{(2L-2)(2L-2\pm 1)} \\
                 \frac{3}{2}-\frac{6}{(2L-2)(2L-2\mp 1)}
\end{array}  \right. &
\begin{array}{c}1,2 \\ 4,3 \end{array}\\
&&&& \\
\mbox{E}_6 & 2\cos \frac{\pi}{12} &
\frac{\pi}{4} \left(1 \mp \frac{1}{12}\right) &
\left\{\begin{array}{l} 1-\frac{6}{12(12 \pm 1)}  \\
                 \frac{3}{2}-\frac{6}{12(12 \mp 1)}
\end{array}  \right. &
\begin{array}{c}1,2 \\ 4,3 \end{array}\\
&&&& \\
\mbox{E}_7 & 2\cos \frac{\pi}{18} &
\frac{\pi}{4} \left(1 \mp \frac{1}{18}\right) &
\left\{\begin{array}{l} 1-\frac{6}{18(18 \pm 1)}  \\
                 \frac{3}{2}-\frac{6}{18(18 \mp 1)}
\end{array}  \right. &
\begin{array}{c}1,2 \\ 4,3 \end{array}\\
&&&& \\
\mbox{E}_8 & 2\cos \frac{\pi}{30} &
\frac{\pi}{4} \left(1 \mp \frac{1}{30}\right) &
\left\{\begin{array}{l} 1-\frac{6}{30(30 \mp 1)}    \\
                 \frac{3}{2}-\frac{6}{30(30 \pm 1)}
\end{array} \right. &
\begin{array}{c}1,2 \\ 4,3 \end{array} \\
\hline
\end{array}
$$
\caption{Central charge of the dilute A-D-E models.}
\label{tab:A.thetable}
\end{table}
%  #] table:

%  #] A-D-E models:

%  #[ Off-critical A model:
\nsection{The off-critical A model}\label{sec:A.offc}
The dilute A$_L$ model (\ref{eq:dA.RSOS})
admits an extension away from criticality while remaining solvable.
In terms of the theta functions of
appendix~\ref{app:A.elliptic},
suppressing the dependence on the nome $p$,
the Boltzmann weights of the off-critical
A$_L$ model are given by \cite{WNS1,WNS2}
%  #[ weights:
\setlength{\mathindent}{0 cm}
\begin{eqnarray}
\lefteqn{\W{}{a}{a}{a}{a}=
\frac{\te(6\la-u)\te(3\la+u)}{\te(6\la)\te(3\la)}}
\nonumber \\  & & \nonumber \\
\lefteqn{\hphantom{\W{}{a}{a}{a}{a}}
-\left(\frac{S(a+1)}{S(a)}\frac{\tv(2a\la-5\la)}{\tv(2a\la+\la)}
      +\frac{S(a-1)}{S(a)}\frac{\tv(2a\la+5\la)}{\tv(2a\la-\la)}\right)
\frac{\te(u)\te(3\la-u)}{\te(6\la)\te(3\la)}}
\nonumber \\ & & \nonumber \\
\lefteqn{\W{}{a}{a}{a}{a\pm 1}=\W{}{a}{a\pm 1}{a}{a}=
\frac{\te(3\la-u)\tv(\pm 2a\la+\la-u)}{\te(3\la)\tv(\pm 2a\la+\la)}}
\nonumber \\ & & \nonumber \\
\lefteqn{\W{}{a\pm 1}{a}{a}{a}=\W{}{a}{a}{a\pm 1}{a}=
\left(\frac{S(a\pm 1)}{S(a)}\right)^{1/2}
\frac{\te(u)\tv(\pm 2a\la-2\la+u)}{\te(3\la)\tv(\pm 2a\la+\la)}}
\nonumber \\ & & \nonumber \\
\lefteqn{\W{}{a}{a\pm 1}{a\pm 1}{a}=\W{}{a}{a}{a\pm 1}{a\pm 1}}
\nonumber \\ & & \nonumber \\
\lefteqn{ \hphantom{\W{}{a}{a\pm 1}{a\pm 1}{a}}
=\left(\frac{\tv(\pm 2a\la+3\la)\tv(\pm 2a\la-\la)}
           {\tv^2(\pm 2a\la+\la)}\right)^{1/2}
\frac{\te(u)\te(3\la-u)}{\te(2\la)\te(3\la)} }
\nonumber \\ & & \label{eq:A.weights} \\
\lefteqn{\W{}{a}{a\mp 1}{a}{a\pm 1}=
\frac{\te(2\la-u)\te(3\la-u)}{\te(2\la)\te(3\la)}}
\nonumber \\ & & \nonumber \\
\lefteqn{\W{}{a\pm 1}{a}{a\mp 1}{a}=
-\left(\frac{S(a-1)S(a+1)}{S^2(a)}\right)^{1/2}
\frac{\te(u)\te(\la-u)}{\te(2\la)\te(3\la)}}
\nonumber \\ & & \nonumber \\
\lefteqn{\W{}{a\pm 1}{a}{a\pm 1}{a}=
\frac{\te(3\la-u)\te(\pm 4a\la+2\la+u)}{\te(3\la)\te(\pm 4a\la+2\la)}
+\frac{S(a\pm 1)}{S(a)}
\frac{\te(u)\te(\pm 4a\la-\la+u)}{\te(3\la) \te(\pm 4a\la+2\la)}}
\nonumber \\ & & \nonumber \\
\lefteqn{\hphantom{\W{}{a\pm 1}{a}{a\pm 1}{a}}=
\frac{\te(3\la+u)\te(\pm 4a\la-4\la+u)}
{\te(3\la)\te(\pm 4a\la-4\la)}}
\nonumber \\ & & \nonumber \\
\lefteqn{\hphantom{\W{}{a\pm 1}{a}{a\pm 1}{a}}+
\left(\frac{S(a\mp 1)}{S(a)}\frac{\te(4\la)}{\te(2\la)}
-\frac{\tv(\pm 2a\la-5\la)}{\tv(\pm 2a\la+\la)} \right)
\frac{\te(u)\te(\pm 4a\la-\la+u)}{\te(3\la) \te(\pm 4a\la-4\la)}}
\nonumber \\ & & \nonumber \\
\lefteqn{S(a)=(-)^{\displaystyle a} \;
\frac{\te(4a\la)}{\tv(2a\la)} \, .}
\nonumber
\end{eqnarray}
%  #] weights:
We note that in the critical limit, $p\to 0$, the {\em crossing
factors} $S(a)$
reduce to the entries of the Perron-Frobenius vector $S_a$:
\setlength{\mathindent}{\mathin}
\begin{equation}
\lim_{p\to 0} \: \frac{S(a)}{S(b)} = \frac{S_a}{S_b} \, .
\end{equation}
We also note that, for later convenience, we have relabelled the
states of the model $a\to L+1-a$ in comparison with those of our
earlier definition of the model in \cite{WNS1}.

The Boltzmann weights (\ref{eq:A.weights})
satisfy the following initial condition and crossing
symmetry:
\begin{eqnarray}
W\left(\hspace{-1mm}\left.\begin{array}{cc}
d & c \\
a & b
\end{array}\right| 0 \right) &=& \delta_{a,c} \\
& &  \nonumber \\
W\left(\hspace{-1mm}\left.\begin{array}{cc}
d & c \\
a & b
\end{array}\right| 3 \la - u \right) &=&
\left(\frac{S(a)S(c)}{S(b)S(d)}\right)^{1/2}
W\left(\hspace{-1mm}\left.\begin{array}{cc}
c & b \\
d & a
\end{array}\right| u \right)  \label{eq:A.crossing}
\end{eqnarray}
and an inversion relation of the form
\begin{eq}
\sum_g
W\left(\hspace{-1.2mm}\left.\begin{array}{cc}
d & g \\
a & b
\end{array}\right| u \right)
W\left(\hspace{-1.2mm}\left.\begin{array}{cc}
d & c \\
g & b
\end{array}\right| \m u \right) =
\frac{\te(2\la-u)\te(3\la-u)\te(2\la+u)\te(3\la+u)}
      {\te^2(2\la)\te^2(3\la)} \: \delta_{a,c} .
\label{eq:A.invrel}
\end{eq}

In equation (\ref{eq:A.branch}) four
different critical branches were defined.
This yields eight regimes for the off-critical A$_L$ model
\begin{equation}
\begin{array}{lcl}
\left.\begin{array}{lr}
\mbox{regime }1^+ \quad & 0<p<1 \\
\mbox{regime }1^- & -1<p<0
\end{array} \right\}
& 0<u<3\la \quad & \la=\frac{\pi}{4}\left(1-\frac{1}{L+1}\right) \\
&&\\
\left.\begin{array}{lr}
\mbox{regime }2^+ \quad & 0<p<1 \\
\mbox{regime }2^- & -1<p<0
\end{array} \right\}
& 0<u<3\la \quad & \la=\frac{\pi}{4}\left(1+\frac{1}{L+1}\right) \\
&&\\
\left.\begin{array}{lr}
\mbox{regime }3^+ \quad & 0<p<1 \\
\mbox{regime }3^- & -1<p<0
\end{array} \right\}
& 3\la-\pi<u<0 \quad & \la=\frac{\pi}{4}\left(1+\frac{1}{L+1}\right) \\
&&\\
\left.\begin{array}{lr}
\mbox{regime }4^+ \quad & 0<p<1 \\
\mbox{regime }4^- & -1<p<0
\end{array} \right\}
& 3\la-\pi<u<0 \quad & \la=\frac{\pi}{4}\left(1-\frac{1}{L+1}\right).
\end{array}
\end{equation}
For regimes $2^{\pm}$ and $3^{\pm}$ we exclude the $L=2$ case
because the model becomes singular.

At criticality all A$_L$ models satisfy the $\Integer_2$ symmetry of the
Dynkin diagram, but the off-critical models, for odd values of $L$,
break this symmetry:
\begin{equation}
W \left(
\begin{array}{cc}
d & c \\
a & b
\end{array}
\right) \neq
W \left(
\begin{array}{cc}
L+1-d & L+1-c \\
L+1-a & L+1-b
\end{array}
\right) \qquad L\mbox{ odd}.
\end{equation}
For $L=3$, if we make the identification $\{1,2,3\}=\{+,0,-\}$,
the model can be viewed as a spin-1 Ising model.
For $p \neq 0$
the up-down symmetry of this model is broken. We can therefore regard
the nome $p$ of the $\vartheta$-functions as a magnetic field.
This is in contrast with the usual role of $p$ as a
temperature-like variable (see, {\em e.g.,} \cite{BaxterBook}).
%? changed back:See below
Also for larger, odd values of $L$
we refer to $p$ as the magnetic
field, even though it is not,
in general, the leading magnetic operator.
%?I inadvertently removed the `also'...it's not that I don't like it,
%?but at the start of the sentence it `sounds funny'
%?We refer to $p$ as the magnetic
%field for the larger, odd values of $L$ also, even though it is not,
%in general, the leading magnetic operator.
%
For odd $L$ the $+$ and $-$ regimes are equivalent because
negating the magnetic field $p$ merely has the effect of relabelling
the heights $a\rightarrow L+1-a$.

The dilute A$_L$ model defined
above is closely related to the Izergin-Korepin
(or A$_2^{(2)}$) SOS model \cite{Kuniba}.
In the latter model the weights are given
by equation (\ref{eq:A.weights})
with $2a\la$ replaced by $2a\la+w_0$,
where $w_0$ is an arbitrary constant, and $a\in \Integer$.
If we set
\begin{eqnarray}
\la&=&\frac{k\pi}{4(L+1)} \qquad  k\in\{1,\ldots L,L+2,\ldots,2L+1\}
\nonumber \\
w_0&=&0   \\
a&\in& \{1,\ldots,L\} \nonumber
\end{eqnarray}
we obtain A$_2^{(2)}$ RSOS models
based on the A$_L$ algebra.
Only when $k=L$ or $L+2$ are these models physical.
Other choices of $k$ correspond to eigenvalues of the adjacency matrix
of A$_L$ which are not the largest. For all odd values of
$k$ the models break the $\Integer_2$ symmetry of the underlying Dynkin
diagram.

Another way to restrict the A$_2^{(2)}$ SOS model is given
by
\begin{eqnarray}
\la&=&\frac{k\pi}{2(L+1)} \qquad  k\in\{1,\ldots L\}
\nonumber \\
w_0&=&\case{1}{4}\ib\ln p  \\
a&\in&\{1,\ldots,L\}. \nonumber
\end{eqnarray}
This possibility has been studied by Kuniba in a more general way
in his study of A$_n^{(2)}$ RSOS models \cite{Kuniba}. However, this
way of restricting the A$_2^{(2)}$ SOS model does
not give rise to symmetry breaking models.

%  #] Off-critical A model:

%  #[ Free energy:
\nsection{The free energy}\label{sec:A.free}
We calculate the free energy or, equivalently, the
partition function per site $\kappa$ of the dilute A model by
the inversion relation method \cite{inv}.
Because the inversion relation (\ref{eq:A.invrel})
is quadratic in $p$ we
can restrict ourselves to the regimes with positive $p$,
and parametrise $p=\exp(-\epsilon)$.

\subsection{Regimes 1$^+$ and 2$^+$}
In regimes $1^+$ and $2^+$ we assume that $\kappa(u)$
is analytic in the strip
$0 < \mbox{Re} (u) < 3\la$,
and may be analytically extended just beyond these boundaries.
The inversion relation (\ref{eq:A.invrel}) implies
\begin{equation}
\kappa(u)\kappa(-u)=
\frac{\te(2\la-u)\te(3\la-u)\te(2\la+u)\te(3\la+u)}
      {\te^2(2\la)\te^2(3\la)}
\label{eq:A.kinv}
\end{equation}
while the crossing symmetry (\ref{eq:A.crossing})
translates to
\begin{equation}
\kappa(u)=\kappa(3 \la -u).
\label{eq:A.cross}
\end{equation}
We make the conjugate modulus transformation (\ref{eq:A.congnome})
and a Laurent expansion of $\ln \kappa(u)$ in powers of
$\exp(-2\pi u/\epsilon)$.
Matching coefficients in equations (\ref{eq:A.kinv}) and
(\ref{eq:A.cross}) we obtain for the free energy
\begin{equation}
\ln \kappa(u) = 2 \sum_{k=-\infty}^{\infty}
\frac{
\cosh (5 \la-\pi) \pi k/\epsilon \,
\cosh \pi \la k/\epsilon \,
\sinh \pi u k/\epsilon \,
\sinh (3\la-u)\pi k/\epsilon}
{k\, \sinh \pi^2 k/\epsilon \,
\cosh 3 \pi \la k/\epsilon}.
\label{eq:FE12}
\end{equation}
We now take the $p\rightarrow 0$ limit in the above expression to
obtain the leading critical singularity. Using the Poisson summation
formula, we find
\begin{equation}
\ln \kappa_{\mbox{\scriptsize sing}} \sim p^{\pi/3\la}.
\label{eq:A.kappasing12}
\end{equation}
If we compare this with
\begin{equation}
\ln \kappa_{\mbox{\scriptsize sing}}  \sim p^{2-\alpha}
\qquad \mbox{or} \qquad
\ln \kappa_{\mbox{\scriptsize sing}}  \sim p^{1+1/\delta}
\label{eq:A.defad}
\end{equation}
we find for the critical exponents $\alpha$ and $\delta$
\begin{equation}
\mbox{regime }1^+:
\left\{\begin{array}{ll}
\alpha=\frac{2(L-2)}{3L} & L\mbox{ even} \\
& \\
\delta=\frac{3L}{L+4} & L\mbox{ odd}
\end{array} \right.
\qquad
\mbox{regime }2^+:
\left\{\begin{array}{ll}
\alpha=\frac{2(L+4)}{3(L+2)} & L\mbox{ even} \\
& \\
\delta=\frac{3(L+2)}{L-2} & L\mbox{ odd}.
\end{array} \right.
\end{equation}
When $L=2$ in regime $1^+$, expression (\ref{eq:A.kappasing12})
for the critical singularity
has to be multiplied by $\ln p$.

\subsection{Regimes 3$^+$ and 4$^+$}
In regimes $3^+$ and $4^+$
the appropriate analyticity strip is $-\pi+ 3\la  < \mbox{Re}(u)< 0$,
and the crossing symmetry becomes
\begin{equation}
\kappa(u)=\kappa(3 \la-\pi -u).
\label{eq:A.cross34}
\end{equation}
Performing the same steps as before
we obtain
\begin{eq}
\ln \kappa(u) = - 2 \sum_{k=-\infty}^{\infty}
\frac{
\cosh (5 \la-\pi) \pi k/\epsilon \,
\cosh \pi \la k/\epsilon \,
\sinh \pi u k/\epsilon \,
\sinh (\pi-3\la+u)\pi k/\epsilon}
{k\, \sinh \pi^2 k/\epsilon \,
\cosh (\pi-3\la) \pi k/\epsilon} \, .
\end{eq}
For the dominant singularity we find, apart from some exceptions we
list below,
\begin{equation}
\ln \kappa_{\mbox{\scriptsize sing}} \sim p^{\pi/(\pi-3\la)}.
\label{eq:A.kappasing34}
\end{equation}
Comparing this with (\ref{eq:A.defad}), we find
for the critical exponents $\alpha$ and $\delta$
\begin{equation}
\mbox{regime }3^+ \! :
\left\{\!\begin{array}{ll}
\alpha=-\frac{2(L+4)}{L-2} & L\mbox{ even} \\
& \\
\delta=\frac{L-2}{3(L+2)} & L\mbox{ odd}
\end{array} \right.
\qquad
\mbox{regime }4^+ \! :
\left\{\!\begin{array}{ll}
\alpha=-\frac{2(L-2)}{L+4} & L\mbox{ even} \\
& \\
\delta=\frac{L+4}{3L} & L\mbox{ odd}.
\end{array} \right.
\end{equation}
For the cases $L=5$ and 8 in regime $3^+$ and $L=2$ in regime $4^+$
equation (\ref{eq:A.kappasing34}) has to be multiplied by $\ln p$.
When $L=3,4,6$ and 14 in regime $3^+$ and $L=8$ in regime $4^+$ the
partition function per site is regular.
%  #] Free energy:

%  #[ LHP:
\nsection{Local height probabilities}\label{sec:A.lhp}
In this section, which forms the main part of our paper,
we calculate the {\em local height probabilities} of the dilute A model
for odd values of $L$.
Since negating the nome $p$ is nothing but a
reversal of the magnetic field,
we can restrict ourselves to the four `+' regimes.
(Recall that for odd $L$ the Boltzmann weights are symmetric under
the transformation $p\to -p$, $a\to L+1-a$.)

%  #[ GC:
\subsection{Groundstate configurations}\label{sec:A.groundstate}
\begin{figure}[hbt]
\centerline{\psfig{file=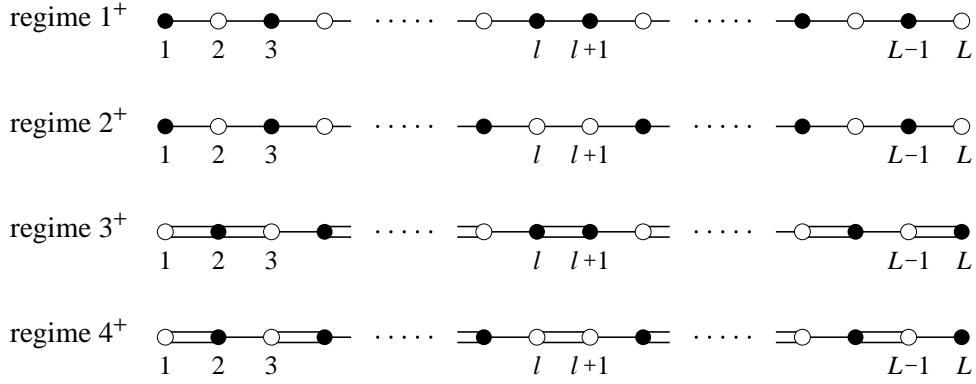,width=13cm}}
\caption{Groundstate configurations for the four different regimes.}
\label{fig:A.groundstate}
\end{figure}

First we describe the groundstate configurations in each of the four
%different
regimes.
Below we depict the set of groundstates by
%means of a
decoration of the
adjacency diagram.
Comparing the various weights in (\ref{eq:A.weights})
in the ordered limit, see appendix~\ref{app:A.H-fun}, we find that the
following types of groundstates occur:
\begin{enumerate}
\item Completely flat or ferromagnetic configurations.
      If the height of this configuration
      is $a$ we will denote this state by a solid circle on
      the adjacency graph $\cal G$
      at node $a$. Conversely, flat configurations
      that do not yield a groundstate will be denoted by an open circle
      on $\cal G$.
\item Antiferromagnetic configurations.
      One sublattice has height $a$ and the other sublattice height
      $b=a\pm 1$. This state, together with the state where the
      heights on the two
      sublattices are interchanged, is denoted by a double bond
      on $\cal G$ between the nodes $a$ and $b$.
\end{enumerate}
If we define the variable $l$ to be
\begin{equation}
l=\left\{
\begin{array}{ll}
2\lfloor \frac{L-1}{4} \rfloor +1 &
\qquad \mbox{regimes $1^+$ and $4^+$}   \\
& \\
2\lfloor \frac{L+1}{4} \rfloor
& \qquad \mbox{regimes $2^+$ and $3^+$}
\end{array} \right.
\label{eq:A.defl}
\end{equation}
where $\lfloor$ $\rfloor$ denotes the integer part,
the groundstates of the four regimes are as shown in
Fig.~\ref{fig:A.groundstate}.
Thus the total number of groundstates is $\case{1}{2}(L+1),
\case{1}{2}(L-1),\case{3}{2}(L+1)$
and $\case{3}{2}(L-1)$ for the regimes $1^+,\ldots,4^+$, respectively.

\subsection{Local height probabilities}
The local height probability $P^{bc}(a)$
is the probability that a given site of the
lattice has height $a$ given that the model is in the
phase indexed by $b$ and $c$.
In the ferromagnetic phases
%we have
$c=b$
and in the antiferromagnetic phases $c=b\pm1$.

The technique of corner transfer matrices (CTMs) which is used to calculate
$P^{bc}(a)$
is well-known and the details of the method are given elsewhere (see, {\em
e.g.,} \cite{BaxterBook,ABF}). Because the weights
(\ref{eq:A.weights}) satisfy the
Yang-Baxter
equation and possess the crossing symmetry (\ref{eq:A.crossing}),
the local height probability in regimes $1^+$ and $2^+$ can be written as
\begin{equation}
P^{bc}(a) = \lim_{m\rightarrow\infty}
\frac{q^{-a^2 \la/\pi} S(a) X_m^{a\,b\,c}(q)}
{\sum_{a=1}^L q^{-a^2 \la/\pi}
S(a) X_m^{a\,b\,c}(q)} \, .
\label{eq:A.lhp}
\end{equation}
Here the variable $q$ is related to $p=\exp(-\epsilon)$ by
\begin{equation}
q=
\e^{-12\pi\la/\epsilon} .
\label{eq:A.qdef}
\end{equation}
The one-dimensional configuration sums $X_m^{a\,b\,c}(q)$ are given by
\begin{equation}
X_m^{\sigma_1\,\sigma_{m+1}\,\sigma_{m+2}}(q)=\sum_{\sigma_2,\ldots,\sigma_m}
q^{\sum_{j=1}^m j H(\sigma_j,\sigma_{j+1},\sigma_{j+2})}.
\label{eq:ODC}
\end{equation}
The weight function $H$ is determined from the Boltzmann weights
in the ordered limit ($p\rightarrow1, u/\epsilon$ fixed). Their behaviour is
\begin{equation}
W \left(\begin{array}{cc}
d & c \\
a & b
\end{array}
\right) \sim \frac{g_a g_c}{g_b g_d}
\; \e^{-2 \pi u H(d,a,b)/\epsilon}
\: \delta_{a,c}
\label{eq:A.Hdef}
\end{equation}
where
\begin{equation}
g_a = \e^{-2 \la u a^2/\epsilon}.
\label{eq:A.gdef}
\end{equation}
The values of the function $H(d,a,b)$ for regimes $1^+$ and $2^+$ are
listed in appendix~\ref{app:A.H-fun}.

{}From (\ref{eq:ODC}) it follows directly that
the one-dimensional configuration sums satisfy
the following recurrence relation:
\begin{equation}
X_m^{a\,b\,c}(q)=q^{m H(b-1,b,c)} X_{m-1}^{a\,b-1\,b}(q)
              +q^{m H(b,b,c)} X_{m-1}^{a\,b\,b}(q)
              +q^{m H(b+1,b,c)} X_{m-1}^{a\,b+1\,b}(q).
\label{eq:A.recurrence}
\end{equation}
The task is to solve this relation, with initial condition
\begin{equation}
X_1^{a\,b\,c}(q)=q^{H(a,b,c)}
\end{equation}
or, equivalently,
\begin{equation}
X_0^{a\,b\,c}(q)=\delta_{a,b}
\label{eq:A.initial}
\end{equation}
and with the conditions
\begin{equation}
X_m^{a\,0\,1}(q)=X_m^{a\,L+1\,L}(q)=0
\label{eq:A.confine}
\end{equation}
which confine the heights in the recursion relation
(\ref{eq:A.recurrence}) to the set $\{1,\ldots,L\}$.

%? Is the following ok? Very nice, apart from the Cst. See my
%?attempt to do without it below.
In regimes $3^+$ and $4^+$, where the spectral parameter $u$ is
negative, the local height probability is given by
\begin{equation}
P^{bc}(a) = \lim_{m\rightarrow\infty}
\frac{q^{a^2 \la/\pi} S(a) X_m^{a\,b\,c}(q)}
{\sum_{a=1}^L q^{a^2 \la/\pi}
S(a) X_m^{a\,b\,c}(q)}
\label{eq:A.lhp34}
\end{equation}
where $q$ is now
\begin{equation}
q=\e^{-4\pi(\pi-3\la)/\epsilon}.
\label{eq:A.qdef34}
\end{equation}
As for the regimes 1$^+$ and 2$^+$, the one-dimensional
configuration sums are expressed in $H$ by (\ref{eq:ODC}).
To determine $H$,
we now need to know the powers
of $\e^{+2\pi u/\epsilon}$ in the ordered limit of the weights.
Relative to regimes 1$^+$ and 2$^+$, after the appropriate
gauge factors $g$ are separated,
and contributions which
cancel in the
ratio in equation (\ref{eq:A.lhp34}) are discarded, $H$ is
effectively replaced
$-H$.
We thus need only take
$q\rightarrow 1/q$ in the regime $1^+$ ($2^+$)
solution of (\ref{eq:A.recurrence}) to
obtain the form of the solution in regime $4^+$ ($3^+$).
Of course, $q$ must still
be given its proper meaning from (\ref{eq:A.qdef34}).

%  #] OPF:
%  #[ Solution:
\subsection{Solution of the recurrence relation}
In this section we derive the solution of
the recursion relation (\ref{eq:A.recurrence})
for each of the four regimes.
In fact, as remarked above, the solution for regimes $3^+$ and
$4^+$ is deduced from the solution for
regimes $1^+$ and $2^+$.

%  #[ Gaussian mult:
\subsubsection*{Gaussian multinomials}
Before we present the solution we need some
preliminaries on the {\em Gaussian multinomials}
$\Mults{m}{k}{l}$,
defined by \cite{Andrews}
\begin{equation}
\Mult{m}{k}{l} = \Mult{m}{k}{l}_q = \frac{(q)_m}{(q)_k (q)_l (q)_{m-k-l}}
\label{eq:A.Gaussm}
\end{equation}
where
\begin{equation}
(q)_m=\prod_{k=1}^m (1-q^k).
\end{equation}
In the limit $q\to 1$ the Gaussian multinomials reduce to ordinary
multinomials
\begin{equation}
\lim_{q\rightarrow 1} \Mult{m}{k}{l}
=\frac{m!}{k! \: l! \: (m \m k \m l)!}
= {m \choose k,l}.
\label{eq:A.multinomial}
\end{equation}
The following identities, which prove useful in
solving the recurrence relation, may be derived by
straightforward manipulation of the definition
\begin{eqnarray}
\Mult{m}{k}{l}
&=& \Mult{m-1}{k}{l}
+ q^{m-k} \Mult{m-1}{k-1}{l}
+ q^{m-k-l} \Mult{m-1}{k}{l-1}
\label{eq:A.Gidena} \\
& & \nonumber \\
&=& \Mult{m-1}{k-1}{l}
+ q^{m-l} \Mult{m-1}{k}{l-1}
+ q^k \Mult{m-1}{k}{l}
\label{eq:A.Gidenb} \\
& & \nonumber \\
&=& \Mult{m-1}{k-1}{l}
+ q^{k+l} \Mult{m-1}{k}{l}
+ q^k \Mult{m-1}{k}{l-1}
\label{eq:A.Gidenc} .
\end{eqnarray}
%  #] Gaussian mult:

%  #[ Regime1:
\smallskip
\subsubsection*{Regime 1$^+$}
As a first step towards the solution of equation
(\ref{eq:A.recurrence}) we consider the
special limit $q=1$.\\ In this limit,
setting $X_m^{a\, b\, c}(1)=X_m^{a\, b\, c}$,
the recurrence relation reduces to the
following combinatorial problem
\begin{equation}
X_m^{a\, b\, c}= X_{m-1}^{a\, b-1\, b}
               + X_{m-1}^{a\, b\, b}+ X_{m-1}^{a\, b+1\, b}.
\label{eq:A.combinatoric}
\end{equation}
The fundamental solution of this equation is
\begin{equation}
X_m^{a\, b\, c}=\sum_{k=0}^{\infty}
{m \choose  k,k \p b}.
\end{equation}
In order to satisfy the initial and confining
conditions (\ref{eq:A.initial}) and (\ref{eq:A.confine})
we must take linear combinations of this solution in the following way:
\begin{equation}
X_m^{a\, b\, c}=\sum_{j,k=-\infty}^{\infty}
\left\{ {m \choose k,k \p 2(L \p 1)j \p a \m b}-
        {m \choose k,k \p 2(L \p 1)j \p a \p b} \right\}.
\label{eq:A.qis1}
\end{equation}

Another piece of information can be gained by considering
the $m\rightarrow \infty$ limit.
It was noted in \cite{Date} that, in this limit,
the one-dimensional configuration sums for the ABF models
in regimes III and IV
become precisely the characters of the related Virasoro
algebra (\ref{eq:A.Virasoro}).

Expecting similar $m \rightarrow \infty$ behaviour for our
configuration sums, we generated large polynomials on the computer
and multiplied these by $Q(q)$. The resulting polynomials,
being very sparse, were then easily identified as Virasoro characters,
with $r=a$ and $s=b$
for $b\leq l$ and $s=b+1$ for
$b>l$, or as linear combinations of two characters.

Guided by these two limiting cases,
replacing $h$ in (\ref{eq:A.Virasoro}) by $L+2$ according to
(\ref{eq:A.Ltoh}), we make the following Ansatz for
the configuration sums, with $b\leq l$,
which yield a single character:
\setlength{\mathindent}{0 cm}
\begin{eqnarray}
\setbox0=\hbox{$\scriptstyle +ab\beta(j,k)$}
\setbox1=\hbox{$\scriptstyle \alpha(j,k)$}
\lefteqn{
X_m^{a\, b\, c} = q^{\gamma \, m} \!
\sum_{j,k=-\infty}^{\infty} \left\{
q^{(L+2)(L+1)j^2+[(L+2)a-(L+1)b]j+\alpha(j,k)}
\Mult{m}{k}{k \p 2(L \p 1)j \p a \m b} \right. } \nonumber \\
& & \nonumber \\
\lefteqn{
\hphantom{X_m^{a\, b\, c} = q^{\gamma \, m} \!
\sum_{j,k=-\infty}^{\infty}}
\hskip -\wd0 \hskip -\wd1
\left. -q^{(L+2)(L+1)j^2+[(L+2)a+(L+1)b]j+a b+\beta(j,k)}
\Mult{m}{k}{k \p 2(L \p 1)j \p a \p b} \right\}. }
\setbox0=\hbox{}
\setbox1=\hbox{}
\end{eqnarray}
For $b>l$ we make a similar Ansatz.
The unknown functions $\alpha$ and $\beta$ are assumed to be
quadratic in their arguments $j$ and $k$.
They depend implicitly on $a$, $b$ and $c$, as does the parameter
$\gamma$.
For the configuration sums that, in the $m\rightarrow\infty$ limit,
yield linear combinations of characters, we make the
appropriate linear combinations of the above Ansatz.

Fitting the Ansatz for small values of $m$ with the correct polynomial, the
coefficients in the functions $\alpha(j,k)$ $\beta(j,k)$ and $\gamma$
can be determined. Before we present the solution, we define
the following auxiliary function:
\setlength{\mathindent}{\mathin}
\begin{eqnarray}
\setbox0=\hbox{$\scriptstyle +as$}
\lefteqn{
F_m^s(b) = q^{(a-s)a/2}
\sum_{j,k=-\infty}^{\infty} } \nonumber \\
& & \nonumber \\
\lefteqn{\qquad \quad \left\{
q^{(L+2)(L+1)j^2+[(L+2)a-(L+1)s]j+k[k+2(L+1)j+a-b]}
\Mult{m}{k}{k \p 2(L \p 1)j \p a \m b} \right. }
\label{eq:A.F} \\
& & \nonumber \\
\lefteqn{\hskip -\wd0 \left. \qquad \quad
-q^{(L+2)(L+1)j^2+[(L+2)a+(L+1)s]j+as+k[k+2(L+1)j+a+b]}
\Mult{m}{k}{k \p 2(L \p 1)j \p a \p b} \right\} } \nonumber
\setbox0=\hbox{}
\end{eqnarray}
where we have suppressed the $a$ dependence. This function
has the following elementary properties
\begin{eqnarray}
\lefteqn{F_m^s(b)=-F_m^{-s}(-b)}  \nonumber \\
\lefteqn{F_m^{L+2+s}(L \p 1 \p b)=-q^{-(L+1)s}
F_m^{L+2-s}(L \p 1 \m b)
\vphantom{\sum_a^a}
}  \label{eq:A.elemprop} \\
\lefteqn{F_m^1(0)=(1 \m q^m) F_{m-1}^1(1).} \nonumber
\end{eqnarray}
It is also convenient to define the four sets
\begin{eqnarray}
s_1 &=& \{1,3,\ldots,l\}  \nonumber \\
s_2 &=& \{l+1,l+3,\ldots,L-1\}  \nonumber \\
s_3 &=& \{2,4,\ldots,l-1\}  \label{eq:A.setreg1} \\
s_4 &=& \{l+2,l+4,\ldots,L\}  \nonumber
\end{eqnarray}
where $l$ is given by (\ref{eq:A.defl}).
With these definitions the solution of the recurrence relation reads
\begin{eq}
\begin{array}{rll}
X_m^{a\,b\,b-1}=q^{mH(b,b,b-1)} \times \hspace{-2.3mm} &
\left\{
\begin{array}{l}
F_m^b(b) \\
q^{\half b} F_m^{b+1}(b) \\
q^{\half b} F_m^{b+1}(b) \\
F_m^b(b) + (1 \m q^m) q^b F_{m \m 1}^{b+2}(b \p1)
\end{array} \right. &
\begin{array}{l}
b \in s_1\setminus\{1\} \\
b \in s_2 \cup \{L \p 1\}  \\
b \in s_3 \\
b \in s_4
\end{array} \\
& & \\
X_m^{a\,b\,b}= q^{mH(b,b,b)} \times \hspace{-2.3mm} &
\left\{
\begin{array}{l}
F_m^{b}(b)
\\
q^{\half b} F_m^{b+1}(b) \\
q^{\half b} F_m^{b+1}(b) +(1 \m q^m) q^{-\half b}
F_{m \m 1}^{b-1}(b \m 1)  \\
F_m^b(b) + (1 \m q^m) q^b F_{m \m 1}^{b+2}(b \p1)
\end{array} \right. &
\begin{array}{l}
b \in s_1 \\
b \in s_2  \\
b \in s_3 \\
b \in s_4
\end{array} \\
& & \\
X_m^{a\,b\,b+1}= q^{mH(b,b,b+1)} \times \hspace{-2.3mm} &
\left\{
\begin{array}{l}
F_m^b(b)
\\
q^{\half b} F_m^{b+1}(b) \\
q^{\half b} F_m^{b+1}(b) +(1 \m q^m) q^{-\half b}
F_{m \m 1}^{b-1}(b \m 1) \\
F_m^b(b)
\end{array} \right. &
\begin{array}{l}
b \in s_1 \\
b \in s_2  \\
b \in s_3\cup \{0\} \\
b \in s_4\setminus\{L\}.
\end{array}
\end{array}
\hspace{-5mm}
\label{eq:A.solnR1}
\end{eq}
We note that in this solution we have included
the terms $X_m^{a\, 0\, 1}$ and $X_m^{a\, L+1\, L}$ which,
according to the confining condition (\ref{eq:A.confine}),
should be identically zero.
Using the simple relations (\ref{eq:A.elemprop})
it follows directly that
this is indeed the case.
The advantage of including these terms in the above way
is that it
enables us to prove the solution without
treating the boundary separately.
{}From the definition (\ref{eq:A.F}) it also follows immediately that
the initial condition (\ref{eq:A.initial}) is satisfied.

Proving that (\ref{eq:A.solnR1})
is indeed the solution of the recurrence relation is now straightforward
but rather tedious. Inserting (\ref{eq:A.solnR1})
into the recursion relation,
using the explicit form of the function $H$ as listed in equation
(\ref{eq:A.HfuncR1}),
we find that only the following four relations need hold:
\begin{eq}
\begin{array}{ll}
F_m^b(b)-F_{m-1}^b(b) \vphantom{\displaystyle \sum_a} &  \\
\qquad
= q^{m-1} \left[F_{m-1}^b(b \m 1)
+ q^{b+1} F_{m-1}^{b+2}(b \p 1)
+q^{-b+1} (1 \m q^{m-1}) F_{m-2}^{b-2}(b \m 2)\right]  &
b \in s_1  \\
F_m^{b+1}(b)-F_{m-1}^{b+1}(b)  \vphantom{\displaystyle \sum_a^a} &  \\
\qquad
= q^{m-1} \left[F_{m-1}^{b+1}(b \p 1)
+ q^{-b+1} F_{m-1}^{b-1}(b \m 1)
+ q^{b+1} (1 \m q^{m-1}) F_{m-2}^{b+3}(b \p 2)\right] &
b \in s_2  \\
F_m^{b+1}(b)-F_{m-1}^{b+1}(b \p 1) \vphantom{\displaystyle \sum_a^a} &  \\
\qquad
= q^{m-1} \left[F_{m-1}^{b+1}(b)
+ q^{-b+1} F_{m-1}^{b-1}(b \m 1)
+ q^{-b} (1 \m q^{m-1}) F_{m-2}^{b-1}(b \m 1)\right] &
b \in s_3  \\
F_m^b(b)-F_{m-1}^b(b \m 1)  \vphantom{\displaystyle \sum_a^a}  & \\
\qquad
= q^{m-1} \left[F_{m-1}^b(b)
+ q^{b+1} F_{m-1}^{b+2}(b \p 1)
+ q^b (1 \m q^{m-1}) F_{m-2}^{b+2}(b \p 1)\right] &
b \in s_4.
\end{array}
\hspace{-5mm}
\end{eq}
The restrictions on $b$ make these special cases of more general expressions,
which we will show to be true for all
values of $b$.
If we widen their applicability in this way,
the relations can in fact be combined to give simpler ones.
In turn, these relations can be further simplified by
requiring that they hold term-by-term in $j$. The resulting pair of
relations are much stronger requirements than the original set, but still
they hold.
Defining the function
\begin{equation}
f_m^s(b)=
\sum_{k=0}^{\infty} \left\{
q^{-\half \gamma s +k(k+\gamma-b)}
\Mult{m}{k}{k+\gamma-b}
-q^{\half \gamma s +k(k+\gamma+b)}
\Mult{m}{k}{k+\gamma+b} \right\}
\label{eq:A.smallf}
\end{equation}
where we have set $2(L+1)j+a=\gamma$,
these two relations are
\begin{eq}
f_m^b(b)-f_m^b(b \m 1)=q^m \left[ q^b f_{m-1}^{b+2}(b \p 1)
- q^{-b+1} f_{m-1}^{b-2}(b \m 2) \right]  \label{eq:A.two1}
\end{eq}
\begin{eq}
f_m^b(b)-f_{m-1}^b(b)=q^{m-1} \left[f_{m-1}^b(b \m 1)
+ q^{b+1} f_{m-1}^{b+2}(b \p 1)
+ q^{-b+1} (1 \m q^{m-1}) f_{m-2}^{b-2}(b \m 2) \right]\!.
\label{eq:A.two2}
\end{eq}

The function $f_m^s$ consists of two polynomials. One, the first term
within the curly braces, has only positive coefficients and the
other, the
second term, has only negative coefficients.
If we demand that the equations (\ref{eq:A.two1}) and
(\ref{eq:A.two2})
be satisfied for the
`positive' and `negative' polynomials independently, and we set
$\gamma\mp b=n$, respectively, this yields
\begin{eqnarray}
\lefteqn{\sum_{k=0}^{\infty} q^{k(k+n)} \Mult{m}{k}{k+n}
-\sum_{k=0}^{\infty} q^{k(k+n+1)} \Mult{m}{k}{k+n+1}}  \nonumber \\
& & \nonumber \\
& &=
q^{m-n}\sum_{k=0}^{\infty} q^{k(k+n-1)} \Mult{m-1}{k}{k+n-1}
-q^{m+n+1}\sum_{k=0}^{\infty} q^{k(k+n+2)} \Mult{m-1}{k}{k+n+2}
\label{eq:A.first}
\end{eqnarray}
for equation (\ref{eq:A.two1}), and
%\newpage
\begin{eqnarray}
\lefteqn{\sum_{k=0}^{\infty} q^{k(k+n)} \Mult{m}{k}{k+n}
-\sum_{k=0}^{\infty} q^{k(k+n)} \Mult{m-1}{k}{k+n}}  \nonumber \\
& & \nonumber \\
& &=
q^{m-1}\sum_{k=0}^{\infty} q^{k(k+n+1)} \Mult{m-1}{k}{k+n+1}
-q^{m-n}\sum_{k=0}^{\infty} q^{k(k+n-1)} \Mult{m-1}{k}{k+n-1}
\label{eq:A.second} \\
& & \nonumber \\
& & +
q^{m+n} (1-q^{m-1}) \sum_{k=0}^{\infty} q^{k(k+n+2)} \Mult{m-2}{k}{k+n+2}
\nonumber
\end{eqnarray}
for equation (\ref{eq:A.two2}).
The proof of these final two equations is now elementary.
Equation (\ref{eq:A.first}) follows immediately from
(\ref{eq:A.Gidena}) and
(\ref{eq:A.Gidenb}) if we set
$l=k+n$ and $l=k+n+1$, respectively. Equation
(\ref{eq:A.second}) follows from
(\ref{eq:A.Gidena}) with $l=k+n$.
%  #] Regime1:

%  #[ Regime2:
\subsubsection*{Regime 2$^+$}
Finding and proving the solution of the recurrence relation for
regime $2^+$
proceeds along similar lines as in regime $1^+$.
To remove any confusion
we will denote the one-dimensional configuration sums
in regime $2^+$ by $Y_m^{a\,b\,c}$.
First of all, the $q=1$ limit (\ref{eq:A.qis1}) is still valid.
Generating large polynomials reveals that in this case the configuration sums
yield Virasoro characters
with $s=a$ and $r=b$ for $b\leq l-1$, and $r=b-1$ for
$b\geq l+2$, or linear combinations of two characters.

We replace $h$ in (\ref{eq:A.Virasoro})
by $L+1$ according to (\ref{eq:A.Ltoh}) and
define the auxiliary function
\begin{eqnarray}
\setbox0=\hbox{$\scriptstyle +ar$}
\lefteqn{
G_m^r(b)= q^{(a-r)a/2}
\sum_{j,k=-\infty}^{\infty} } \nonumber \\
& & \nonumber \\
\lefteqn{\quad \qquad
\left\{ q^{(L+1)Lj^2-[(L+1)r-La]j+k[k+2(L+1) j+a-b]}
\Mult{m}{k}{k \p 2 (L \p 1) j \p a \m b} \right.}
\label{eq:A.G} \\
& & \nonumber \\
\lefteqn{\quad \qquad \left.
\hskip -\wd0
-q^{(L+1)Lj^2+[(L+1)r+La]j+ar+k[k+2(L+1)j+a+b]}
\Mult{m}{k}{k \p 2 (L \p 1) j \p a \p b} \right\} }
\setbox0=\hbox{}
\nonumber
\end{eqnarray}
which satisfies the following simple relations
\begin{eqnarray}
\lefteqn{G_m^r(b)=-G_m^{-r}(-b)}  \nonumber \\
\lefteqn{G_m^{L+r}(L \p 1 \p b)=-q^{-(L+1)r}
G_m^{L-r}(L+1-b)
\vphantom{\sum_a^a}}  \label{eq:A.elemprop2} \\
\lefteqn{G_m^1(0)=(1 \m q^m) G_{m-1}^1(1).} \nonumber
\end{eqnarray}
Again we define four sets
\begin{eqnarray}
t_1 &=& \{1,3,\ldots,l-1\}  \nonumber \\
t_2 &=& \{l+2,l+3,\ldots,L-1\}  \nonumber \\
t_3 &=& \{2,4,\ldots,l\}  \label{eq:A.setreg2} \\
t_4 &=& \{l+1,l+3,\ldots,L\}.  \nonumber
\end{eqnarray}
The solution of the recurrence relation then reads
\begin{eq}
\begin{array}{rll}
Y_m^{a\,b\,b-1}=q^{mH(b,b,b-1)} \times \hspace{-2.3mm} &
\left\{ \!\!
\begin{array}{l}
G_m^b(b) \\
q^{-\half b} G_m^{b-1}(b) \\
q^{-\half b} G_m^{b-1}(b) +(1 \m q^m) q^{\half b} G_{m-1}^{b+1}(b \p1)  \\
\hphantom{q^{-\half b} G_m^{b-1}(b)}-(1 \m q^m) q^{-1}
G_{m-1}^1(1) \delta_{b,2} \\
G_m^b(b)
\end{array} \right. &
\begin{array}{l}
b \in t_1\setminus\{1\} \\
b \in t_2 \cup \{L \p 1\}  \\  \\
b \in t_3 \\
b \in t_4
\end{array} \\
& & \\
Y_m^{a\,b\,b}=q^{mH(b,b,b)} \times \hspace{-2.3mm} &
\left\{ \!\!
\begin{array}{l}
G_m^b(b) \\
q^{-\half b} G_m^{b-1}(b) \\
q^{-\half b} G_m^{b-1}(b) +(1 \m q^m) q^{\half b} G_{m-1}^{b+1}(b \p 1) \\
G_m^b(b) +(1 \m q^m) q^{-b} G_{m-1}^{b-2}(b \m 1)
\end{array} \right. &
\begin{array}{l}
b \in t_1 \\
b \in t_2  \\
b \in t_3 \\
b \in t_4
\end{array} \\
& & \\
Y_m^{a\,b\,b+1}=q^{mH(b,b,b+1)} \times \hspace{-2.3mm} &
\left\{ \!\!
\begin{array}{l}
G_m^b(b) \\
q^{-\half b} G_m^{b-1}(b) \\
q^{-\half b} G_m^{b-1}(b) \\
G_m^b(b) +(1 \m q^m) q^{-b} G_{m-1}^{b-2}(b \m 1)
\end{array} \right. &
\begin{array}{l}
b \in t_1 \\
b \in t_2 \\
b \in t_3 \\
b \in t_4\setminus\{L\}.
\end{array}
\end{array}
\hspace{-5mm}
\label{eq:A.solnR2}
\end{eq}
As in regime $1^+$ we have extended the solution to include the term
$Y_m^{a\,L+1\,L}$, which, using (\ref{eq:A.elemprop2}),
indeed yields zero.
However, in contrast to regime $1^+$,
we can no longer extend the above solution
to the term $Y_m^{a\,0\,1}$ which according to
(\ref{eq:A.solnR2}) would not be zero.
Therefore, in proving the recurrence relation we have to treat cases that
involve this boundary term separately.
This `irregularity' is however compensated by the exceptional
terms $Y_{m}^{a\,2\,1}$ of (\ref{eq:A.solnR2})
and $H(1,2,1)$ of (\ref{eq:A.HfuncR2}).
As a result, inserting the above solution into the recurrence relation,
using (\ref{eq:A.elemprop2}),
we again obtain only four relations that should hold:
\begin{eq}
\begin{array}{ll}
G_m^b(b)-G_{m-1}^b(b) \vphantom{\displaystyle \sum_a} & \\
\qquad = q^{m-1}\! \left[G_{m-1}^b(b \p 1)
+ q^{-b+1} G_{m-1}^{b-2}(b \m 1)
+ q^{b+1} (1 \m q^{m-1}) G_{m-2}^{b+2}(b \p 2)\right] &
b \in t_1  \\
G_m^{b-1}(b)-G_{m-1}^{b-1}(b) \vphantom{\displaystyle \sum_a^a}& \\
\qquad
= q^{m-1}\! \left[G_{m-1}^{b-1}(b \m 1)
+ q^{b+1} G_{m-1}^{b+1}(b \p 1)
+ q^{-b+1} (1 \m q^{m-1}) G_{m-2}^{b-3}(b \m 2)\right] &
b \in t_2  \\
G_m^{b-1}(b)-G_{m-1}^{b-1}(b \m 1) \vphantom{\displaystyle \sum_a^a}& \\
\qquad
= q^{m-1}\! \left[G_{m-1}^{b-1}(b)
+ q^{b+1} G_{m-1}^{b+1}(b \p 1)
+ q^b (1 \m q^{m-1}) G_{m-2}^{b+1}(b \p 1)\right] &
b \in t_3  \\
G_m^b(b)-G_{m-1}^b(b \p 1) \vphantom{\displaystyle \sum_a^a}& \\
\qquad
= q^{m-1}\! \left[G_{m-1}^b(b)
+ q^{-b+1} G_{m-1}^{b-2}(b \m 1)
+ q^{-b} (1 \m q^{m-1}) G_{m-2}^{b-2}(b \m 1)\right] &
b \in t_4.
\end{array}
\end{eq}

Again these equations hold for all values of $b$ and we drop the
restrictions on $b$. Doing so, the four equations can be combined
to give simpler equations.
As before, these are true term-by-term in $j$.
Setting $2 (L+1) j + a = \gamma$, we find
\begin{diseq}
f_m^b(b)-f_m^b(b \p 1)=q^m \left[ q^{-b} f_{m-1}^{b-2}(b \m 1)
- q^{b+1} f_{m-1}^{b+2}(b \p 2) \right]
\end{diseq}

\begin{eq}
f_m^b(b)-f_{m-1}^b(b)=q^{m-1} \left[f_{m-1}^b(b \p 1)
+ q^{-b+1} f_{m-1}^{b-2}(b \m 1)
+ q^{b+1} (1 \m q^{m-1}) f_{m-2}^{b+2}(b \p 2) \right]
\end{eq}
with $f_m^s$ defined in (\ref{eq:A.smallf}).
Finally, making the same splitting into
`positive' and `negative' polynomials as before, now setting
$\gamma\mp b=-n$, yields precisely the equations
(\ref{eq:A.first}) and (\ref{eq:A.second}).
%  #] Regime2:
%  #[ Regime3&4:
\subsubsection*{Regimes 3$^+$ and 4$^+$}
As explained previously,
the form of the solution of the recurrence relation for
regimes $3^+$
and $4^+$ can simply be obtained by replacing $q$ with $1/q$ in
(\ref{eq:A.solnR2}) and (\ref{eq:A.solnR1}),
respectively. It should again be stressed
that the precise meaning of the variable $q$ in the various regimes is
not related, and is given by equation (\ref{eq:A.qdef}) in regimes $1^+$
and $2^+$, and by (\ref{eq:A.qdef34})in regimes 3$^+$ and 4$^+$.
%  #] Regime3&4:
%  #] Solution:
%  #[ ThermoLimit:
\subsection{Thermodynamic limit}\label{sec:A.thermo}
In this section, using the solutions
for the one-dimensional configuration sums
$X_m^{a\, b\, c}$ and $Y_m^{a\, b\, c}$, we obtain expressions
for the local height probabilities $P^{bc}(a)$.

\subsubsection*{Regime 1$^+$}
As described in section~\ref{sec:A.groundstate},
in regime 1$^+$ we have only ferromagnetic groundstate
configurations.
Hence we need to calculate
\begin{eq}
P^{bb}(a) = \lim_{m\rightarrow\infty}
\frac{q^{-a^2 \la/\pi} S(a) X_m^{a\,b\,b}(q)}
{\sum_{a=1}^L q^{-a^2 \la/\pi}
S(a) X_m^{a\,b\,b}(q)}  \qquad b=1,3,\ldots,l,l+1,l+3,\ldots,L-1.
\label{eq:A.pbbreg1}
\end{eq}

{}From the solution (\ref{eq:A.solnR1})
for the one-dimensional configuration sums, we find that we have to
consider the $m\to\infty$ limit of the auxiliary function $F_m^s$
defined in (\ref{eq:A.F}).
To do so, we need the result
\begin{equation}
\lim_{m\to\infty}
\sum_{k=-\infty}^\infty q^{k(k+a)}\Mult{m}{k}{k+a}_q
=\frac{1}{Q(q)}
\label{eq:A.limgpoly}
\end{equation}
which holds for arbitrary fixed $a$.
To establish this we take the limit
inside the sum and use the elementary result
\begin{equation}
\lim_{m\to\infty} \Mult{m}{k}{k+a}_q
=\lim_{m\to\infty}\;
\frac{(q)_m}
{(q)_k \, (q)_{k+a} \, (q)_{m-2k-a}}
=\frac{1}{(q)_k \, (q)_{k+a}}
\end{equation}
to obtain
\begin{equation}
\lim_{m\to\infty} \sum_{k=-\infty}^\infty
q^{k(k+a)}\Mult{m}{k}{k+a}_q
=\sum_{k=0}^\infty
\frac{q^{k(k+a)}}
{(q)_k\,  (q)_{k+a}}
=\frac{1}{Q(q)}\, .
\end{equation}
Here the last equality follows from the $q$-analogue of Kummer's theorem
\cite{Andrews}
\begin{equation}
\sum_{k=0}^\infty
\frac{q^{k(k-1)}z^k}
{(1-q)\ldots (1-q^{k})(1-z)(1-zq)\ldots (1-zq^{k-1})}
=\prod_{k=0}^\infty \frac{1}{1-zq^k}
\end{equation}
by setting $z=q^{a+1}$.
{}From the above considerations we conclude that,
in the $m\to\infty$ limit,
the auxiliary function
$F_m^s$ yields the Virasoro characters defined in equation
(\ref{eq:A.Virasoro})
\setlength{\mathindent}{0 cm}
\begin{eqnarray}
\lefteqn{
\lim_{m\to\infty} F_m^s(b)=
\frac{q^{\half(a-s)a}}{Q(q)}
\sum_{j=-\infty}^{\infty}
\left\{
q^{(L+2)(L+1)j^2+[(L+2)a-(L+1)s]j} \m
q^{(L+2)(L+1)j^2+[(L+2)a+(L+1)s]j+as}
\right\} }\nonumber \\
\lefteqn{ \hphantom{\lim_{m\to\infty} F_m^s(b) } \vphantom{\sum^a_a}
=q^{\half(a-s)a-\Delta_{a,s}^{(L+2)}+c/24}\:\chi_{a,s}^{(L+2)}(q)   } \\
\lefteqn{ \hphantom{\lim_{m\to\infty} F_m^s(b) }
\sim q^{a^2\la/\pi} \:\chi_{a,s}^{(L+2)}(q). } \nonumber
\end{eqnarray}
In the last step we have used
$\la=\pi L/ (4(L+1))$, and have
omitted terms independent of $a$.

Substituting the appropriate elements of the
solution (\ref{eq:A.solnR1}), the $X_m^{a\, b\, b}$, into the expression for
local height probabilities,
and using the limiting behaviour of $F_m^s$,
we find
\setlength{\mathindent}{\mathin}
\begin{equation}
P^{bb}(a)=
\frac{S(a) \chi_{a,s}^{(L+2)}(q)}
{\sum_{a=1}^L S(a) \chi_{a,s}^{(L+2)}(q)}
\end{equation}
where
\begin{equation}
s=\left\{
\begin{array}{ll}
b   & \qquad b = 1,3,\ldots,l  \\
b+1 & \qquad b = l+1,l+3,\ldots,L-1.
\end{array} \right.
\end{equation}
We note that $s$ takes the values $1,3,\ldots,L$.

By performing the conjugate modulus transformation (\ref{eq:A.congnome})
we rewrite the Virasoro characters as
\begin{equation}
\chi_{r,s}^{(h)} (q) = \frac{q^{-c/24-1/4h(h-1)}}{Q(q)}
\left(\frac{\epsilon'}{\pi}\right)^{1/2}
\left[ \td\! \left(\case{\pi}{2}
\left(\case{r}{h-1}\m\case{s}{h}\right),t\right)-
\td\! \left(\case{\pi}{2}
\left(\case{r}{h-1}\p\case{s}{h}\right),t\right)
\right]
\end{equation}
where
\begin{equation}
q^{h(h-1)} = \e^{-\pi^2/\epsilon'} \qquad \mbox{and}\qquad
t = \e^{-\epsilon'}.
\end{equation}
As a result the
local height probabilities can be written in the form
\begin{equation}
P^{bb}(a)=
\frac{S(a)}{N_1(s)} \left[
\td\!\left(\case{\pi}{2}\left(\case{a}{L+1} \m
\case{s}{L+2}\right),t \right) -
\td\!\left(\case{\pi}{2}\left(\case{a}{L+1} \p
\case{s}{L+2}\right),t \right)\right]
\label{eq:A.lhpreg1}
\end{equation}
where the normalisation factor in the denominator
is given by
\begin{eqnarray}
N_1(s)&=&\sum_{a=1}^L S(a) \left[
\td\left(\case{\pi}{2}
\left(\case{a}{L+1} \m \case{s}{L+2}\right),t \right) -
\td\left(\case{\pi}{2}
\left(\case{a}{L+1} \p \case{s}{L+2}\right),t \right)\right]
\nonumber\\
&=&\sum_{a=0}^{2L+1} S(a) \,
\td\left(\case{\pi}{2}
\left(\case{a}{L+1} \m \case{s}{L+2}\right),t \right).
\label{eq:A.N1}
\end{eqnarray}
We recall that the nome $q$ is defined in terms
of the nome $p$ as
$q = \exp(-12\pi\la/\epsilon)$.
This yields the following relation between the nomes $p$ and
$t$:
\begin{equation}
p=t^{3L(L+2)}.
\label{eq:A.pnomet}
\end{equation}

{}From the definition of the crossing factor
\begin{equation}
S(a)=(-1)^a \frac{\te\left(4a\la,p\right)}
{\tv\left(2a\la,p\right)}
\label{eq:A.newcross}
\end{equation}
and the simple identity
\begin{equation}
\frac{\te(2u,p)}{\tv(u,p)}=
\frac{\te(2u,p^2)\td(u,p)}{p^{1/4}Q(p^2)Q(p^4)}
\label{eq:A.alternS}
\end{equation}
we find an alternative form for $S$:
\begin{equation}
S(a) =
\te\!\left(\case{a\pi}{L+1},t^{6L(L+2)}\right)
\td\!\left(\case{a\pi L}{2(L+1)},t^{3L(L+2)}\right)
\label{eq:A.altS}
\end{equation}
where we have again neglected $a$-independent factors.
Substituting this back into the definition of the normalisation factor,
we find that the summation in $N_1(s)$ can actually be performed to give
\begin{eq}
\frac{S(a)}{N_1(s)}=\frac{ t^{-L(2L+3)/2}\:
\te\!\left(\case{a\pi}{L+1},t^{6L(L+2)}\right)
\td\!\left(\case{a\pi L}{2(L+1)},t^{3L(L+2)}\right)
\tt\!\left(\case{s\pi}{L+2},t^{6L(L+1)}\right)}
{2(L + 1)
Q\left(t^{12L(L+1)}\right)
Q\left(t^{12L(L+2)}\right)
\te\!\left(\case{\pi}{6},t^{2(L+1)(L+2)}\right)
\te\!\left(\case{2s\pi}{L+2},t^{6L(L+1)}\right)} \, .
\label{eq:A.N1den}
\end{eq}
The proof of this {\em denominator identity},
being rather technical and lengthy,
is given in appendix~\ref{app:A.den1}.

\subsubsection*{Regime 2$^+$}
The working for regime 2$^+$ is very similar to
that of regime 1$^+$.
Again we have only ferromagnetic groundstates, and we are interested
in calculating
\begin{eq}
P^{bb}(a) =  \lim_{m\rightarrow\infty}
\frac{q^{-a^2 \la/\pi} S(a) Y_m^{a\,b\,b}(q)}
{\sum_{a=1}^L q^{-a^2 \la/\pi}
S(a) Y_m^{a\,b\,b}(q)}  \; \quad b=1,3,\ldots,l-1,l+2,l+4,\ldots,L-1.
\label{eq:A.pbbreg2}
\end{eq}

{}From the solution (\ref{eq:A.solnR2}) for the one-dimensional
configuration sums we see that we have to take the limit
$m\to\infty$ of the auxiliary function $G_m^r$ defined in
(\ref{eq:A.G}). Using the result (\ref{eq:A.limgpoly}) yields
\begin{eqnarray}
\lefteqn{
\lim_{m\to \infty}G_m^r(b)=
\frac{q^{\half(a-r)a}}{Q(q)}
\sum_{j=-\infty}^{\infty}
\left\{
q^{(L+1)Lj^2-[(L+1)r-La]j} \m
q^{(L+1)Lj^2+[(L+1)r+La]j+ar}
\right\} } \nonumber \\
\lefteqn{ \hphantom{\lim_{m\to\infty} G_m^s(b) } \vphantom{\sum^a_a}
=q^{\half(a-r)a-\Delta_{r,a}^{(L+1)}+c/24} \: \chi_{r,a}^{(L+1)}(q) } \\
\lefteqn{ \hphantom{\lim_{m\to\infty} G_m^s(b) }
\sim q^{a^2\la/\pi} \:\chi_{r,a}^{(L+1)}(q) } \nonumber
\end{eqnarray}
where we have used that $\la=\pi(L+2)/(4(L+1))$.

Substituting the elements $Y_m^{a\, b\, b}$ of the
solution (\ref{eq:A.solnR2}),
and using the limiting behaviour of $G_m^r$,
we find that the local height probabilities are given by
\begin{equation}
P^{bb}(a)=
\frac{S(a) \chi_{r,a}^{(L+1)}(q)}
{\sum_{a=1}^L S(a) \chi_{r,a}^{(L+1)}(q)}
\end{equation}
where
\begin{equation}
r=\left\{
\begin{array}{ll}
b   & \qquad b = 1,3,\ldots,l-1  \\
b-1 & \qquad b = l+2,l+4,\ldots,L-1.
\end{array} \right.
\end{equation}
We note that $r$ takes the values $1,3,\ldots,L-2$.

If we use the conjugate modulus form of the Virasoro characters,
we can rewrite this as
\begin{equation}
P^{bb}(a)=
\frac{S(a)}{N_2(r)} \left[
\td\!\left(\case{\pi}{2}\left(\case{r}{L} \m
\case{a}{L+1}\right),t \right) -
\td\!\left(\case{\pi}{2}\left(\case{r}{L} \p
\case{a}{L+1}\right),t \right)\right]
\label{eq:A.lhpreg2}
\end{equation}
where the normalisation factor in the denominator
is
\begin{equation}
N_2(r)=\sum_{a=0}^{2L+1} S(a) \,
\td\left(\case{\pi}{2}
\left(\case{r}{L} \m \case{a}{L+1}\right),t \right).
\label{eq:A.N2}
\end{equation}
{}From the relation between the nomes $p$ and $q$
we find that $p$ and $t$ are again related
as in equation (\ref{eq:A.pnomet}). Inserting the
form (\ref{eq:A.altS}) for the crossing factor $S$ into the
expression for the normalisation factor $N_2$, we can carry out the
summation over $a$. The result, proved in
appendix~\ref{app:A.den2}, is
\begin{eq}
\frac{S(a)}{N_2(r)}=\frac{ t^{-(L+2)(2L+1)/2} \:
\te\!\left(\case{a\pi}{L+1},t^{6L(L+2)}\right)
\td\!\left(\case{a\pi L}{2(L+1)},t^{3L(L+2)}\right)
\tt\!\left(\case{r\pi}{L},t^{6(L+1)(L+2)}\right)}
{2(L + 1)
Q\left(t^{12L(L+2)}\right)
Q\left(t^{12(L+1)(L+2)}\right)
\te\!\left(\case{\pi}{ 6},t^{2L(L+1)}\right)
\te\!\left(\case{2r\pi}{L},t^{6(L+1)(L+2)}\right)} \, .
\nonumber
\label{eq:A.N2den}
\end{eq}

\subsubsection*{Regime 3$^+$}
The solution of the recurrence relation for regime 3$^+$ is
obtained by replacing $q$ by $1/q$ in the solution (\ref{eq:A.solnR2})
for regime 2$^+$.
We therefore
need to consider the auxiliary function
$G_m^r$, with $q$ replaced by $1/q$.
The effect of this replacement on the Gaussian
multinomials is given by
\begin{equation}
\Mult{m}{k}{l}_{1/q} = q^{k^2+l^2+kl-(k+l)m} \Mult{m}{k}{l}_q .
\label{eq:A.qto1onq}
\end{equation}
Applying this to $G_m^r$, we find
\setlength{\mathindent}{0 cm}
\begin{eqnarray}
\setbox0=\hbox{$\scriptstyle -ar$}
\lefteqn{
G_m^r(b)=q^{(r-a)a/2}\sum_{j,k=-\infty}^\infty} \nonumber \\
\lefteqn{
\hphantom{G_m^s(b)}
\times \left\{
q^{-(L+1)Lj^2+[(L+1)r-La]j
+k^2+[k+2(L+1)j+a-b]^2
-m[2k+2(L+1)j+a-b]} \!
\Mults{m}{k}{k+2(L+1)j+a-b} \right.} \\
& & \nonumber \\
\lefteqn{
\hphantom{G_m^s(b)\times}
\hskip -\wd0 \,
\left.
- q^{-(L+1)Lj^2-[(L+1)r+La]j-ar
+k^2+[k+2(L+1)j+a+b]^2
-m[2k+2(L+1)j+a+b]} \!
\Mults{m}{k}{k+2(L+1)j+a+b}
\right\} . } \nonumber
\setbox0=\hbox{}
\end{eqnarray}
In the first term we replace $k$ by
$-k-(L+1)j+(m-\mu_a-a+b)/2$ and in the second term we replace $k$ by
$-k-(L+1)j+(m-\mu_a-a-b)/2$, where $\mu_a=0, 1$ is
determined by
the requirement that
\begin{equation}
(m-\mu_a-a+b)/2 \in \Integer .
\end{equation}
The subscript of $\mu$, indicating its dependence on $a$, is included
for later convenience.
After simplification, we thus obtain
\begin{eqnarray}
\setbox0=\hbox{$\scriptstyle +as$}
\lefteqn{
G_m^{2b-s}(b)=q^{(b^2+\mu_a-m^2-as)/2}
\sum_{j,k=-\infty}^\infty q^{2k^2+2\mu_a k} }
\nonumber \\
\lefteqn{\hphantom{G_m^{2b-s}(b)}
\times\left\{
q^{(L+1)(L+2)j^2+[(L+2)a-(L+1)s]j}
\Mults{m}
{(m- \mu_a - a + b)/2 - k - (L+1)j}
{(m- \mu_a + a - b)/2 - k + (L+1)j} \right.}
\label{eq:A.invG}\\
& & \nonumber \\
\lefteqn{\hphantom{G_m^{2b-s}(b) \times} \hskip -\wd0 \,
\left.
 - q^{(L+1)(L+2)j^2+[(L+2)a+(L+1)s]j+as}
\Mults{m}
{(m - \mu_a - a - b)/2 - k -(L+1)j}
{(m - \mu_a + a + b)/2 - k +(L+1)j}
\right\}. } \nonumber
\setbox0=\hbox{}
\end{eqnarray}
\setlength{\mathindent}{\mathin}

\paragraph*{Antiferromagnetic phases}
In contrast to the previous two regimes, we now have antiferromagnetic
as well as ferromagnetic phases.
At first, we restrict our attention to the antiferromagnetic
groundstates, and calculate
\begin{equation}
P^{bb+1}(a) = \lim_{m\rightarrow\infty}
\frac{q^{a^2 \la/\pi} S(a) Y_m^{a\,b\,b+1}(q^{-1})}
{\sum_{a=1}^L q^{a^2 \la/\pi}
S(a) Y_m^{a\,b\,b+1}(q^{-1})}  \qquad b=1,2,4,6,\ldots,L-1.
\label{eq:A.pbbreg3a}
\end{equation}
It follows from the solution (\ref{eq:A.solnR2}) that we have to
find expressions for the function $G_m^r$ of equation
(\ref{eq:A.invG})
in the $m\to\infty$ limit.
To do so we use the result
\begin{eqnarray}
\lim_{m\to\infty} \Mult{2m}{m-a}{m-b}&=&\lim_{m\to\infty}
\frac{(q)_{2m}}{(q)_{m-a} \, (q)_{m+a}}\:
\frac{(q)_{m+a}}{(q)_{m-b} \, (q)_{a+b}}
\nonumber \\
&=&
\frac{1}{Q(q)}\:
\frac{1}{(q)_{a+b}}
\qquad\qquad a+b\ge 0 .
\label{eq:A.limitgp}
\end{eqnarray}
We therefore conclude that, in the limit $m\to\infty$,
the auxiliary function $G_m^{2b-s}$
is given by
a product of Virasoro characters
\begin{eqnarray}
\lefteqn{
\lim_{m\to\infty}q^{m^2/2}G_m^{2b-s}(b)=
q^{(b^2+\mu_a-as)/2}\frac{1}{Q(q)}
\sum_{k=0}^\infty \frac{q^{2k^2+2\mu_a k}}{(q)_{2k+\mu_a}} }
\nonumber\\
\lefteqn{ \qquad \qquad
\times \! \! \sum_{j=-\infty}^{\infty}
\left\{
q^{(L+2)(L+1)j^2+[(L+2)a-(L+1)s]j} \m
q^{(L+2)(L+1)j^2+[(L+2)a+(L+1)s]j+as}\right\} } \nonumber \\
\lefteqn{
\hphantom{\lim_{m\to\infty}q^{m^2/2}G_m^{2b-s}(b)}
= q^{\,(b^2-as)/2-\Delta_{a,s}^{(L+2)}+c/24+1/48}
\chi_{\mu_a/2}(q) \: \chi_{a,s}^{(L+2)}(q)
\vphantom{\sum_a^a} } \label{eq:A.limG3} \\
\lefteqn{
\hphantom{\lim_{m\to\infty}q^{m^2/2}G_m^{2b-s}(b)}
\sim q^{-a^2\la/\pi}
\chi_{\mu_a/2}(q) \: \chi_{a,s}^{(L+2)}(q)  }
\nonumber
\end{eqnarray}
with $\la = \pi(L+2)/(4(L+1))$. In the last step we have used
the following
Rogers-Ramanujan identity
\cite{Slater} for the
$c=\case{1}{2}$ characters $\chi_0$ and $\chi_{1/2}$:
\begin{equation}
\sum_{k=0}^{\infty}
\frac{q^{2k^2+2\mu_a k}}{(q)_{2k+\mu_a}}
=q^{1/48-\mu_a/2}\chi_{\mu_a+1,1}^{(4)}(q)
\equiv q^{1/48-\mu_a/2} \chi_{\mu_a/2}(q).
\end{equation}

In contrast to the ferromagnetic phases treated so far,
the local height probabilities for antiferromagnetic phases
depend on whether $m$ is taken to infinity
through odd or even values.
We therefore write $P^{bb+1}_{\sigma}(a)$, where
$\sigma$ is defined to be the parity of $m+b$
\begin{equation}
\sigma = m+b \bmod 2 .
\end{equation}
For $\mu_a$ this gives
\begin{equation}
\mu_a=\sigma +a \bmod 2 .
\end{equation}
Replacing $q$ by $1/q$ in the solution
(\ref{eq:A.solnR2}) for the one-dimensional
configuration sums and using the result
(\ref{eq:A.limG3}), we find
\begin{equation}
P^{bb+1}_\sigma(a)=
\frac{S(a) \chi_{\mu_a/2}(q) \chi_{a,s}^{(L+2)}(q)}
{\sum_{a=1}^L S(a) \chi_{\mu_a/2}(q) \chi_{a,s}^{(L+2)}(q)}
\end{equation}
where
\begin{equation}
s=\left\{
\begin{array}{ll}
b   & \qquad b=1 \\
b+1 & \qquad b=2,4,\ldots,L-1.
\end{array} \right.
\end{equation}
We note that $s$ takes the values $1,3,\ldots,L$.

Performing a conjugate modulus transformation,
 we find that the local height
probabilities are given by
\begin{equation}
P^{bb+1}_{\sigma}(a)=
\frac{S(a)\chi_{\mu_a/2}}{N_3(s)}
\left[\td\!\left(\case{\pi}{2}\left(\case{a}{L+1} \m
\case{s}{L+2}\right),t \right) \m
\td\!\left(\case{\pi}{2}\left(\case{a}{L+1}\p
\case{s}{L+2}\right),t\right)\right]
\label{eq:A.lhpreg3anti}
\end{equation}
where the normalisation factor is defined as
\begin{equation}
N_3(s)
=\sum_{a=0}^{2L+1} S(a)\, \chi_{\mu_a/2} \,
\td\!\left(\case{\pi}{2}
\left(\case{a}{L+1}\m \case{s}{L+2}\right),t \right).
\end{equation}
{}From the relation between the nome $p=\exp(-\epsilon)$ and
the nome $q=\exp(-4\pi(\pi-3\la)/\epsilon)$
we obtain
\begin{equation}
p=t^{L^2-4}.
\end{equation}
We take the conjugate modulus transformation (\ref{eq:A.congnome})
of the $c=\case{1}{2}$
characters after first rewriting them using
the formula \cite{Slater}
\begin{equation}
\chi_{\mu_a/2}(q)={q^{\mu_a/2-1/48} \over Q(q^2)} E(-q^{3-2\mu_a},q^8).
\end{equation}
Furthermore, we rewrite the crossing factor $S$
by using the identity (\ref{eq:A.alternS}) as well as the
relation between the nomes $p$ and $t$. Substituting this
into the defining relation of the
normalisation factor $N_3$, we can perform the summation, yielding
\begin{eqnarray}
\lefteqn{
\frac{S(a)\chi_{\mu_a/2}}{N_3(s)}=
\frac{t^{- (L - 1)(L + 2)/4}
\td\!\left(\case{(1+2\mu_a)\pi}{8},t^{(L+1)(L+2)/4}\right)}
{2(L + 1)
Q\left(t^{4(L^2-4)}\right)
Q\left(t^{2(L+1)(L+2)}\right)} } \nonumber \\
& & \nonumber \\
& & \qquad \qquad \qquad \qquad \qquad \times \:
\frac{
\te\!\left(\case{a\pi}{L+1},t^{2(L^2-4)}\right)
\td\!\left(\case{a\pi L}{2(L+1)},t^{L^2-4}\right)}
{\td\!\left(\case{\pi}{4},t^{(L+1)(L+2)}\right)
\te\!\left(\case{s\pi}{L+2},t^{(L-2)(L+1)}\right)} \, .
\label{eq:A.N3denanti}
\end{eqnarray}
The proof of the denominator identity is presented in
appendix~\ref{app:A.den3}.

\paragraph*{Ferromagnetic phases}
For the ferromagnetic phases of regime 3$^+$
the local height probabilities are given by
\begin{equation}
P^{bb}(a) = \lim_{m\rightarrow\infty}
\frac{q^{a^2 \la/\pi} S(a) Y_m^{a\,b\,b}(q^{-1})}
{\sum_{a=1}^L q^{a^2 \la/\pi}
S(a) Y_m^{a\,b\,b}(q^{-1})}  \qquad b=2,4,\ldots,l,l+1,l+3,\ldots,L.
\label{eq:A.pbbreg3b}
\end{equation}
{}From the expressions for $Y_m^{a\,b\,b}$ in
equation (\ref{eq:A.solnR2}) it follows that we have to
consider the following combinations of the
function $G_m^r$ in the $m\to\infty$ limit:
\begin{eqnarray}
\lefteqn{ \vphantom{\sum_a}
G_{m}^{b-1}(b) - q^{-m-b} G_{m-1}^{b+1}(b+1)
\qquad \quad b=2,4,\ldots,l }
\nonumber \\
\lefteqn{
G_{m}^{b}(b) - q^{-m+b} G_{m-1}^{b-2}(b-1)
\qquad \quad \; \,  b=l+1,l+3,\ldots,L . }
\nonumber
\end{eqnarray}
Here $G_m^r$ is given by equation (\ref{eq:A.invG}) and we have neglected
terms that have an extra factor $q^m$, which do not contribute in the large
$m$ limit.
We again take $m\to\infty$ through even or odd values of $m$
with $a$ and $b$
fixed and $\mu_a=0,1$ chosen accordingly.
For this we need the result
\begin{eqnarray}
\lefteqn{
\lim_{m\to\infty}q^{-m}\left\{ \Mult{2m}{m-a-b}{m-a+b} \m
\Mult{2m-1}{m-a-b-1}{m-a+b}\right\} } \nonumber \\
& & \nonumber \\
\lefteqn{\qquad \qquad \qquad \qquad
=\lim_{m\to\infty}q^{-m}
\left\{1 \m \frac{1-q^{m-a-b}}{1-q^{2m}}\right\}
\Mult{2m}{m-a-b}{m-a+b} }
\nonumber \\
& & \nonumber \\
\lefteqn{\qquad \qquad \qquad \qquad
=\lim_{m\to\infty}
q^{-a-b}\Mult{2m}{m-a-b}{m-a+b} }
\label{eq:A.pairlim}\\
& & \nonumber \\
\lefteqn{\qquad \qquad \qquad \qquad
= \frac{q^{-a-b}}{Q(q)\,(q)_{2a}} }
\nonumber
\end{eqnarray}
where we have used equation (\ref{eq:A.limitgp}).
If we apply this limit to the combination of $G_m^r$ functions,
it follows that
\begin{eqnarray}
\lefteqn{
\lim_{m\to\infty}
q^{(m^2-m+b)/2} \left[
G_{m}^{b-1}(b) - q^{-m-b} G_{m-1}^{b+1}(b+1)
\right] }
\nonumber \\
\lefteqn{ \qquad =
q^{(b-a)b/2}\frac{1}{Q(q)}
\sum_{k=0}^{\infty}
\frac{q^{2k^2+(2\mu_a-1) k}}
{(q)_{2k+\mu_a}} }
\nonumber\\
\lefteqn{ \qquad \quad \times  \! \!
\sum_{j=-\infty}^{\infty}
\left\{q^{(L+2)(L+1)j^2+[(L+2)a-(L+1)b]j}
-q^{(L+2)(L+1)j^2+[(L+2)a+(L+1)b]j+ab}\right\} }
\label{eq:A.nomudep}\\
\lefteqn{
\qquad = q^{(b-a)b/2-\Delta_{a,b}^{(L+2)}+c/24-1/24}
\chi_{1/16}(q) \chi_{a,b}^{(L+2)}(q) }
\vphantom{\sum_a^a}
\nonumber \\
\lefteqn{ \qquad
\sim q^{-a\la/\pi} \chi_{a,b}^{(L+2)}(q). }
\nonumber
\end{eqnarray}
To obtain this result, we have used the
Rogers-Ramanujan identity for the
$c=\case{1}{2}$ character $\chi_{1/16}$:
\begin{equation}
\sum_{k=0}^\infty
\frac{q^{2k^2+(2\mu_a-1) k}}
{(q)_{2k+\mu_a}}
=\frac{Q(q^2)}{Q(q)}
=q^{-1/24}\chi_{1,2}^{(4)}(q)
\equiv q^{-1/24}\chi_{1/16}(q).
\label{eq:A.RR16}
\end{equation}
The left-hand side of this identity does not depend on the
value of $\mu_a$, and hence the result (\ref{eq:A.nomudep})
is independent of the parity of $m$. For a ferromagnetic phase
this is indeed what one would expect.
Similarly we find that
\begin{equation}
\lim_{m\to\infty}q^{(m^2-m-b)/2} \left[
G_{m}^{b}(b) - q^{-m+b} G_{m-1}^{b-2}(b-1) \right]
\sim q^{-a\la/\pi} \chi_{a,b+1}^{(L+2)}(q).
\end{equation}

Replacing $q$ by $1/q$ in the solution (\ref{eq:A.solnR2}) for
$Y_m^{a\,b\,c}$, and
substituting the above results, we get
\begin{equation}
P^{bb}(a) =
\frac{S(a) \chi_{a,s}^{(L+2)} (q)}
{\sum_{a=1}^L S(a) \chi_{a,s}^{(L+2)} (q)}
\end{equation}
where
\begin{equation}
s=\left\{
\begin{array}{ll}
b   & \qquad b=2,4,\ldots,l \\
b+1 & \qquad b=l+1,l+3,\ldots,L.
\end{array} \right.
\end{equation}
We note that $s$ takes the values $2,4,\ldots,L+1$.

If we continue as for the antiferromagnetic phases, we arrive at
the final result
\begin{equation}
P^{bb}(a)=
\frac{S(a)}{N_4(s)}
\left[\td\!\left(\case{\pi}{2}\left(\case{a}{L+1} \m
\case{s}{L+2}\right),t \right) \m
\td\!\left(\case{\pi}{2}\left(\case{a}{L+1}\p
\case{s}{L+2}\right),t\right)\right]
\label{eq:A.lhpreg3ferro}
\end{equation}
where the term occurring in the denominator reads
\begin{equation}
N_4(s)=\sum_{a=0}^{2L+1} S(a) \,
\td\!\left(\case{\pi}{2}
\left(\case{a}{L+1} \m \case{s}{L+2}\right),t \right).
\end{equation}
Again the summation over $a$ can be carried out, yielding
\begin{eqnarray}
\lefteqn{
\frac{S(a)}{N_4(s)}=
\frac{t^{-L(L+2)/2} \:
\te\!\left(\case{\pi}{4},t^{(L+1)(L+2)}\right)
\te\!\left(\case{\pi}{4},t^{(L-2)(L+1)}\right)}
{4(L + 1)
Q^2\left(t^{4(L+1)(L+2)}\right)
Q\left(t^{4(L^2-4)}\right)}} \nonumber \\
& & \nonumber \\
& & \qquad \qquad \qquad \qquad \qquad \times \:
\frac{
\te\!\left(\case{a\pi}{L+1},t^{2(L^2-4)}\right)
\td\!\left(\case{a\pi L}{2(L+1)},t^{L^2-4}\right)}
{\te\!\left(\case{s\pi}{L+2},t^{2(L-2)(L+1)}\right)
\td\!\left(\case{s\pi}{L+2},t^{2(L-2)(L+1)}\right)} \, .
\label{eq:A.N4den}
\end{eqnarray}
A proof of the denominator identity leading to this result is given in
appendix~\ref{app:A.den3}.

\subsubsection*{Regime 4$^+$}
The solution of the recurrence relation for regime 4$^+$ is
obtained by replacing $q$ by $1/q$ in the solution (\ref{eq:A.solnR1})
for regime 1$^+$.
Consequently we
need to consider the auxiliary function
$F_m^r$, with $q$ replaced with $1/q$.
Using the inversion formula (\ref{eq:A.qto1onq}) and
carrying out the same sequence of transformations as for
regime 3$^+$, we obtain
\begin{eqnarray}
\setbox0=\hbox{$\scriptstyle +ar$}
\lefteqn{
F_m^{2b-r}(b)=q^{(b^2+\mu_a-m^2-ar)/2}
\sum_{j,k=-\infty}^\infty q^{2k^2+2\mu_a k} }
\nonumber \\
\lefteqn{\hphantom{F_m^{2b-r}(b)}
\times\left\{
q^{(L+1)Lj^2-[(L+1)r-La]j}
\Mults{m}
{(m- \mu_a - a + b)/2 - k - (L+1)j}
{(m- \mu_a + a - b)/2 - k + (L+1)j} \right.}
\label{eq:A.invF}\\
& & \nonumber \\
\lefteqn{\hphantom{F_m^{2b-r}(b) \times} \hskip -\wd0 \,
\left.
- q^{(L+1)Lj^2+[(L+1)r+La]j+ar}
\Mults{m}
{(m - \mu_a - a - b)/2 - k -(L+1)j}
{(m - \mu_a + a + b)/2 - k +(L+1)j}
\right\}. } \nonumber
\setbox0=\hbox{}
\end{eqnarray}
\paragraph*{Antiferromagnetic phases}
We again have antiferromagnetic as well as ferromagnetic pha\-ses.
Both types admit
treatment akin to that applied in regime 3$^+$. In the following we
therefore leave out some of the details.
We begin by treating the antiferromagnetic
groundstates and calculate
\begin{equation}
P^{bb+1}(a) = \lim_{m\rightarrow\infty}
\frac{q^{a^2 \la/\pi} S(a) X_m^{a\,b\,b+1}(q^{-1})}
{\sum_{a=1}^L q^{a^2 \la/\pi}
S(a) X_m^{a\,b\,b+1}(q^{-1})}  \qquad b=1,3\ldots,L-2.
\label{eq:A.pbbreg4a}
\end{equation}
{}From the solution (\ref{eq:A.solnR1}) we see that we have to
obtain the $m\to\infty$ limit of the
function $F_m^s$ in equation
(\ref{eq:A.invF}).
We find that in this limit we get a product of Virasoro characters
\begin{eqnarray}
\lefteqn{
\lim_{m\to\infty}q^{m^2/2}F_m^{2b-r}(b)
\sim q^{-a^2\la/\pi}
\chi_{\mu_a/2}(q) \: \chi_{r,a}^{(L+1)}(q)  }
\nonumber
\end{eqnarray}
where $\la = \pi L/(4(L+1))$.
For the local height probabilities this gives
\begin{equation}
P^{bb+1}_\sigma(a)=
\frac{S(a) \chi_{\mu_a/2}(q) \chi_{r,a}^{(L+1)}(q)}
{\sum_{a=1}^L S(a) \chi_{\mu_a/2}(q) \chi_{r,a}^{(L+1)}(q)}
\label{eq:prodstruc}
\end{equation}
where
\begin{equation}
r=b \qquad b=1,3,\ldots,L-2.
\end{equation}
After a conjugate modulus transformation this gives
\begin{equation}
P^{bb+1}_{\sigma}(a)=
\frac{S(a)\chi_{\mu_a/2}}{N_5(r)}
\left[\td\!\left(\case{\pi}{2}\left(
\case{r}{L} \m \case{a}{L+1}
\right),t \right) \m
\td\!\left(\case{\pi}{2}\left(
\case{r}{L} \p \case{a}{L+1}
\right),t\right)\right]
\label{eq:A.lhpreg4anti}
\end{equation}
with the following normalisation factor:
\begin{equation}
N_5(r)
=\sum_{a=0}^{2L+1} S(a)\, \chi_{\mu_a/2} \,
\td\!\left(\case{\pi}{2}
\left(\case{r}{L}\m\case{a}{L+1} \right),t \right) .
\end{equation}
{}From the relation between the nome $p$ and $q$ we get
\begin{equation}
p=t^{L(L+4)}.
\end{equation}
Using the result of appendix~\ref{app:A.den4} for the
normalisation factor $N_5$, we finally obtain
\begin{eq}
\frac{S(a)\chi_{\mu_a/2}}{N_5(r)}=
\frac{t^{- L(L + 3)/4}
\td\!\left(\case{(1+2\mu_a)\pi}{8},t^{L(L+1)/4}\right)
\te\!\left(\case{a\pi}{L+1},t^{2L(L+4)}\right)
\td\!\left(\case{a\pi L}{2(L+1)},t^{L(L+4)}\right)}
{2(L \p 1)
Q\left(t^{4L(L+4)}\right)
Q\left(t^{2L(L+1)}\right)
\td\!\left(\case{\pi}{4},t^{L(L+1)}\right)
\te\!\left(\case{r\pi}{L},t^{(L+1)(L+4)}\right)} \, .
\label{eq:A.N5den}
\end{eq}

\paragraph*{Ferromagnetic phases}
For the ferromagnetic phases of regime 4$^+$,
the local height probabilities are
\begin{equation}
P^{bb}(a) = \lim_{m\rightarrow\infty}
\frac{q^{a^2 \la/\pi} S(a) X_m^{a\,b\,b}(q^{-1})}
{\sum_{a=1}^L q^{a^2 \la/\pi}
S(a) X_m^{a\,b\,b}(q^{-1})}  \quad b=2,4,\ldots,l-1,l+2,l+4,\ldots,L.
\label{eq:A.pbbreg4b}
\end{equation}
{}From the solution for the configuration sums $X_m^{a\,b\,b}$, as listed
in equation (\ref{eq:A.solnR1}), it follows that we have to take the
the $m\to\infty$ limit of the following combinations of
the function $F_m^s$:
\begin{eqnarray}
\lefteqn{ \vphantom{\sum_a}
F_{m}^{b+1}(b) - q^{-m+b} F_{m-1}^{b-1}(b-1)
\qquad \; \; b=2,4,\ldots,l-1 }
\nonumber \\
\lefteqn{
F_{m}^{b}(b) - q^{-m-b} F_{m-1}^{b+2}(b+1)
\qquad \quad \, b = l+2,l+4,\ldots,L. }
\end{eqnarray}
where $F_m^r$ is given by equation (\ref{eq:A.invF}).
The infinite limit can easily be taken using
(\ref{eq:A.pairlim}) and (\ref{eq:A.RR16}), and we arrive at
\begin{eqnarray}
\lefteqn{
\lim_{m\to\infty}
q^{(m^2-m-b)/2} \left[
F_{m}^{b+1}(b) - q^{-m+b} F_{m-1}^{b-1}(b-1)
\right]
\sim q^{-a\la/\pi} \chi_{b,a}^{(L+1)}(q) }
\nonumber \\
& & \nonumber \\
\lefteqn{
\lim_{m\to\infty}
q^{(m^2-m+b)/2} \left[
F_{m}^{b}(b) - q^{-m-b} F_{m-1}^{b+2}(b+1)
\right]
\sim q^{-a\la/\pi} \chi_{b-1,a}^{(L+1)}(q). }
\end{eqnarray}
Again we observe that the dependence on the parity of $m$ has dropped out.

Replacing $q$ by $1/q$ in the solution (\ref{eq:A.solnR1}) for
$X_m^{a\,b\,b}$ and
substituting the above results, we get
\begin{equation}
P^{bb}(a) =
\frac{S(a) \chi_{r,a}^{(L+1)} (q)}
{\sum_{a=1}^L S(a) \chi_{r,a}^{(L+1)} (q)}
\end{equation}
where
\begin{equation}
r=\left\{
\begin{array}{ll}
b   & \qquad b=2,4,\ldots,l-1 \\
b-1 & \qquad b=l+2,l+4,\ldots,L.
\end{array} \right.
\end{equation}
We note that $r$ takes the values $2,4,\ldots,L-1$.

Taking a conjugate modulus transformation, we obtain
the final result
\begin{equation}
P^{bb}(a)=
\frac{S(a)}{N_6(r)}
\left[\td\!\left(\case{\pi}{2}\left(\case{r}{L} \m
\case{a}{L+1}\right),t \right) \m
\td\!\left(\case{\pi}{2}\left(\case{r}{L}\p
\case{a}{L+1}\right),t\right)\right]
\label{eq:A.lhpreg4ferro}
\end{equation}
with the following normalisation factor:
\begin{equation}
N_6(r)=\sum_{a=0}^{2L+1} S(a) \,
\td\!\left(\case{\pi}{2}
\left(\case{r}{L} \m \case{a}{L+1}\right),t \right).
\end{equation}
Performing the sum over $a$, using the denominator identity for $N_6$ in
appendix~\ref{app:A.den4}, yields
\begin{eqnarray}
\lefteqn{
\frac{S(a)}{N_6(r)}=
\frac{t^{- L(L+2)/2}
\te\!\left(\case{\pi}{4},t^{L(L+1)}\right)
\te\!\left(\case{\pi}{4},t^{(L+1)(L+4)}\right)}
{4(L + 1)
Q^2\left(t^{4L(L+1)}\right)
Q\left(t^{4L(L+4)}\right)} } \nonumber \\
& & \nonumber \\
& &
\qquad \qquad \qquad \qquad \qquad \times \:
\frac{
\te\!\left(\case{a\pi}{L+1},t^{2L(L+4)}\right)
\td\!\left(\case{a\pi L}{2(L+1)},t^{L(L+4)}\right)}
{\te\!\left(\case{r\pi}{L},t^{2(L+1)(L+4)}\right)
\td\!\left(\case{r\pi}{L},t^{2(L+1)(L+4)}\right)}\, .
\label{eq:A.N6den}
\end{eqnarray}

%  #] ThermoLimit:
%  #] LHP:

%  #[ exponenten:
\nsection{Order Parameters and Critical Exponents}
Following Huse~\cite{Huse},
we define generalised order
parameters in terms of the local height
probabilities obtained in
the previous section:
\begin{equation}
R_{k,\pm}^{b c}=\sum_{a=1}^L \:
\frac{
\sin\!\left(\frac{(k+1)a\pi}{L+1}\right)}{
\sin\!\left(\frac{a\pi}{L+1}\right)} \:
P^{bc}_{\pm}(a)
\label{eq:GOP}
\end{equation}
where $k=0,1,2,\ldots,L-1$ and
\begin{equation}
P^{bc}_{\pm}(a) = \frac{1}{2}
\left[P_1^{bb+1}(a)\pm P_0^{bc}(a)\right].
\label{eq:clhp}
\end{equation}
For $k=0$ these order parameters are trivial
because
$R_{0,+}^{bc}=1$ and $R_{0,-}^{bc}=0$. For ferromagnetic phases
there is no distinction between the two sublattice
probabilities, and so $R_{k,-}^{bb}=0$.

We are not able in general to perform the sum in (\ref{eq:GOP}).
To determine the
associated critical exponents we therefore expand the
local height probabilities for
the four regimes in powers of $t$, and hence of $p$.
In this process the critical values of the local
height probabilities may also obtained, and are given by
\begin{equation}
P_{\rm crit}^{bc}(a)=\frac{2}{L+1} \:
\sin^2 \! \left(\case{a\pi}{L+1}\right)
\end{equation}
in all regimes, and are independent of the boundary spins $b$
and $c$.

Using the exponents obtained via the free energy,
we extract the conformal weights corresponding to the
critical exponents of the generalised order parameters.

\subsection{Off-critical perturbation}
We first identify the scaling field corresponding to the elliptic
nome $p$.
{}From the scaling relations
\begin{equation}
\Delta_p = \frac{1-\alpha}{2-\alpha}
\qquad \qquad \qquad \Delta_p = \frac{1}{1+\delta}
\end{equation}
and the exponents $\alpha$ and $\delta$ obtained in
section~\ref{sec:A.free},
we find that the conformal weight of the perturbing field
is
\begin{equation}
\Delta_p=
\left\{\begin{array}{ll}
\Delta_{2,1}^{(L+2)}=\frac{L+4}{4(L+1)}& \vphantom{\displaystyle \sum_a^a}
\qquad \mbox{regime 1$^{\pm}$} \\
\Delta_{1,2}^{(L+1)}=\frac{L-2}{4(L+1)}& \vphantom{\displaystyle \sum_a^a}
\qquad \mbox{regime 2$^{\pm}$} \\
\Delta_{2,1}^{(4)}+\Delta_{2,1}^{(L+2)}
 =\frac{1}{2}+\frac{L+4}{4(L+1)}& \vphantom{\displaystyle \sum_a^a}
\qquad \mbox{regime 3$^{\pm}$} \\
\Delta_{2,1}^{(4)}+\Delta_{1,2}^{(L+1)}
 =\frac{1}{2}+\frac{L-2}{4(L+1)}& \vphantom{\displaystyle \sum_a^a}
\qquad \mbox{regime 4$^{\pm}$}.
\end{array} \right.
\label{eq:A.deltap}
\end{equation}
We therefore conclude that the nome
$p$ corresponds to a perturbation in the direction of
the spinless operator $\phi_p$, with
\begin{equation}
\phi_p=
\left\{ \begin{array}{ll}
\phi_{2,1}^{(L+2)}=\phi_{L-1,L+1}^{(L+2)}&
\vphantom{\displaystyle \sum_a^a}
\qquad \mbox{regime 1$^{\pm}$} \\
\phi_{1,2}^{(L+1)}=\phi_{L-1,L-1}^{(L+1)}&
\vphantom{\displaystyle \sum_a^a}
\qquad \mbox{regime 2$^{\pm}$} \\
 \phi_{2,1}^{(4)} \, \phi_{2,1}^{(L+2)}
=\phi_{2,1}^{(4)} \, \phi_{L-1,L+1}^{(L+2)}&
\vphantom{\displaystyle \sum_a^a}
\qquad \mbox{regime 3$^{\pm}$} \\
 \phi_{2,1}^{(4)} \, \phi_{1,2}^{(L+1)}
=\phi_{2,1}^{(4)} \, \phi_{L-1,L-1}^{(L+1)}&
\vphantom{\displaystyle \sum_a^a}
\qquad \mbox{regime 4$^{\pm}$}.
\end{array} \right.
\end{equation}
Here $\phi^{(h)}_{r,s}$ denotes the operator of the unitary minimal model
with central charge
\begin{equation}
c=1-\frac{6}{h(h-1)} \, .
\end{equation}

\subsection{Order parameters for regime 1$^+$}
Because this is a ferromagnetic regime, the only non-zero order parameters
are
\begin{equation}
R_{k,+}^{bb}\equiv R_k^{bb}=\sum_{a=1}^L \:
\frac{
\sin\!\left(\frac{(k+1)a\pi}{L+1}\right)}{
\sin\!\left(\frac{a\pi}{L+1}\right)} \:
P^{bb}(a)
\label{eq:A.orderHuse}
\end{equation}
where $k=0,1,2,\ldots,L-1$.
To evaluate the associated
critical exponents, we expand equation (\ref{eq:A.N1den}) using the
representation (\ref{eq:A.thetasums}) of the
$\vartheta$-functions as a series of trigonometric functions.
In the $t\to 0$ limit this yields
\begin{equation}
\frac{S(a)}{N_1(s)}=\frac{1}{2(L + 1) t} \:
\frac{
\sin\!\left(\frac{a\pi}{L+1}\right)}{
\sin\!\left(\frac{s\pi}{L+2}\right)}
\left[1
+O\!\left(t^{3 L(L+2)}\right)\right].
\end{equation}
Expanding also the $\td$-functions in the expression
(\ref{eq:A.lhpreg1})
for the local height probabilities, and using
a simple trigonometric identity for the
difference of cosines, gives
\begin{equation}
P^{bb}(a)=
\frac{2}{L+1}\:
\frac{
\sin\!\left(\frac{a\pi}{L+1}\right)}{
\sin\!\left(\frac{s\pi}{L+2}\right)}
\sum_{n=1}^{\infty} t^{n^2-1}
\sin\!\left(\case{a\pi n}{L+1}\right) \,
\sin\!\left(\case{s\pi n}{L+2}\right)
+ O\!\left(t^{3L(L+2)}\right).
\end{equation}
Setting $n=1$ we obtain the aforementioned result for the
critical values of the local height
probabilities (\ref{eq:clhp}).

If we substitute the above expansion
into the expression for the order parameters,
then use the orthogonality relation
\begin{equation}
\frac{2}{L+1}\ \sum_{a=1}^L
\sin\!\left(\case{a\pi n}{L+1}\right) \,
\sin\!\left(\case{a\pi m}{L+1}\right)=
\sum_{k=-\infty}^{\infty} \left[
\delta_{n-m,2k(L+1)}-
\delta_{n+m,2k(L+1)}
\right]
\label{eq:A.ortho}
\end{equation}
and interchange the
order of summation, we find
\begin{equation}
R_k^{bb}=
\frac{1}{
\sin\!\left(\frac{s\pi}{L+2}\right)} \:
\sum_{m=-\infty}^{\infty}
t^{[k+1+2m(L+1)]^2-1}\sin\!\left(\case{[k+1+2m(L+1)]s\pi}{L+2}\right)
+ O\!\left(t^{3L(L+2)}\right).
\end{equation}
To leading order we thus find
\begin{equation}
R_k^{bb}\sim
\frac{
\sin\!\left(\frac{(k+1)s\pi}{L+2}\right)}{
\sin\!\left(\frac{s\pi}{L+2}\right)} \:
t^{(k+1)^2-1}
\end{equation}
Hence, for $k\geq 1$, the order parameters vanish at
criticality with a power law behaviour
\begin{equation}
R_k^{bb}\sim p^{1/\delta_k}
\label{eq:A.powerlaw}
\end{equation}
where the critical exponents $\delta_k$ are given by
\begin{equation}
\delta_k=\frac{3L(L+2)}{(k+1)^2-1}  \qquad k=1,2,\ldots,L-1.
\end{equation}
Here we have used the fact that $p=t^{3L(L+2)}$.
To find the corresponding conformal weights $\Delta_k$
we use the scaling relation
\begin{equation}
\Delta_k=\frac{1-\Delta_p}{\delta_k} .
\label{eq:A.scaling}
\end{equation}
{}From this relation we get
\begin{equation}
\Delta_k=
\frac{(k+1)^2-1}{4(L+2)(L+1)} = \Delta_{k+1,k+1}^{(L+2)}
\qquad  k=1,2,\ldots,L-1.
\end{equation}
Thus we have obtained all `diagonal' weights of the Kac table
(\ref{eq:A.Kactable}).

\subsection{Order parameters for regime 2$^+$}
To obtain the generalised order parameters for regime 2$^+$
we closely follow the previous working.
The non-zero order parameters are again given by
(\ref{eq:A.orderHuse}), with $k=0,\ldots,L-1$.
Expanding the results
(\ref{eq:A.lhpreg2}) and (\ref{eq:A.N2den}) for
the local height probabilities, we find
\begin{equation}
R_k^{bb}=
\frac{1}{
\sin\!\left(\frac{r\pi}{L}\right)} \:
\sum_{m=-\infty}^{\infty}
t^{[k+1+2m(L+1)]^2-1}\sin\!\left(\case{[k+1+2m(L+1)]r\pi}{L}\right)
+ O\!\left(t^{3L(L+2)}\right).
\label{fred}
\end{equation}
For $k=1,\ldots,L-2$, the leading term is obtained for
$m=0$. When $k=L-1$, however, the amplitude of this term vanishes
and we need the next-to-leading term $m=-1$. Consequently we get
\begin{eqnarray}
R_k^{bb} & \sim &
\frac{
\sin\!\left(\frac{(k+1)r\pi}{L}\right)}{
\sin\!\left(\frac{r\pi}{L}\right)} \: t^{(k+1)^2-1}
\qquad k=1,\ldots,L-2 \nonumber \\
& & \nonumber \\
R_{L-1}^{bb} & \sim &
2\cos\!\left(\case{r\pi}{L}\right) \: t^{(L+2)^2-1}
\end{eqnarray}
where we have used the fact that $r$ is odd.
{}From this we can read off the exponents
\begin{eqnarray}
\delta_k&=&\frac{3L(L+2)}{(k+1)^2-1}  \qquad k=1,2,\ldots,L-2.
\nonumber \\
& & \nonumber \\
\delta_{L-1} &=& \frac{3L(L+2)}{(L+2)^2-1}.
\end{eqnarray}
To find the corresponding conformal weights $\Delta_k$
we again use the scaling relation (\ref{eq:A.scaling}).
In addition to the
diagonal weights of the Kac table this yields the
conformal weight $\Delta_{2,1}^{(L+1)}$:
\begin{eqnarray}
\Delta_k &=&
\frac{(k+1)^2-1}{4L(L+1)} = \Delta_{k+1,k+1}^{(L+1)}
\qquad  k=1,2,\ldots,L-2 \nonumber \\
& & \nonumber \\
\Delta_{L-1} &=&
\frac{(L+2)^2-1}{4L(L+1)} = \Delta_{2,1}^{(L+1)}.
\end{eqnarray}

\subsection{Order parameters for regime 3$^+$}
\subsubsection*{Antiferromagnetic phases}
Using the identities
\begin{eqnarray}
\td(\case{1}{8}\pi,p) +
\td(\case{3}{8}\pi,p) &=&
2\,\vartheta_{3}(\case{1}{4}\pi,p^4)
\nonumber \\
\td(\case{1}{8}\pi,p) -
\td(\case{3}{8}\pi,p) &=&
2\,\vartheta_{1}(\case{1}{4}\pi,p^4)
\vphantom{\sum^a}
\end{eqnarray}
and the result (\ref{eq:A.N3denanti}), it follows from equation
(\ref{eq:A.lhpreg3anti}) that
\begin{equation}
\frac{P^{b b+1}_{-}(a)}{P^{b b+1}_{+}(a)} =
(-1)^{a+1} \:
\frac
{\te\!\left(\case{\pi}{4},t^{(L+1)(L+2)}\right)}
{\td\!\left(\case{\pi}{4},t^{(L+1)(L+2)}\right)}\, .
\end{equation}
For the order parameters (\ref{eq:GOP}) this yields the relation
\begin{eqnarray}
R_{L-1-k,-}^{b b+1}& = &
\frac
{\te\!\left(\case{\pi}{4},t^{(L+1)(L+2)}\right)}
{\td\!\left(\case{\pi}{4},t^{(L+1)(L+2)}\right)}\:
R^{bb+1}_{k,+}
\label{ted}   \\
& & \nonumber \\
&\sim & \sqrt{2} \: t^{(L+1)(L+2)/4} \:
R^{bb+1}_{k,+}.
\label{ted2}
\end{eqnarray}
Consequently we can restrict our attention to
$R^{bb+1}_{k,+}$.

As for the regimes 1$^+$ and 2$^+$ we first expand
the normalisation factor (\ref{eq:A.N3denanti}).
We note, however, that we now need
more terms in the expansion:
\begin{equation}
\frac{S(a)}{N_3(s)}(\chi_0+\chi_{1/2}) =
\frac{1}{(L+1)t} \:
\frac{
\sin\!\left(\frac{a\pi}{L+1}\right)}{
\sin\!\left(\frac{s\pi}{L+2}\right)}
\left[1+2 \, t^{L^2-4} \cos\!\left(\case{a\pi L}{L+1}\right)
+O\!\left(t^{2(L-2)(L+1)}\right)\right].
\end{equation}
Expanding the $\td$-functions appearing in the local height
probabilities (\ref{eq:A.lhpreg3anti}) we arrive at
\begin{eqnarray}
\lefteqn{
P_{+}^{bb+1}(a)=
\frac{2}{L+1}\:
\frac{
\sin\!\left(\frac{a\pi}{L+1}\right)}{
\sin\!\left(\frac{s\pi}{L+2}\right)}
\sum_{n=1}^{\infty} t^{n^2-1}
\sin\!\left(\case{a\pi n}{L+1}\right) \,
\sin\!\left(\case{s\pi n}{L+2}\right)
\left[1+2 t^{L^2-4} \cos\!\left(\case{a\pi L}{L+1}\right)
\right] } \nonumber \\
& & \nonumber \\
& & \hspace{7cm} +O\!\left(t^{2(L-2)(L+1)}\right)
\end{eqnarray}
If we apply the orthogonality relation (\ref{eq:A.ortho}), after
substituting this expression in the definition of the order
parameters, we obtain
\begin{eqnarray}
\lefteqn{R_{k,+}^{bb+1}=
\frac{1}{
\sin\!\left(\frac{s\pi}{L+2}\right)} \:
\sum_{m=-\infty}^{\infty} \left\{
t^{[k+1+2m(L+1)]^2-1}\sin\!\left(\case{[k+1+2m(L+1)]s\pi}{L+2}\right)
\right. }
\nonumber \\
& & \qquad \qquad \quad \;
+t^{[k+1-L+2m(L+1)]^2+L^2-5}
\sin\!\left(\case{[k+1-L+2m(L+1)]s\pi}{L+2}\right)
\vphantom{\sum^a}
\label{hannah} \\
& & \nonumber \\
& &\qquad \qquad \quad \; \left.
+t^{[k+1+L+2m(L+1)]^2+L^2-5}
\sin\!\left(\case{[k+1+L+2m(L+1)]s\pi}{L+2}\right) \right\}
+O\!\left(t^{2(L-2)(L+1)}\right) .
\nonumber
\end{eqnarray}
Except for $k=L-1$, the leading term comes from the first term
within the summation, with $m=0$.
Defining the exponents $\delta_{k,+}$ as in equation
(\ref{eq:A.powerlaw}), we can read off
\begin{equation}
\delta_{k,+}=\frac{L^2-4}{(k+1)^2-1} \qquad k=1,2,\ldots,L-2.
\label{eq:onemissing}
\end{equation}
When $k=L-1$ the first and the third term, with $m=0$ and $m=-1$,
respectively, are of the same order. Using the fact that $s$ is odd
we find that $R_{L-1,+}^{bb+1}$ vanishes.
Because of the symmetry (\ref{ted}) this is as it should be.

Using the relation (\ref{ted2}) we readily obtain the exponents
\begin{equation}
\delta_{k,-}=
\frac{L^2-4}{(L+2)(L+1)/4+(L-k)^2-1} \qquad k=1,2,\ldots,L-2.
\end{equation}
{}From the scaling relation (\ref{eq:A.scaling}) and the
conformal weight $\Delta_p$ in equation (\ref{eq:A.deltap}),
we thus find
\begin{equation}
\left. \begin{array}{lll}
\Delta_{k,+}=&\!\!\! &\!\!\!\frac{(k+1)^2-1}{4(L+2)(L+1)}=
\Delta_{1,1}^{(4)}+\Delta_{k+1,k+1}^{(L+2)} \\ & &\\
\Delta_{k,-}=\frac{1}{16}&\!\!\!+&\!\!\!\frac{(L-k)^2-1}{4(L+2)(L+1)}=
\Delta_{2,2}^{(4)}+ \Delta_{L-k,L-k}^{(L+2)}
\end{array} \right\}
\qquad  k=1,2,\ldots,L-2.
\end{equation}
We conclude that we have obtained all but one of the diagonal
weights of the relevant Kac table, shifted by the
diagonal $c=\case{1}{2}$ weights.
It is this structure,
which we observe also in Regime 4$^+$ below, together with the
products of Virasoro characters which we found in the local
height probabilities for these two regimes, which led to the
proposal (\ref{mipf}) for the modular invariant partition function
for the corresponding critical branches.

\subsubsection*{Ferromagnetic phases}
The derivation of the order parameters for the ferromagnetic
phases of regime 3$^+$ proceeds along similar lines to
the calculation for the antiferromagnetic phases.

The non-zero order parameters are again given by
(\ref{eq:A.orderHuse}) and, using the results
(\ref{eq:A.lhpreg3ferro}) and (\ref{eq:A.N4den}),
are found to be
\begin{eqnarray}
\lefteqn{
R_k^{bb}=
\frac{1}{\sin\!\left(\frac{s\pi}{L+2}\right)} \:
\sum_{m=-\infty}^{\infty} \left\{
t^{[k+1+2m(L+1)]^2-1}\sin\!\left(\case{[k+1+2m(L+1)]s\pi}{L+2}\right)
\right. } \nonumber \\
& & \qquad \qquad \; \;+ t^{[k+1-L+2m(L+1)]^2+L^2-5}
\sin\!\left(\case{[k+1-L+2m(L+1)]s\pi}{L+2}\right)
\vphantom{\sum^a} \\
& & \nonumber \\
& & \qquad \qquad \; \;\left. +t^{[k+1+L+2m(L+1)]^2+L^2-5}
\sin\!\left(\case{[k+1+L+2m(L+1)]s\pi}{L+2}\right) \right\}
+ O\!\left(t^{2(L-2)(L+1)} \right). \nonumber
\end{eqnarray}
For $k=1,\ldots,L-2$ the leading term comes from the first term
within the curly braces, with $m=0$:
\begin{equation}
R_k^{bb}\sim
\frac{
\sin\!\left(\frac{(k+1)s\pi}{L+2}\right)}{
\sin\!\left(\frac{s\pi}{L+2}\right)} \:
t^{(k+1)^2-1} \qquad k=1,\ldots,L-2.
\end{equation}
For $k=L-1$ the first term with $m=0$ and the third term with $m=-1$
are of equal order. Using the fact that $s$ is even gives
\begin{equation}
R_{L-1}^{bb}\sim
-4 \cos\!\left(\case{s\pi}{L+2}\right) \:
t^{L^2-1}.
\end{equation}
{}From these two equations we extract the exponents
\begin{equation}
\delta_k=\frac{L^2-4}{(k+1)^2-1}  \qquad k=1,2,\ldots,L-1.
\end{equation}
which are the exponents $\delta_{k,+}$ found in (\ref{eq:onemissing}),
plus the `missing' exponent $\delta_{L-1,+}$.

\subsection{Order parameters for regime 4$^+$}
\subsubsection*{Antiferromagnetic phases}
{}From the results (\ref{eq:A.lhpreg3anti}) and
(\ref{eq:A.N5den}) for the local height probabilities, and the
symmetry relation
\begin{eqnarray}
R_{L-1-k,-}^{b b+1}& = &
\frac
{\te\!\left(\case{\pi}{4},t^{L(L+1)}\right)}
{\td\!\left(\case{\pi}{4},t^{L(L+1)}\right)}\:
R^{bb+1}_{k,+}
\nonumber \\
& & \nonumber \\
&\sim & \sqrt{2} \: t^{L(L+1)/4} \:
R^{bb+1}_{k,+}
\end{eqnarray}
we obtain the
following results for the order parameters in the small field limit:
\begin{eqnarray}
\lefteqn{
R_{k,+}^{bb+1}\sim
\frac{
\sin\!\left(\case{(k+1)r\pi}{L}\right)}{
\sin\!\left(\case{r\pi}{L}\right)} \:
t^{(k+1)^2-1} }\nonumber \\
& & \nonumber \\
\lefteqn{
R_{k,-}^{bb+1}\sim
\sqrt{2}\,t^{L(L+1)/4} \:
\frac{
\sin\!\left(\case{(L-k)r\pi}{L}\right)}{
\sin\!\left(\case{r\pi}{L}\right)} \:
t^{(L-k)^2-1}.}
\end{eqnarray}
{}From this we read off the critical exponents
\begin{eqnarray}
\lefteqn{
\delta_{k,+}=\frac{L(L+4)}{(k+1)^2-1} } \nonumber \\
& & \nonumber \\
\lefteqn{
\delta_{k,-}=
\frac{L(L+4)}{L(L+1)/4+(L-k)^2-1} }
\end{eqnarray}
where $k=1,2,\ldots,L-2$.
We apply the scaling relation (\ref{eq:A.scaling}) to the above
results to find
\begin{equation}
\left. \begin{array}{lll}
\Delta_{k,+}=&\!\!\! &\!\!\!\frac{(k+1)^2-1}{4(L+1)L}=
\Delta_{1,1}^{(4)}+\Delta_{k+1,k+1}^{(L+1)} \\ & &\\
\Delta_{k,-}=\frac{1}{16}&\!\!\!+&\!\!\!\frac{(L-k)^2-1}{4(L+1)L}=
\Delta_{2,2}^{(4)}+\Delta_{L-k,L-k}^{(L+1)}
\end{array} \right\}
\qquad  k=1,2,\ldots,L-2.
\end{equation}
We have thus obtained the diagonal
weights of the Kac table, with $h=L+1$, shifted by the
diagonal $c=\case{1}{2}$ weights.

\subsubsection*{Ferromagnetic phases}
{}From the expression for the order parameters which survive in the
ferromagnetic
phases (\ref{eq:A.orderHuse}), and the results for the local
height probabilities
(\ref{eq:A.lhpreg4ferro}) and (\ref{eq:A.N6den}),
we have
\begin{equation}
R_k^{bb}\sim
\frac{
\sin\!\left(\frac{(k+1)r\pi}{L}\right)}{
\sin\!\left(\frac{r\pi}{L}\right)} \:
t^{(k+1)^2-1} \qquad  k=1,2,\ldots,L-2.
\end{equation}
{}From this we obtain a subset of the exponents found for the
antiferromagnetic phases. We are also able to determine that
\begin{equation}
R_{L-1}^{bb} \sim -
4\cos\!\left(\case{r\pi}{L}\right) \: t^{(L+2)^2-1}
\end{equation}
which corresponds to the conformal weight
$\Delta_{1,1}^{(4)}+\Delta_{2,1}^{(L+1)}$.
%  #] exponenten:

%  #[ phase diagram:
\nsection{The phase diagram}\label{sec:A.phdiag}
The study of the solvable manifolds spanned by the elliptic nome and
the spectral parameter reveals some aspects of the role these manifolds
play in a larger parameter space. In particular the off-critical
regimes support coexistence between a specified number of phases, each
associated with a groundstate, detailed in
section~\ref{sec:A.groundstate}.
The critical branches are the critical points terminating these
coexistence lines.
In the following we discuss some aspects of the phase
diagram of the dilute A models.
The discussion is not limited to the solvable manifolds,
but is restricted to the region where the weights are
$\Integer_2$ symmetric.

When the $\Integer_2$ symmetry $a \rightarrow L+1-a$ is obeyed,
the weights permit a duality transformation.
This transformation is a direct generalisation of the orbifold
duality, relating the A$_{2k+1}$ and
D$_{k+2}$ Temperley-Lieb models \cite{Fendley},
to the dilute A and D models.
Since the A$_3$ and D$_3$ Dynkin diagrams are identical,
we have a dual symmetry of the phase diagram in the case $L=3$.
Because the phase diagram for $L>3$ is complicated by
the large number of states of the model, and by the absence of a
dual symmetry, we
limit detailed discussion to
the case $L=3$.
Only afterwards do we mention which conclusions are valid for
general $L$.

The duality transformation for the case $L=3$ can be performed by
separating each degree of freedom into an Ising-like variable
$\sigma=\pm 1$, that
discriminates between the states $a=1$ and $a=3$, and the
variable $s=0,1$, that decides whether a state is $a=2$, so that
$a=2+s\sigma$.
For a given configuration of $s$, the Ising variables live on the
sites where $s=1$.
Since these Ising variables take the same value on neighbouring sites,
there is effectively only one Ising variable in each cluster of $s=1$
sites that are mutually connected (directly or indirectly) by nearest
neighbour links.
Such clusters form an irregular lattice, with interactions between them
wherever they are linked via a second neighbour bond.
On this lattice an ordinary Kramers-Wannier duality transformation can
be performed which results in new Ising variables on the dual lattice,
formed in the same way, but by the clusters where $s=0$. The
$s$-variables, fixed in the procedure so far, are now each
replaced by $1-s$.
The resulting  duality transformation on the weights is
\begin{eqnarray}
\lefteqn{\Wt{}{1}{1}{1}{1} = \W{}{2}{2}{2}{2}  \qquad \qquad \qquad
\Wt{}{2}{2}{2}{2} = \W{}{1}{1}{1}{1}} \nonumber \\
\lefteqn{\Wt{}{1}{1}{1}{2} = \W{}{2}{2}{2}{1}  \qquad \qquad \qquad
\Wt{}{2}{1}{1}{1} = \W{}{1}{2}{2}{2} 2^{1/2}} \nonumber \\
\lefteqn{\Wt{}{2}{2}{2}{1} = \W{}{1}{1}{1}{2} \qquad \qquad \qquad
\Wt{}{1}{2}{2}{2} = \W{}{2}{1}{1}{1} 2^{-1/2}} \nonumber \\
\lefteqn{\Wt{}{1}{2}{2}{1} = \W{}{2}{1}{1}{2}} \nonumber \\
\lefteqn{ \Wt{}{2}{1}{2}{3} =
           \W{}{1}{2}{1}{2} - \W{}{3}{2}{1}{2} } \\
\lefteqn{ \Wt{}{3}{2}{1}{2} = {\displaystyle{1}\over\displaystyle{2}}
   \left[ \W{}{2}{1}{2}{1} - \W{}{2}{1}{2}{3} \right] }
            \nonumber \\
\lefteqn{ \Wt{}{2}{1}{2}{1} =
           \W{}{1}{2}{1}{2} + \W{}{3}{2}{1}{2} } \nonumber \\
\lefteqn{ \Wt{}{1}{2}{1}{2} =  {\displaystyle{1}\over\displaystyle{2}}
   \left[ \W{}{2}{1}{2}{1} + \W{}{2}{1}{2}{3} \right] }
            \nonumber
\end{eqnarray}
where the weights are symmetric under exchange of the states 1 and
3, and the transformation rules are invariant under
reflection in the diagonals of the elementary squares.

\begin{figure}[hbt]
\centerline{\psfig{file=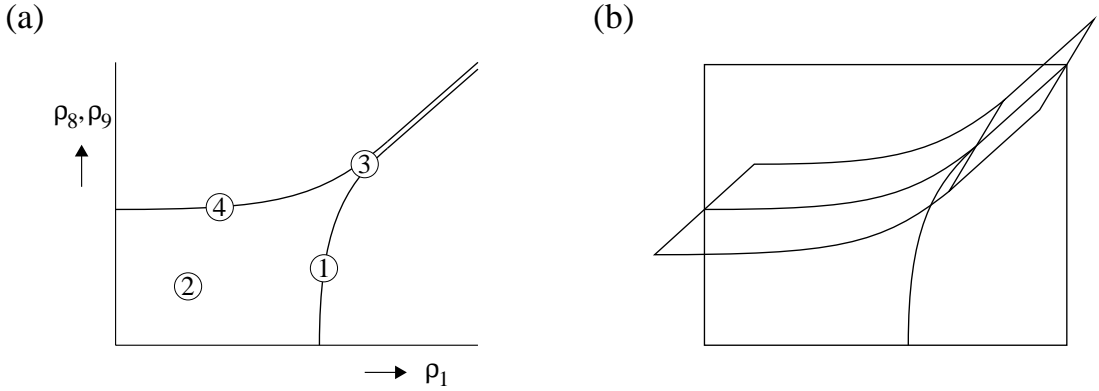}}
\caption{
The phase diagram for $L=3$. (a) The self-dual subspace. The encircled
numbers represent the four solvable critical branches; branches 1 and 4
describe the transition of the entire curves
on which they are located, and branch 2 describes the transition
across the whole lower
left section of the diagram. The double curve marks a
first-order transition.
(b) The phase diagram with an additional parameter that reverses
sign under duality.
In the extreme limits (front and back) the model reduces to
the interacting hard square model.
}
\label{fig:PD}
\end{figure}

The self-dual subspace contains all four solvable critical branches.
It constitutes a transition between a phase symmetric under
exchange of the states 1 and 3 and one in which this symmetry
is spontaneously broken.
The nature of this phase transition is not the same everywhere
in the self-dual subspace.
The phase structure within this subspace, which is equivalent to
the phase diagram of the O$(n)$ loop model \cite{BN},
is indicated in Fig.~\ref{fig:PD}.
It may be parametrised in terms of the weights $\rho_1,\ldots,\rho_9$.
If we keep the weights $\rho_2,\ldots,\rho_7$ fixed,
we may describe the
self-dual parameter space in terms of the weight $\rho_1$ of the
ferromagnetic states and the weights
$\rho_8$ and $\rho_9$ of the antiferromagnetic
states.

When $\rho_1$ as well as $\rho_8$ and $\rho_9$ are small, the
transition (across the self-dual space) is of second order, in the
Ising universality class. It is described by the solvable critical
point of branch 2.
The character of the phases is ferromagnetic.
When the weight $\rho_1$ is sufficiently large the
transition is of first order. These regions are separated by an Ising
tricritical transition governed by branch 1.

When, on the other hand, $\rho_8$ and $\rho_9$ are
sufficiently large and
$\rho_1$ is relatively small, antiferromagnetic
configurations, in which one of the sublattices takes the value
$a=2$ while the
other sublattice fluctuates between $a=1$ and $a=3$, are
favoured.
There will be
a phase separation between regions where one or the other sublattice has
$a=2$. The onset of this phase separation is an ordinary Ising
transition, which appears to be independent of the transition
across the self-dual subspace. It is described by the critical point of
branch 4. The independence of these transitions is reflected in the
product structure of the order parameters on this branch,
see {\em e.g.}, equation~(\ref{eq:prodstruc}), and in the value of the
central charge $c=1$, see
Table~\ref{tab:A.thetable}.
It implies that the transition across the self-dual subspace is of the
same (Ising) universality class on either side of the Ising transition
within the subspace.

When the ferromagnetic weight $\rho_1$ and antiferromagnetic weights
$\rho_8$, $\rho_9$ are all large relative to the other weights, the system
must make a choice between the first-order regime, in which the
ferromagnetic configurations dominate, and the critical
antiferromagnetic regime. From the fact that the surface tension between
these phases can be made arbitrarily large, we conclude that the
transition between the two regimes is first order.
The point where this first order transition is met by the other
two transitions within the subspace is the critical point of branch
3.

We further observe that the transition between the ferromagnetic and
antiferromagnetic region of the self-dual subspace, persists away
from the self-dual subspace, see Fig.~\ref{fig:PD}b.
In fact, when the weights are taken
far from self-duality, the
model may be reduced to the interacting hard square model
\cite{ihs1,ihs2,Huse2,ihs3},
in which the  particles are represented by $a=1$ or $a=3$, and the
empty sites by  $a=2$. Of course, by dual symmetry, the roles are
reversed in the  opposite
extreme.
The hard-square model is known to have an Ising
critical and tricritical transition into a sublattice-ordered phase.
We expect the critical transition to join continuously with branch 4 and
the tricritical transition with branch 3.

We now see that the critical point of branch 3 plays a
central
role. It sits at the intersection of two phase-transition manifolds,
{\em i.e.,}
the ferro-antiferromagnetic transition and the self-dual subspace.
Furthermore, it sits at the line where both transition sheets turn
first  order.

For general $L$ we expect the same topology of the phase diagram.
Though there is no longer a dual
symmetry that maps the weight space into itself, the subspace given by
(\ref{eq:dA.RSOS}) still plays the role of a phase transition.
The universality class of the transition across this subspace varies with
$L$, and is that of regime III-IV of the $L$-state ABF
model for branch~2, and of the $L+1$-state ABF model for branch~1.
The transition between ferro- and antiferromagnetic regions
remains Ising-like for all $L$.
Where the transition manifolds intersect, the universality class is
that of a direct product of an Ising and an ABF model (except where
they are first order).
Again, for general $L$ there are directions in the parameter space in
which the model can be reduced to the interacting hard square model.

To conclude we remark that
with the phase diagram proposed above we do not do full justice to the
fact that the critical behaviour of branch~3 is that of the product of
an Ising model and branch~1.

%  #] phase diagram:

%  #[ Ising model:
\section{An Ising model in a field}\label{sec:Isfi}
Because the model in regime $2^+$ is in the universality class of the
Ising model in a magnetic field, we take a special interest in this
case, and collect the results in this section.

As we have already observed, we can make the identification
$\{1,2,3\}=\{+,0,-\}$ to relate the heights of the dilute A$_3$
model to the states of a spin-1 Ising model.
This particular model has the restriction that
a `$+$' and a `$-$' spin may not be adjacent on the lattice.
Because the
nome $p$ breaks the up-down symmetry, it plays the role of the
magnetic field. In Regime $2^+$ we see from
Table~\ref{tab:A.thetable} that the central charge of the
model is $c=\case{1}{2}$.

Using result (\ref{eq:FE12}) we can rewrite the free energy as
\begin{equation}
\ln \kappa(u) = 2\sum_{k=-\infty}^{\infty} \int_{-\infty}^{\infty}
\frac{
\cosh 9 \pi x \:
\cosh 5 \pi x \:
\sinh 16 \pi u x\:
\sinh (15\pi-16 u)x}
{x \: \sinh 16\pi x \:
\cosh 15 \pi x}
\; \e^{32 \is \epsilon k x} \: dx .
\end{equation}

The magnetisation of this Ising model can be expressed as a linear
combination of the local state probabilities:
\begin{equation}
m\equiv\langle + \rangle - \langle - \rangle=
P^{11}(1)-P^{11}(3)=R_{1}^{11}/\sqrt{2}. \label{magnet}
\end{equation}
Analogously, we can define a density for the spin-1 model, which is the
probability that a site is occupied by a spin of either sign,
\begin{equation}
\rho\equiv\langle + \rangle + \langle - \rangle=
P^{11}(1)+P^{11}(3)=\frac{1}{2}(1+R_{2}^{11}). \label{density}
\end{equation}
When we group the following combination of elliptic functions:
\begin{equation}
K=\frac{
\te\!\left(\case{\pi}{3},t^{8}\right)
\te\!\left(\case{\pi}{4},t^{90}\right)}
{\te\!\left(\case{\pi}{3},t^{4}\right)N_2(r=1)}
 \sim\frac{1}{4\sqrt{6}} + O(t^{24})
\end{equation}
and make some simplifications, the magnetisation and density are
\begin{eqnarray}
m & = & 4\sqrt{3} \,t^3 K \left\{
E\left(-t^{54},t^{144}\right)
E\left(-t^{90},t^{240}\right)
-t^{24}
E\left(-t^{18}, t^{144}\right)
E\left(-t^{30}, t^{240}\right) \right\}
\nonumber\\
& & \nonumber \\
\rho & = & 2{\sqrt{6}} K
%\left\{
%E\left(-t^{72},t^{144}\right)
%E\left(-t^{120},t^{240}\right)
%-t^{48}
%E\left(-1,t^{144}\right)
%E\left(-1,t^{240}\right)\right\}    \\
%&=&
%2{\sqrt{6}} K
\left\{
\td\left(0,t^{72}\right)
\td\left(0,t^{120}\right)
-
\tt\left(0,t^{72}\right)
\tt\left(0,t^{120}\right)\right\} .
\end{eqnarray}
The leading order behaviour is
\begin{equation}
\begin{array}{rcccl}
\ln \kappa_{\mbox{\scriptsize sing}} & \sim & & & p^{16/15} \\
&&&& \\
m&\sim& t^3 &=&p^{1/15}\\
&&&&\\
 \rho-\case{1}{2} & \sim &t^{24}&=&p^{8/15}
\end{array}
\end{equation}
The first two expressions both lead to the exponent $\delta=15$.
The last exponent is related by scaling to the conformal
weights of the energy and spin operators:
$ \Delta_{1,2}/(1-\Delta_{2,1}) = 8/15$.
%  #] Ising model:

%  #[ Summary:
\nsection{Summary and discussion}\label{sec:A.sumres}
In this paper we have considered the infinite hierarchy
of dilute A models. These models, which belong to the family
of dilute A-D-E models, can be viewed as spin-1 generalisations
of the solid-on-solid models of Andrews, Baxter and Forrester
\cite{ABF}.
For any integer $L\geq 2$, the dilute A model defines a
$L$-state solid-on-solid model,
with heights labelled by the Dynkin diagram of the
classical Lie algebra A$_L$.
At the critical point, the Boltzmann weights (\ref{eq:A.weights})
of this solid-on-solid
model obey the $\Integer_2$ symmetry of the underlying Dynkin diagram,
but away from the critical point this symmetry is broken for
odd values of $L$.

For each value of $L$ the model has four different critical branches.
Two of these branches provide new
realisations of the unitary minimal series and the two other
branches can be viewed as the direct product of this
same series and a $c=\case{1}{2}$ model.
Away from criticality the four branches yield eight distinct regimes.

For all regimes we have calculated the free energy. From this
we have obtained critical exponents $\alpha$, for even $L$, and
exponents $\delta$, for odd $L$. Among these results is the
magnetic exponent $\delta=15$ of the Ising model.

The main part of this paper is concerned, however, with the
calculation of the generalised order parameters for the
symmetry-breaking  models of the dilute A hierarchy. This calculation
involved the evaluation of new sums-of-products identities for
theta functions of the Rogers-Ramanujan type.
{}From the resulting expressions for the order parameters we extracted
a set of critical exponents $\delta_k$.
For the models corresponding to the unitary minimal series, these
exponents correspond to the diagonal conformal weights of the
relevant Kac table.
For the models corresponding to a direct product of a
unitary minimal model and a $c=\case{1}{2}$ model, the
critical exponents yield the diagonal weights of the Kac table
of the minimal model incremented by either one of the diagonal weights
of the $c=\case{1}{2}$ model.

We have not yet succeeded in computing the order parameters
for the non-symmetry-breaking dilute A models, obtained
for even values of $L$.
In this case the Boltzmann weights no longer
enjoy the diagonal property (\ref{eq:A.Hdef})
and an additional diagonalisation is required.
We hope to report on these `even' members of
the dilute A hierarchy in a future publication.

Another open problem is the relation between the symmetry-breaking
spin-1 Ising model, corresponding to the dilute A$_3$ models in
the regimes 2$^{+}$ and 2$^{-}$,
and the integrable field theory of the critical
Ising model in a magnetic field,
found by Zamolodchikov \cite{Zamol,Fateev}.
As shown by Smirnov \cite{Smirnov},
Zamolodchikov's $S$-matrix corresponds to an RSOS
projection of the Izergin-Korepin R-matrix.
Hence the dilute A$_3$ model provides a likely candidate
for describing the corresponding solvable lattice model.
The precise relation between Zamolodchikov's $S$-matrix,
which has a structure related to the Lie algebra E$_8$,
and the dilute A$_3$ model remains, however, unclear.
By studying the excitation spectrum of
the dilute A models, and in particular the A$_3$ case, we hope
to establish the connection, if present, in the near future.

%  #] Summary:

%  #[ Acknowledgements:
\section*{Acknowledgements}
This work has been supported by the Stichting voor Fundamenteel
Onderzoek der Materie (FOM) and the Australian Research Council.

%  #] Acknowledgements:

%  #[ Appendix:
\appendix
%\setcounter{section}{0}
%\renewcommand{\thesection}{\arabic{chapter}.\Alph{section}}
%  #[ AppendixA:
\nsection{The function $H$}\label{app:A.H-fun}
In this appendix we give the weights (\ref{eq:A.weights}) in a form
which is more suitable for considering the ordered or strong-field
limit, and hence for obtaining the weight
function $H$ defined in equation (\ref{eq:A.Hdef}).

We first define the new variables
\begin{equation}
w=\e^{-2\pi u/\epsilon} \quad \mbox{and} \quad x=\e^{-2\pi \la/\epsilon}.
\end{equation}
In terms of these variables, the ordered limit is given by
$x \rightarrow 0$ with $w$ fixed.
Using the conjugate nome expressions
(\ref{eq:A.congnome}) of the theta functions and setting
\begin{equation}
E\left( u,x^{\pi/\la} \right) = E(u)
\end{equation}
the Boltzmann weights in conjugate modulus
parametrisation read
\setlength{\mathindent}{0 cm}
\begin{eqnarray}
\lefteqn{\W{}{a}{a}{a}{a}=
\frac{E(x^6 w^{-1})E(x^3 w)}{E(x^6)E(x^3)} + x^3
\frac{E(w)E(x^3 w^{-1})}{E(x^6)E(x^3)}}
\nonumber \\  & & \nonumber \\
\lefteqn{\hphantom{\W{}{a}{a}{a}{a}}
\times \left(x^2 \frac{E(-x^{2a})E(x^{4(a+1)})E(-x^{2a-5})}
{E(-x^{2(a+1)})E(x^{4a})E(-x^{2a+1})}
      + x^{-2} \frac{E(-x^{2a})E(x^{4(a-1)})E(-x^{2a+5})}
{E(-x^{2(a-1)})E(x^{4a})E(-x^{2a-1})}\right)}
\nonumber \\ & & \nonumber \\
\lefteqn{\W{}{a}{a}{a}{a\pm 1}=\W{}{a}{a\pm 1}{a}{a}=
\frac{g_a}{g_{a\pm 1}}
w \frac{E(x^3 w^{-1})E(-x^{\pm 2a+1}w^{-1})}
{E(x^3)E(-x^{\pm 2a+1})}}
\nonumber \\ & & \nonumber \\
\lefteqn{\W{}{a\pm 1}{a}{a}{a}=\W{}{a}{a}{a\pm 1}{a}}
\nonumber \\ & & \nonumber \\
\lefteqn{\hphantom{\W{}{a\pm 1}{a}{a}{a}}=
\frac{g_{a\pm 1}}{g_a}
w^{-1} x^3\left(-x^{\mp 1}\frac{E(-x^{2a})E(x^{4(a\pm 1)})}
{E(-x^{2(a \pm 1)})E(x^{4a})}\right)^{1/2}
\frac{E(w)E(-x^{\pm 2a-2} w)}{E(x^3)E(-x^{\pm 2a+1})}}
\nonumber \\ & & \nonumber \\
\lefteqn{\W{}{a}{a\pm 1}{a\pm 1}{a}=\W{}{a}{a}{a\pm 1}{a\pm 1}}
\nonumber \\ & & \nonumber \\
\lefteqn{\hphantom{\W{}{a}{a\pm 1}{a\pm 1}{a}}=
x \left(\frac{E(-x^{\pm 2a+3})E(-x^{\pm 2a-1})}
           {E^2(-x^{\pm 2a+1})}\right)^{1/2}
\frac{E(w)E(x^3 w^{-1})}{E(x^2)E(x^3)}}
\label{eq:A.conjweights} \\ & & \nonumber \\
\lefteqn{\W{}{a}{a\mp 1}{a}{a\pm 1}=
\frac{g_a^2}{g_{a+1} g_{a-1}}
w\frac{E(x^2 w^{-1})E(x^3 w^{-1})}{E(x^2)E(x^3)}}
\nonumber \\ & & \nonumber \\
\lefteqn{\W{}{a\pm 1}{a}{a\mp 1}{a}=
-\frac{g_{a+1} g_{a-1}}{g_a^2}
x^2\left(\frac{E^2(-x^{2a})E(x^{4(a-1)})E(x^{4(a+1)})}
{E(-x^{2(a+1)})E(-x^{2(a-1)})E^2(x^{4a})}\right)^{1/2}
\frac{E(w)E(xw^{-1})}{E(x^2)E(x^3)}}
\nonumber \\ & & \nonumber \\
\lefteqn{\W{}{a\pm 1}{a}{a\pm 1}{a}=
\frac{g_{a\pm 1}^2}{g_a^2}
\frac{E(x^3 w^{-1})E(x^{\pm 4a+2} w)}{E(x^3)E(x^{\pm 4a+2})}}
\nonumber \\ & & \nonumber \\
\lefteqn{\hphantom{\W{}{a\pm 1}{a}{a\pm 1}{a}}
-\frac{g_{a\pm 1}^2}{g_a^2} w^{-1} x^{3 \mp 1}
\frac{E(-x^{2a})E(x^{4(a\pm 1)})E(w)E(x^{\pm 4a-1} w)}
{E(-x^{2(a \pm 1)})E(x^{4a})E(x^3) E(x^{\pm 4a+2})}}
\nonumber \\
& & \nonumber \\
\lefteqn{\hphantom{\W{}{a\pm 1}{a}{a\pm 1}{a}}
=\frac{g_{a\pm 1}^2}{g_a^2} w^{-1}
\frac{E(x^3 w)E(x^{\pm 4a-4} w)}{E(x^3)E(x^{\pm 4a-4})}
-\frac{g_{a\pm 1}^2}{g_a^2} w^{-1}
\frac{E(w) E(x^{\pm 4a-1} w)}{E(x^3)E(x^{\pm 4a-4})} }
\nonumber \\ & & \nonumber \\
\lefteqn{\hphantom{\W{}{a\pm 1}{a}{a\pm 1}{a}}
\times
\left( x^{\pm 1 -1}
\frac{E(-x^{2a}) E(x^{4(a \mp 1)}) E(x^4)}
{E(-x^{2(a \mp 1)} E(x^{4a}) E(x^2)}
+x^3
\frac{E(-x^{\pm 2a-5})}{E(-x^{\pm 2a+1})} \right) } \nonumber
\end{eqnarray}
where we have omitted an overall normalisation factor $\exp(2u(3\la-u)/
\epsilon)$ and $g_a$ is defined in equation (\ref{eq:A.gdef}).
In this alternative representation of the Boltzmann weights, we can
readily
extract their
leading behaviour in the ordered limit.
To do so, we make use of the simple properties
of the function $E$ as listed in equation (\ref{eq:A.limEfun})
of appendix~\ref{app:A.elliptic}.
It is clear from the definition of $x$ that the result depends on
the value of $\la$.
For regime $1^+$, using definition (\ref{eq:A.setreg1}), we get
\setlength{\mathindent}{\mathin}
\begin{eqnarray}
H(b,b,b)&=&\left\{ \begin{array}{ll}
0 & \quad b\in s_1 \\
0 & \quad b\in s_2 \\
1 & \quad b\in s_3 \\
1 & \quad b\in s_4
\end{array} \right. \nonumber \\
& & \nonumber \\
H(b+1,b,b)=H(b,b,b+1)&=&\left\{ \begin{array}{ll}
\half(b+1) & \quad b\in s_1 \\
\half b & \quad b\in s_2 \\
\half(b+2) & \quad b\in s_3 \\
\half(b+1) & \quad b\in s_4 \setminus\{L\}
\end{array} \right. \nonumber \\
& & \nonumber \\
H(b-1,b,b)=H(b,b,b-1)&=&\left\{ \begin{array}{ll}
-\half(b-1) & \quad b\in s_1\setminus\{1\}  \\
-\half(b-2) & \quad b\in s_2 \\
-\half(b-2) & \quad b\in s_3 \\
-\half(b-3) & \quad b\in s_4
\end{array} \right. \nonumber \\
& & \label{eq:A.HfuncR1} \\
H(b+1,b,b+1)&=&\left\{ \begin{array}{ll}
b+1 & \quad b\in s_1 \\
b & \quad b\in s_2 \\
b+1 & \quad b\in s_3 \\
b+1 & \quad b\in s_4 \setminus\{L\}
\end{array} \right.   \nonumber  \\
& & \nonumber \\
H(b-1,b,b-1)&=&\left\{ \begin{array}{ll}
-(b-1) & \quad b\in s_1 \setminus\{1\} \\
-(b-2) & \quad b\in s_2 \\
-(b-2) & \quad b\in s_3 \\
-(b-2) & \quad b\in s_4
\end{array} \right. \nonumber \\
& & \nonumber \\
H(b \pm 1,b,b \mp 1)&=&\begin{array}{ll} 1 & \quad b\in\{2,3,\ldots,L-1 \}
\end{array}. \nonumber
\end{eqnarray}
To facilitate the proof of solution (\ref{eq:A.solnR1})
we include the following
two terms to the above list: $H(0,1,1)=0$ and $H(L+1,L,L)=\half(L+1)$.

For regime $2^+$, using definition (\ref{eq:A.setreg2}), we get
\begin{eqnarray}
H(b,b,b)&=&\left\{ \begin{array}{ll}
0 & \quad b\in t_1\\
0 & \quad b\in t_2\\
1 & \quad b\in t_3\\
1 & \quad b\in t_4
\end{array} \right. \nonumber \\
& & \nonumber \\
H(b+1,b,b)=H(b,b,b+1)&=&\left\{ \begin{array}{ll}
\half(b+1) & \quad b\in t_1 \\
\half(b+2) & \quad b\in t_2 \\
\half(b+2) & \quad b\in t_3 \\
\half(b+3) & \quad b\in t_4 \setminus\{L\}
\end{array} \right. \nonumber \\
& & \nonumber \\
H(b-1,b,b)=H(b,b,b-1)&=&\left\{ \begin{array}{ll}
-\half(b-1) & \quad b\in t_1 \setminus\{1\}\\
-\half b    & \quad b\in t_2 \\
-\half(b-2) & \quad b\in t_3 \\
-\half(b-1) & \quad b\in t_4
\end{array} \right. \nonumber \\
& & \label{eq:A.HfuncR2} \\
H(b+1,b,b+1)&=&\left\{ \begin{array}{ll}
b+1 & \quad b\in t_1 \\
b+2 & \quad b\in t_2 \\
b+2 & \quad b\in t_3 \\
b+2 & \quad b\in t_4 \setminus\{L\}
\end{array} \right.   \nonumber  \\
& & \nonumber \\
H(b-1,b,b-1)&=&\left\{ \begin{array}{ll}
-(b-1) & \quad b \in t_1 \setminus\{1\}\\
- b    & \quad b \in t_2 \\
-(b-1) +\delta_{b,2}& \quad b \in t_3 \\
-(b-1) & \quad b\in t_4
\end{array} \right. \nonumber \\
& & \nonumber \\
H(b \pm 1,b,b \mp 1)&=&\begin{array}{ll} 1 & \quad b\in\{2,\ldots,L-1 \}
\end{array}. \nonumber
\end{eqnarray}
Now we only add the single term $H(0,1,1)=0$.

%  #] AppendixA:

%  #[ AppendixB:
\nsection{Denominator identities}
\label{app:A.denom}
In this appendix we prove the six denominator identities
used in section~\ref{sec:A.thermo} to normalise the local height
probabilities.
All identities involve sums over products
of elliptic functions similar
to the sums-of-products identities of Rogers and Ramanujan.

For brevity we shall adopt the convention that sums over
$j,k,l,m$ and $n$ always run over $\Integer$.
%  #[ denom1:
\subsection{Regime 1$^+$}\label{app:A.den1}
For regime 1$^+$ the denominator identity reads
\setlength{\mathindent}{0 cm}
\begin{eqnarray}
\lefteqn{N_1 \equiv
\sum_{a=0}^{2L+1}
\te\!\left(\case{a\pi}{L+1},t^{6L(L+2)}\right)
\td\!\left(\case{a\pi L}{2(L+1)},t^{3L(L+2)}\right)
\td\!\left(\case{\pi}{2}\left(\case{a}{L+1}
\m \case{s}{L+2}\right),t\right)}
\label{eq:A.denom1} \\
\lefteqn{
\hphantom{N_1}=
2(L + 1) \, t^{L(2L+3)/2}
Q\left(t^{12L(L+2)}\right)
Q\left(t^{12L(L+1)}\right)
\te\!\left(\case{\pi}{6},t^{2(L+1)(L+2)}\right)
\frac{
\te\!\left(\case{2s\pi}{L+2},t^{6L(L+1)}\right)}
{\tt\!\left(\case{s\pi}{L+2},t^{6L(L+1)}\right)} }
\nonumber
\end{eqnarray}
where $s=1,3,5,\ldots, L$ and $L=3,5,7,\ldots$.

\smallskip
{\em Proof:}
To prove this identity we start by recasting the
the left-hand side
using the representation (\ref{eq:A.thetasums}) of the
$\vartheta$-functions as infinite sums:
\setlength{\mathindent}{\mathin}
\begin{eqnarray}
\lefteqn{
N_1=-\ib t^{3L(L+2)/2}\sum_{a=0}^{2L+1}
\sum_{j,k,\ell}
(-1)^j \,
\e^{\pi \is a (2j+1+kL+\ell) /(L+1)} \,
\e^{-\pi \is s \ell/(L+2)} }
\nonumber \\
& & \qquad \qquad \qquad \qquad \qquad \qquad \times \:
\vphantom{\sum^a}
t^{6L(L+2)(j^2+j)+3L(L+2)k^2+\ell^2} .
\end{eqnarray}
Summing the roots of unity over $a$, we find that the above
expression vanishes unless
$\ell=-2j-1-kL+2m(L+1)$, with $m\in\Integer$. Hence we get
\begin{eqnarray}
\lefteqn{
N_1=-2 \ib (L+1) \, t^{3L(L+2)/2}
\sum_{j,k,m} (-1)^{j+k} \,
\e^{\pi \is s(2j+1-2k+2m)/(L+2)} }
\nonumber\\
& & \qquad \qquad \qquad \qquad \qquad \qquad \times \:
\vphantom{\sum^a}
t^{6L(L+2)(j^2+j)+3L(L+2)k^2+[2j+1+kL-2m(L+1)]^2}
\end{eqnarray}
where we have used the fact that $s$ is odd.

We now carry out a sequence of transformations to diagonalise the
quadratic exponent of $t$.
First we make the substitution $m\to m-j+k$ followed by $k\to k+2j$.
Then we replace the sum over $k$ by a sum over $\ell$ and $\alpha$
by setting $k=3\ell+\alpha$, where $\alpha$ has to be summed over
$0,1$ and 2. Similarly, we replace the sum over $m$ by a sum over
$n$ and $\beta$ by setting $m=3n+\beta$.
Finally we make the replacements $j\to j-\ell$ and $\ell\to \ell-n$.
After all these transformations we obtain
\begin{eqnarray}
\lefteqn{
N_1=-2 \ib (L+1) \, t^{(3L^2+6L+2)/2}
\sum_{\alpha,\beta=0}^2
\sum_{j,\ell,n}
(-1)^{j+\alpha} \, \e^{\pi \is s(6n+2\beta+1)/(L+2)} }
\nonumber\\
& & \quad \qquad \times \:
t^{4(\alpha+\beta)(\alpha+\beta-1)
+2L(5\alpha^2+4\beta^2+6\alpha\beta-\alpha-2\beta)
+4L^2(\alpha^2+\beta^2+\alpha\beta)}
\vphantom{\sum_a^a}
\nonumber\\
& & \quad \qquad \times \:
t^{6L(L+2)(3j+1+2\alpha)j+
6(L+1)(L+2)(3\ell-1+2\alpha+2\beta)\ell+
6L(L+1)(3n+1+2\beta)n}
\vphantom{\sum_a^a}
\nonumber\\
\lefteqn{\hphantom{N_1}=
-2 \ib (L+1) \, t^{(3L^2+6L+2)/2}} \\
& & \qquad \quad \times
\sum_{\alpha,\beta=0}^2 (-1)^{\alpha} \,
t^{4(\alpha+\beta)(\alpha+\beta-1)
+2L(5\alpha^2+4\beta^2+6\alpha\beta-\alpha-2\beta)
+4L^2(\alpha^2+\beta^2+\alpha\beta)}
\nonumber\\
& & \qquad \quad \times \:
\e^{\pi \is s(2\beta+1)/(L+2)}
E\!\left(-\e^{6\pi \is s/(L+2)} \, t^{12L(L+1)(2+\beta)},
t^{36L(L+1)}\right)
\vphantom{\sum_a^a}
\nonumber \\
& & \qquad \quad \times \:
E\!\left(t^{12L(L+2)(2+\alpha)},t^{36L(L+2)}\right)
E\!\left(-t^{12(L+1)(L+2)(1+\alpha+\beta)},t^{36(L+1)(L+2)}\right).
\vphantom{\sum^a}
\nonumber
\end{eqnarray}
where, in the second step,
we have used the definition (\ref{eq:A.Efunc})
of the function $E$.
The terms with $\alpha=1$ are identically zero and
the two terms with $\beta=1$, $\alpha\neq 1$ cancel.
The remaining four terms factorise to give
\begin{eqnarray}
\lefteqn{
N_1=2 \ib (L+1)  \, t^{(3L^2+6L+2)/2} \,
\e^{-\pi \is s /(L+2)}
Q\left(t^{12L(L+2)}\right)
\vphantom{\sum_a}}
\nonumber\\
\lefteqn{\qquad \qquad \times \!
\left[
E\!\left(-t^{12(L+1)(L+2)},t^{36(L+1)(L+2)}\right)
-t^{4(L+1)(L+2)}
E\!\left(-1,t^{36(L+1)(L+2)}\right) \right]
\vphantom{\sum_a^a} }
 \nonumber \\
\lefteqn{
\qquad \qquad \times \!
\left[
E\!\left(-\e^{6\pi \is s/(L+2)} \,
t^{12L(L+1)},t^{36L(L+1)}\right) \right.
\vphantom{\sum_a^a} }
\label{eq:A.lhsntt}\\
\lefteqn{
\qquad \qquad \qquad \qquad \qquad \qquad
\left. -\e^{2\pi \is s/(L+2)}
E\!\left(-\e^{-6\pi \is s/(L+2)} \, t^{12L(L+1)},
t^{36L(L+1)}\right)
 \right] .
\vphantom{\sum^a} } \nonumber
\end{eqnarray}
We now note the following two elliptic function identities:
\begin{eqnarray}
\lefteqn{ \vphantom{\sum_a^a}
E\left(-x^3 p,p^3\right)
- x E\left(-x^{-3} p,p^3\right)
=\frac{ Q(p) \, E\left(x^2,p\right)}{
E\left(-x,p\right)}} \nonumber \\
\lefteqn{ \vphantom{\sum_a^a}
E\left(-p^3,p^9\right)
-p E\left(-1,p^9\right) = \e^{\pi \is /3}
E\left(\e^{\pi \is/3},p\right).}
\label{eq:A.Eid}
\end{eqnarray}
To prove the first identity we note that the
function $f(x)$, defined as the ratio of the
left-hand side over the right-hand side,
has the following periodicity property:
$f(px)=f(x)$.
Since the function $E(x,p)$ is analytic in $0<|x|<\infty$,
the only possible poles in the period annulus
$p\leq |x| \leq 1$ are the zeros
$x=\pm 1$ and $x=\pm p^{1/2}$ of the function $E(x^2,p)$.
Using formula (\ref{eq:A.limEfun}) to manipulate the function $E$
it is readily verified that these poles have zero residue.
Hence, by Liouville's theorem, $f$ is constant.
Setting $x=-p^{1/3}$ shows that this constant is one.
The second identity is actually a corollary of the first.
Replacing $p$ by $p^3$, then setting $x=p^2$ and again using
formula (\ref{eq:A.limEfun}), yields
\begin{eqnarray}
\lefteqn{
E\left(-p^3,p^9\right) -p E\left(-1,p^9\right) =
-p \: \frac{E\left(p^4,p^3\right) Q\left(p^3\right)}
{E\left(-p^2,p^3\right)} } \nonumber \\
\lefteqn{ \hphantom{
E\left(-p^3,p^9\right) -p E\left(-1,p^9\right)}
= \prod_{n=1}^{\infty} \left( 1-p^n+p^{2n} \right) } \\
\lefteqn{ \hphantom{
E\left(-p^3,p^9\right) -p E\left(-1,p^9\right)}
= \e^{\pi \is /3} E\left(\e^{\pi \is/3},p\right).
\vphantom{\sum^a}} \nonumber
\end{eqnarray}
Applying
both identities and the relation (\ref{eq:A.thetasasE}) of
appendix~\ref{app:A.elliptic},
we obtain the desired right-hand side of (\ref{eq:A.denom1}).

%  #] denom1:

%  #[ denom2:
\subsection{Regime 2$^+$}\label{app:A.den2}
For regime 2$^+$ we have to prove the
following identity:
\setlength{\mathindent}{0 cm}
\begin{eqnarray}
\lefteqn{
N_2\equiv\sum_{a=0}^{2L+1}
\te\!\left(\case{a\pi}{L+1},t^{6L(L+2)}\right)
\td\!\left(\case{a\pi L}{2(L+1)},t^{3L(L+2)}\right)
\td\!\left(\case{\pi}{2}\left(\case{r}{L}
\m\case{a}{L+1}\right),t\right)}
\label{eq:A.denom2} \\
%& & \nonumber \\
\lefteqn{
%\vphantom{\sum^a}
\hphantom{N_2}=
2(L \p 1) t^{(L+2)(2L+1)/2}
Q\left(t^{12L(L+2)}\right)
Q\left(t^{12(L+1)(L+2)}\right)
\te\!\left(\case{\pi}{6},t^{2L(L+1)}\right)
\frac{
\te\!\left(\case{2r\pi}{L},t^{6(L+1)(L+2)}\right)}{
\tt\!\left(\case{r\pi}{L},t^{6(L+1)(L+2)}\right)}}
\nonumber
\end{eqnarray}
where $r=1,3,5,\ldots, L-2$ and $L=3,5,7,\ldots$.
\smallskip

{\it Proof:}
The proof of this denominator identity is
similar to that presented for regime 1$^+$.
Using the representation of the $\vartheta$-functions as
infinite sums yields
\setlength{\mathindent}{\mathin}
\begin{eq}
N_2=-\ib t^{3L(L+2)/2} \sum_{a=0}^{2L+1}
\sum_{j,k,\ell}
(-1)^j
\e^{\pi \is a (2j+1+kL-\ell)/(L+1)}
\e^{\pi \is r \ell/L}
t^{6L(L+2)(j^2+j)+3L(L+2)k^2+\ell^2}\! .
\end{eq}
Performing the sum over $a$ gives
$\ell=2j+1+kL+2m(L+1)$, where $m$ assumes integer values.
Thus we find
\begin{eqnarray}
\lefteqn{
N_2=- 2 \ib (L+1) \, t^{3L(L+2)/2}
\sum_{j,k,m} (-1)^{j+k} \,
\e^{\pi \is r(2j+1+2m)/L} }
\nonumber\\
& & \qquad \qquad \qquad \qquad \qquad \qquad \times \:
\vphantom{\sum^a}
t^{6L(L+2)L(j^2+j)+3L(L+2)k^2+[2j+1+kL+2m(L+1)]^2}
\end{eqnarray}
where we have used the fact that $r$ is odd.

As before, we carry out a sequence of transformations to diagonalise
the quadratic exponent of $t$.
First we make the replacements $m\to m-j$ and $k\to k+2j$.
Then we set $k=3\ell+\alpha$ and $m=3n+\beta$ and finally
we substitute $j\to j-\ell$ followed by $\ell\to \ell-n$.
After making these transformations we obtain
\begin{eqnarray}
\lefteqn{
N_2=-2 \ib (L + 1) \, t^{(3L^2+6L+2)/2}
\sum_{\alpha,\beta=0}^2 (-1)^{\alpha} \,
t^{4\beta(\beta+1)+
2L(3\alpha^2+4\beta^2+2\alpha\beta+\alpha+2\beta)+
4L^2(\alpha^2+\beta^2+\alpha\beta)} }
\nonumber\\
& & \qquad \quad \times \:
\e^{\pi \is r(2\beta+1)/L}
E\!\left(-\e^{6\pi \is r/L} \,
t^{12(L+1)(L+2)(2+\beta)},
t^{36(L+1)(L+2)}\right)
\vphantom{\sum_a^a} \\
& & \qquad \quad \times \:
E\!\left(t^{12L(L+2)(2+\alpha)},t^{36L(L+2)}\right)
E\!\left(-t^{12L(L+1)(1+\alpha+\beta)},t^{36L(L+1)}\right) .
\vphantom{\sum^a}
\nonumber
\end{eqnarray}
The terms with $\alpha=1$ vanish and the terms with $\beta=1$ cancel.
The remaining four terms can be written as
\begin{eqnarray}
\lefteqn{
N_2=2 \ib (L+1) \,
t^{(3L^2+6L+2)/2} \,
\e^{-\pi \is r/L}
Q\left(t^{12L(L+2)}\right)
\vphantom{\sum_a} }
\nonumber\\
\lefteqn{\qquad \qquad \times \!
\left[ E\!\left(-t^{12L(L+1)},t^{36L(L+1)}\right)-t^{4L(L+1)}
E\!\left(-1,t^{36L(L+1)}\right) \right]
\vphantom{\sum_a^a} } \\
\lefteqn{\qquad \qquad\times \!
\left[
E\!\left(-\e^{6\pi \is r/L} \,
t^{12(L+1)(L+2)},t^{36(L+1)(L+2)}\right) \right.
\vphantom{\sum_a^a}}
\nonumber \\
\lefteqn{\qquad \qquad \qquad \qquad
\qquad \qquad \left.
- \e^{2\pi \is r/L}
E\!\left(-\e^{-6\pi \is r/L} \, t^{12(L+1)(L+2)},
t^{36(L+1)(L+2)}\right)
\right].
\vphantom{\sum^a} } \nonumber
\end{eqnarray}
If we again apply the two identities of
equation (\ref{eq:A.Eid})
and use the relation (\ref{eq:A.thetasasE}),
we obtain the required right-hand
side of equation (\ref{eq:A.denom2}).
%  #] denom2:

%  #[ denom3:
\subsection{Regime 3$^+$}\label{app:A.den3}
\subsubsection*{Antiferromagnetic phase}
For the antiferromagnetic phases of regime 3$^+$
the denominator identity is given by
\setlength{\mathindent}{0 cm}
\begin{eqnarray}
\lefteqn{N_3\equiv
\sum_{a=0}^{2L+1}
\td\!\left(\case{(1+2\mu_a)\pi}{8},t^{(L+1)(L+2)/4}\right)
\te\!\left(\case{a\pi}{L+1},t^{2(L^2-4)}\right)
\td\!\left(\case{a\pi L}{2(L+1)},t^{L^2-4}\right)
\td\!\left(\case{\pi}{2}\left(
\case{a}{L+1} \m
\case{s}{L+2} \right),t\right)}
\nonumber \\
& & \nonumber \\
\lefteqn{\hphantom{N_3}=
%\qquad\qquad\qquad\qquad=
2(L + 1) \, t^{(L-1)(L+2)/4}
Q\left(t^{4(L^2-4)}\right)
Q\left(t^{2(L+1)(L+2)}\right)
%\vphantom{\sum_a^a}
} \label{eq:A.denom3}\\
& & \nonumber \\
\lefteqn{
%\vphantom{\sum^a}
\qquad\qquad\qquad\qquad\qquad\qquad\qquad\qquad\quad \times
\td\!\left(\case{\pi}{4},t^{(L+1)(L+2)}\right)
\te\!\left(\case{s\pi}{L+2},t^{(L+1)(L-2)}\right)} \nonumber
\end{eqnarray}
where $\mu_a=a+\sigma \bmod 2$ with $\sigma=0,1$;
$s=1,3,5,\ldots, L$ and $L=3,5,7,\ldots$.

\smallskip
{\it Proof:}
We split the summation on the
left-hand side depending on the parity of $a$.
Using the fact that $\td(u+\case{1}{2}\pi,p)=
\tv(u,p)=\td(u,-p)$, we get
\begin{eqnarray}
\lefteqn{
N_3=
\td\!\left(\case{\pi}{8},
(-1)^{\sigma}t^{(L+1)(L+2)/4}\right)
\sum_{a=0}^{L}
\te\!\left(\case{2a\pi}{L+1},t^{2(L^2-4)}\right)
\td\!\left(\case{a\pi L}{L+1},t^{L^2-4}\right)
\td\!\left(\case{\pi}{2}\left(
\case{2a}{L+1} \m
\case{s}{L+2} \right),t\right)} \nonumber \\
%& & \nonumber \\
\lefteqn{\hphantom{N_3} +
\td\!\left(\case{3\pi}{8},
(-1)^{\sigma}t^{(L+1)(L+2)/4}\right)
\sum_{a=0}^{L}
\te\!\left(\case{(2a+1)\pi}{L+1},t^{2(L^2-4)}\right) } \\
%& & \nonumber \\
\lefteqn{
\qquad \qquad \qquad \qquad \qquad \qquad \qquad \qquad
\times \:
\td\!\left(\case{(2a+1)\pi L}{2(L+1)},t^{L^2-4}\right)
\td\!\left(\case{\pi}{2}\left(
\case{2a+1}{L+1} \m
\case{s}{L+2} \right),t\right). } \nonumber
\end{eqnarray}
We now represent the theta-functions with $a$-dependent arguments by
series to obtain
\begin{eqnarray}
\lefteqn{
N_3=-\ib t^{(L^2-4)/2}
\sum_{a=0}^L \sum_{j,k,\ell}
(-1)^j \,
\e^{2\pi \is a(2j+1+kL+\ell)/(L+1)} \,
\e^{-\pi \is s\ell/(L+2)} \,
t^{(L^2-4)(2j^2+2j+k^2)+\ell^2}}
\nonumber\\
\lefteqn{ \vphantom{\sum^a}
\qquad \times \left[
\td\!\left(\case{\pi}{8},(-1)^{\sigma}t^{(L+1)(L+2)/4}\right)+
\e^{\pi \is (2j+1+kL+\ell)/(L+1)}
\td\!\left(\case{3\pi}{8},(-1)^{\sigma}t^{(L+1)(L+2)/4}\right)
 \right]. }
\end{eqnarray}
Performing the sum over $a$, we find that
$\ell=-2j-1-kL+m(L+1)$ and hence
\setlength{\mathindent}{\mathin}
\begin{eqnarray}
\lefteqn{
N_3= -\ib(L+1) \, t^{(L^2-4)/2}
\sum_{j,k,m} (-1)^{j+k} \,
\e^{\pi \is s (2j+1-2k+m)/(L+2)} }
\nonumber\\
& & \qquad \qquad \quad \times \:
\vphantom{\sum_a^a}
t^{(L^2-4)(2j^2+2j+k^2)+[2j+1+kL-m(L+1)]^2} \\
& & \qquad \qquad \quad \times \:
\left[
(-1)^m
\td\!\left(\case{\pi}{8},
(-1)^{\sigma}t^{(L+1)(L+2)/4}\right)
+
\td\!\left(\case{3\pi}{8},
(-1)^{\sigma}t^{(L+1)(L+2)/4}\right)\right]
\nonumber
\end{eqnarray}
where we have used the fact that $s$ is odd.

Again we carry out a sequence of transformations to
diagonalise the quadratic exponent of $t$.
First we make the replacements $m\to m-2j+2k$ and $k\to k+2j$,
followed by
the substitutions $k=3\ell+\alpha$ and $m=4n+\beta$. Finally
we set $j\to j-\ell$ and $\ell\to \ell-n$
to obtain
\setlength{\mathindent}{0 cm}
\begin{eqnarray}
\lefteqn{
N_3 = -\ib(L+1) \, t^{(L^2-2)/2}
\sum_{\alpha=0}^2 \sum_{\beta=0}^3 (-1)^{\alpha}\,
t^{\beta^2+4\alpha\beta-4\alpha-2\beta
+2L(2\alpha^2+\beta^2+3\alpha\beta-\alpha-\beta)+
L^2(2\alpha^2+\beta^2+2\alpha\beta)}} \nonumber \\
& & \quad \qquad
\times \left[ (-1)^{\beta}
\td\!\left(\case{\pi}{8},
(-1)^{\sigma}t^{(L+1)(L+2)/4}\right)+
\td\!\left(\case{3\pi}{8},
(-1)^{\sigma}t^{(L+1)(L+2)/4}\right)\right]
\vphantom{\sum_a^a}
\nonumber \\
& & \quad \qquad
\times \:
\e^{\pi \is s(\beta+1)/(L+2)}
E\!\left(-\e^{4\pi \is s/(L+2)} \,
t^{2(L+1)(L-2)(3+\beta)},
t^{8(L+1)(L-2)}\right)
\vphantom{\sum_a^a}
\\
& & \quad \qquad
\times \:
E\!\left(t^{4(L^2-4)(2+\alpha)},t^{12(L^2-4)}\right)
E\!\left(-t^{2(L+1)(L+2)(5+4\alpha+3\beta)},
t^{24(L+1)(L+2)}\right).
\vphantom{\sum^a}
\nonumber
\end{eqnarray}
In this expression the terms with $\alpha=1$ vanish
and the terms with $\beta=1$ and $\beta=3$ cancel.
Using the simple identity
\setlength{\mathindent}{\mathin}
\begin{equation}
\td\!\left(\case{\pi}{8},p\right)+
\td\!\left(\case{3\pi}{8},p\right)
= 2 \td\!\left(\case{\pi}{4},p^4\right)
\label{eq:A.simpleident}
\end{equation}
which follows directly from (\ref{eq:A.thetasums}),
the remaining four terms factorise as
\setlength{\mathindent}{0 cm}
\begin{eqnarray}
\lefteqn{ \vphantom{\sum_a}
N_3 =
2\ib(L+1) \, \e^{-\pi \is s/(L+2)} \, t^{(L^2-2)/2}
Q\left(t^{4(L^2-4)}\right)
\td\!\left(\case{\pi}{4},t^{(L+1)(L+2)}\right)}
\nonumber\\
\lefteqn{\vphantom{\sum_a^a}
\qquad \times
\left[ E\!\left(-t^{10(L+1)(L+2)},t^{24(L+1)(L+2)}\right)
- t^{2(L+1)(L+2)}
E\!\left(-t^{2(L+1)(L+2)},t^{24(L+1)(L+2)}\right) \right] }
\nonumber \\
\lefteqn{\vphantom{\sum_a^a}
\qquad \times
\left[ E\!\left(-\e^{4\pi \is s/(L+2)} \, t^{2(L+1)(L-2)},
t^{8(L+1)(L-2)}\right)
\right. }
\label{eq:A.fred} \\
\lefteqn{\vphantom{\sum_a^a}
\qquad \qquad \qquad \qquad \qquad \qquad \left.
-\e^{2\pi \is s/(L+2)}
E\!\left(-\e^{-4\pi \is s/(L+2)} \,
t^{2(L+1)(L-2)},t^{8(L+1)(L-2)}\right) \right].}\nonumber
\end{eqnarray}
We now note the identities
\setlength{\mathindent}{\mathin}
\begin{eqnarray}
\lefteqn{ \vphantom{\sum_a^a}
E\left(-x^2 p ,p^4\right)
- x E\left(-x^{-2} p ,p^4\right)
= E\left(x,p\right)}
\nonumber \\
\lefteqn{ \vphantom{\sum^a}
E\left(-p^5,p^{12}\right) - p E\left(-p,p^{12}\right)=Q(p).}
\label{eq:A.useful}
\end{eqnarray}
The first one follows directly from Liouville's theorem and the second
one can be obtained from the first upon setting $x=p$ after
replacing $p$ by $p^3$.
If we apply these two identities, as well as
equation (\ref{eq:A.thetasasE}),
we find precisely the right-hand side of
equation (\ref{eq:A.denom3}). This completes the proof.

\subsubsection*{Ferromagnetic phase}
The denominator identity for the ferromagnetic phases in
regime 3$^+$ reads
\setlength{\mathindent}{0 cm}
\begin{eqnarray}
\lefteqn{N_4\equiv
\sum_{a=0}^{2L+1}
\te\!\left(\case{a\pi}{L+1},t^{2(L^2-4)}\right)
\td\!\left(\case{a\pi L}{2(L+1)},t^{L^2-4}\right)
\td\!\left(\case{\pi}{2}\left(
\case{a}{L+1} \m \case{s}{L+2}\right),t\right) }
\label{eq:A.denom3b} \\
\lefteqn{
\hphantom{N_4}=
4(L\p 1) t^{L(L+2)/2}
Q^2\!\left(t^{4(L+1)(L+2)}\right)
Q\!\left(t^{4(L^2-4)}\right)
\frac{
\te\!\left(\case{s\pi}{L+2},t^{2(L+1)(L-2)}\right)
\td\!\left(\case{s\pi}{L+2},t^{2(L+1)(L-2)}\right)}{
\te\!\left(\case{\pi}{4},t^{(L+1)(L-2)}\right)
\te\!\left(\case{\pi}{4},t^{(L+1)(L+2)}\right) }}
\nonumber
\end{eqnarray}
where $s=2,4,6,\ldots,L+1$ and $L=3,5,7,\ldots$.

\smallskip
{\it Proof:}
We begin by writing the sum over the $\vartheta$-functions
in the left-hand side as
\setlength{\mathindent}{\mathin}
\begin{eq}
N_4=-\ib t^{(L^2-4)/2}
\sum_{a=0}^{2L+1}
\sum_{j,k,\ell}
(-1)^j \,
\e^{\pi \is a (2j+1+kL+\ell)/(L+1)}  \,
\e^{-\pi \is s\ell/(L+2)} \,
t^{(L^2-4)(2j^2+2j+k^2)+\ell^2}.
\end{eq}
Summing over $a$ yields $\ell=-2j-1-kL+2m(L+1)$
and hence, using the fact that $s$ is even, we get
\begin{eqnarray}
\lefteqn{
N_4=-2 \ib (L+1) \,t^{(L^2-4)/2}
\sum_{j,k,m} (-1)^j \,
\e^{\pi \is s(2j+1-2k+2m)/(L+2)} }
\nonumber\\
& & \qquad \qquad \qquad \qquad \qquad \qquad
\times \:
\vphantom{\sum^a}
t^{(L^2-4)(2j^2+2j+k^2)+[2j+1+kL-2m(L+1)]^2}.
\end{eqnarray}

In order to diagonalise the
quadratic exponent of $t$, we make the
replacements $m\to m-j+k$ and $k\to k+2j$ followed by
the substitutions
$k=3\ell+\alpha$ and $m=2n+\beta$.
Further replacing
$j\to j-\ell$ and $\ell\to \ell-n$ leads to the result
\begin{eqnarray}
\lefteqn{
N_4=-2\ib (L+1) \, t^{(L^2-2)/2}
\sum_{\alpha=0}^2 \sum_{\beta=0}^1
t^{4(\beta^2+2\alpha\beta-\alpha-\beta)
+2L(2\alpha^2+4\beta^2+6\alpha\beta-\alpha-2\beta)
+2L^2(\alpha^2+2\beta^2+2\alpha\beta)} } \nonumber \\
\lefteqn{\vphantom{\sum_a^a} \qquad
\qquad  \times
\e^{\pi \is s (2\beta+1)/(L+2)}
E\!\left(\e^{4\pi \is s/(L+2)}\,t^{2(L+1)(L-2)(3+2\beta)},
t^{8(L+1)(L-2)}\right) } \\
\lefteqn{\qquad \vphantom{\sum_a^a}
\qquad \times
E\!\left(t^{4(L^2-4)(2+\alpha)},t^{12(L^2-4)}\right)
E\!\left(t^{2(L+1)(L+2)(5+4\alpha+6\beta)},t^{24(L+1)(L+2)}\right).}
\nonumber
\end{eqnarray}
The terms with $\alpha=1$ vanish and
the other terms may be combined to give
\begin{eqnarray}
\lefteqn{ \vphantom{\sum_a}
N_4=
2 \ib (L+1) \, \e^{-\pi \is s/(L+2)} \, t^{(L^2-2)/2}
Q\left(t^{4(L^2-4)}\right) }\nonumber \\
\lefteqn{\vphantom{\sum_a^a}
\qquad \times
\left[
E\!\left(t^{10(L+1)(L+2)},t^{24(L+1)(L+2)}\right) +
t^{2(L+1)(L+2)}
E\!\left(t^{2(L+1)(L+2)},t^{24(L+1)(L+2)}\right)
\right]}
\nonumber \\
\lefteqn{\vphantom{\sum_a^a}
\qquad \times
\left[
E\!\left(\e^{4\pi \is s/(L+2)}\,t^{2(L+1)(L-2)},t^{8(L+1)(L-2)}\right)
\right. } \\
\lefteqn{\vphantom{\sum_a^a}
\qquad \qquad \qquad \qquad \qquad \qquad \left.
-\e^{2\pi \is s/(L+2)} E\!\left(\e^{-4\pi \is s/(L+2)}\,
t^{2(L+1)(L-2)},t^{8(L+1)(L-2)}\right) \right].} \nonumber
\end{eqnarray}
If we multiply this expression with
the prefactors outside the sum
in equation (\ref{eq:A.denom3b}) and
use the following identities
\begin{eqnarray}
\lefteqn{\vphantom{\sum_a^a}
E\left(x^2 p,p^4\right)
- x E\left(x^{-2} p,p^4\right) =
\sqrt{2} \, \e^{-\pi \is/4} \:
\frac{E\left(x,p^2\right) E\left(-xp,p^2\right)}
{E(i,p)} }
\nonumber \\
\lefteqn{
E\left(p^5,p^{12}\right) + p E\left(p,p^{12}\right)
= \sqrt{2} \, \e^{-\pi \is/4} \:
\frac{Q^2\left(p^2\right)}
{E(i,p)}}
\label{eq:A.handy}
\end{eqnarray}
we find precisely the right-hand side of
equation (\ref{eq:A.denom3b}).
The proof of these two relations
is similar to those obtained earlier.
The first identities follows
from application of Liouville's theorem and the second
identity is a consequence of the first.
Replacing the nome $p$ by $p^3$ and then setting $x=p^2$ yields
%\begin{eqnarray}
%\lefteqn{
%E\left(p^5,p^{12}\right) + p E\left(p,p^{12}\right) } \nonumber \\
%& & \qquad =
%\sqrt{2} \, \e^{-\pi \is/4} \:
%Q(p^2) \prod_{n=1}^{\infty} \frac{
%\left(1+p^{6n-5}\right)
%\left(1+p^{6n-1}\right)
%\left(1+p^{3n}\right)}
%{\left(1+p^{6n}\right)}
%\times \frac{
% \left(1+p^{6n-3}\right)}
%{\left(1+p^{6n-3}\right)} \nonumber \\
%& & \qquad =
%\sqrt{2} \, \e^{-\pi \is/4}  \: Q(p^2)
%\prod_{n=1}^{\infty} \left(1+p^{2n-1}\right) \\
%& & \qquad =
%\sqrt{2} \, \e^{-\pi \is/4} \:
%\frac{\left[Q\left(p^2\right) \right]^2}
%{E(i,p)}. \nonumber
%\end{eqnarray}
\begin{eqnarray}
\lefteqn{
E\left(p^5,p^{12}\right) + p E\left(p,p^{12}\right)
= \sqrt{2} \, \e^{-\pi \is/4} \:
Q(p^2) \:\frac{E(-p^5,p^6)}{E(i,p^3)} } \nonumber \\
\lefteqn{
\hphantom{E\left(p^5,p^{12}\right) + p E\left(p,p^{12}\right)}
=
Q(p^2) \prod_{n=1}^{\infty} \frac{
\left(1+p^{6n-5}\right)
\left(1+p^{6n-1}\right)
\left(1+p^{3n}\right)}
{\left(1+p^{6n}\right)} }\nonumber \\
\lefteqn{
\hphantom{E\left(p^5,p^{12}\right) + p E\left(p,p^{12}\right)}
= Q(p^2)
\prod_{n=1}^{\infty} \left(1+p^{2n-1}\right) }\\
\lefteqn{
\hphantom{E\left(p^5,p^{12}\right) + p E\left(p,p^{12}\right)}
=
\sqrt{2} \, \e^{-\pi \is/4} \:
\frac{Q^2\left(p^2\right)}
{E(i,p)}.} \nonumber
\end{eqnarray}

%  #] denom3:

%  #[ denom4:
\subsection{Regime 4$^+$}\label{app:A.den4}
\subsubsection*{Antiferromagnetic phase}
The denominator identity in the antiferromagnetic phase of regime
4$^+$ is given by
\setlength{\mathindent}{0 cm}
\begin{eqnarray}
\lefteqn{N_5\equiv
\sum_{a=0}^{2L+1}
\td\!\left(\case{(1+2\mu_a)\pi}{8},t^{L(L+1)/4}\right)
\te\!\left(\case{a\pi}{L+1},t^{2L(L+4)}\right)
\td\!\left(\case{a\pi L}{2(L+1)},t^{L(L+4)}\right)
\td\!\left(\case{\pi}{2}\left(
\case{r}{L} \m \case{a}{L+1} \right),t\right)}
\nonumber \\
& & \nonumber \\
\lefteqn{
\hphantom{N_5}=
2(L+1) \,t^{L(L+3)/4} Q\left(t^{4L(L+4)}\right)
Q\left(t^{2L(L+1)}\right)
\td\!\left(\case{\pi}{4},t^{L(L+1)}\right)
\te\!\left(\case{r\pi}{L},t^{(L+1)(L+4)}\right)}
\label{eq:A.denom4}
\end{eqnarray}
where $\mu_a=a+\sigma \bmod 2$, with $\sigma=0,1$;
$r=1,3,5,\ldots, L-2$ and $L=3,5,7,\ldots$.

\smallskip
{\it Proof:}
Following the proof for the antiferromagnetic phase
of regime 3$^+$,
we split the summation on the
left-hand side
depending on the parity of $a$.
Representing the theta functions with
$a$-dependent arguments by series yields
\setlength{\mathindent}{\mathin}
\begin{eqnarray}
\lefteqn{
N_5 = -\ib t^{L(L+4)/2}
\sum_{a=0}^L
\sum_{j,k,\ell}
(-1)^j \,
\e^{2\pi \is a(2j+1+kL-\ell)/(L+1)} \,
\e^{\pi \is r\ell/L} \,
t^{L(L+4)(2j^2+2j+k^2)+\ell^2}}
\nonumber\\
\lefteqn{\vphantom{\sum_a^a} \qquad  \times
\left[
\td\!\left(\case{\pi}{8},(-1)^{\sigma}t^{L(L+1)/4}\right) +
\e^{\pi \is (2j+1+kL-\ell)/(L+1)}
\td\!\left(\case{3\pi}{8},(-1)^{\sigma}t^{L(L+1)/4}\right)
\right]. }
\end{eqnarray}
Summing the roots of unity over $a$ we see that
$\ell=2j+1+kL+m(L+1)$ and as a result we have
\setlength{\mathindent}{0 cm}
\begin{eqnarray}
\lefteqn{
N_5=-\ib(L+1) \, t^{L(L+4)/2}
\sum_{j,k,m} (-1)^{j+k} \,
\e^{\pi \is r (2j+1+m)/ L} \,
t^{L(L+4)(2j^2+2j+k^2)+[2j+1+kL+m(L+1)]^2} }
\nonumber\\
& & \qquad \qquad \qquad \qquad \times
\left[
(-1)^m
\td\!\left(\case{\pi}{8},
(-1)^{\sigma}t^{L(L+1)/4}\right)+
\td\!\left(\case{3\pi}{8},
(-1)^{\sigma}t^{L(L+1)/4}\right)\right]
\end{eqnarray}
where we have used the fact that $r$ is odd.

We diagonalise the
quadratic exponent of $t$ by carrying out the
following sequence of replacements.
We first set $m\to m-2j$ and $k\to k+2j$. We then replace
$k=3\ell+\alpha$ and $m=4n+\beta$ and
finally we substitute $j\to j-\ell$ followed by $\ell\to \ell-n$.
We then have
\setlength{\mathindent}{\mathin}
\begin{eqnarray}
\lefteqn{
N_5= -\ib (L+1) \, t^{(L^2+4L+2)/2}
\sum_{\alpha=0}^2 \sum_{\beta=0}^3
(-1)^{\alpha} \,
t^{\beta(\beta+2)+
2L(2\alpha^2+\beta^2+\alpha\beta+\alpha+\beta)+
L^2(2\alpha^2+\beta^2+2\alpha\beta)}} \nonumber \\
& & \qquad \quad
\times\left[ (-1)^{\beta}
\td\!\left(\case{\pi}{8},
(-1)^{\sigma}t^{L(L+1)/4}\right)+
\td\!\left(\case{3\pi}{8},
(-1)^{\sigma}t^{L(L+1)/4}\right)\right]
\vphantom{\sum_a^a}
\nonumber \\
& & \qquad \quad
\times \:
\e^{\pi \is r(\beta+1)/L}
E\!\left(- \e^{4\pi \is r/L}\, t^{2(L+1)(L+4)(3+\beta)},
t^{8(L+1)(L+4)}\right)
\vphantom{\sum_a^a}
\\
& & \qquad \quad
\times \:
E\!\left(t^{4L(L+4)(2+\alpha)},t^{12L(L+4)}\right)
E\!\left(-t^{2L(L+1)(5+4\alpha+3\beta)},
t^{24L(L+1)}\right).
\vphantom{\sum^a}
\nonumber
\end{eqnarray}
As before, the terms with $\alpha=1$ vanish
and the terms with $\beta=1$ and $\beta=3$ cancel.
Using the relation (\ref{eq:A.simpleident}), the remaining four
terms can be written as
\begin{eqnarray}
\lefteqn{  \vphantom{\sum_a}
N_5 =
2 \ib(L+1) \, \e^{-\pi \is r/L} \, t^{(L^2+4L+2)/2}
Q\left(t^{4L(L+4)}\right)
\td\!\left(\case{\pi}{4},t^{L(L+1)}\right)}
\nonumber\\
\lefteqn{\qquad \qquad \times \!
\left[
E\!\left(-t^{10L(L+1)},t^{24L(L+1)}\right) - t^{2L(L+1)}
E\!\left(-t^{2L(L+1)},t^{24L(L+1)}\right) \right]
\vphantom{\sum_a^a}} \nonumber \\
\lefteqn{\qquad \qquad
\times \!
\left[ E\!\left(-\e^{4\pi \is r /L} \, t^{2(L+1)(L+4)},
t^{8(L+1)(L+4)}\right) \right.
\vphantom{\sum_a^a} }  \\
\lefteqn{ \qquad \qquad \qquad \qquad \qquad \qquad
\left.
-\e^{2\pi \is r/L}
E\!\left(-\e^{-4\pi \is r/L} \,
t^{2(L+1)(L+4)},t^{8(L+1)(L+4)}\right) \right].
\vphantom{\sum^a}} \nonumber
\end{eqnarray}
After application of the identities (\ref{eq:A.useful}),
we obtain the required right-hand side of (\ref{eq:A.denom4}).

\subsubsection*{Ferromagnetic phase}
The denominator identity for
the ferromagnetic phase in regime 4$^+$ is
\setlength{\mathindent}{0 cm}
\begin{eqnarray}
\lefteqn{
N_6\equiv
\sum_{a=0}^{2L+1}
\te\!\left(\case{a\pi}{L+1},t^{2L(L+4)}\right)
\td\!\left(\case{a\pi L}{2(L+1)},t^{L(L+4)}\right)
\td\!\left(\case{\pi}{2}\left(
\case{r}{L} \m \case{a}{L+1}\right),t\right)}
\label{eq:A.denom4b} \\
%& & \nonumber \\
\lefteqn{\hphantom{N_6}=
4(L+1) \, t^{L(L+2)/2}
Q^2\left(t^{4L(L+1)}\right)
Q\left(t^{4L(L+4)}\right)
\frac{
\te\!\left(\case{r\pi}{L},t^{2(L+1)(L+4)}\right)
\td\!\left(\case{r\pi}{L},t^{2(L+1)(L+4)}\right)}{
\te\!\left(\case{\pi}{4},t^{L(L+1)}\right)
\te\!\left(\case{\pi}{4},t^{(L+1)(L+4)}\right) }}
\nonumber
\end{eqnarray}
where $L=3,5,7,\ldots$, and $r=2,4,6,\ldots,L-1$.

\smallskip
{\it Proof:}
We rewrite the sum over the $\vartheta$-functions
in the left-hand side as
\setlength{\mathindent}{\mathin}
\begin{eq}
N_6=-\ib t^{L(L+4)/2}
\sum_{a=0}^{2L+1} \sum_{j,k,\ell}
(-1)^j \,
\e^{\pi \is a(2j+1+kL-\ell)/(L+1)} \,
\e^{\pi \is r\ell/L} \,
t^{L(L+4)(2j^2+2j+k^2)+\ell^2}.
\end{eq}
By summing over the roots of unity, this vanishes unless
$\ell=2j+1+kL+2m(L+1)$.
As a result we find
\begin{eq}
N_6=-2\ib(L + 1)  t^{L(L+4)/2}
\! \sum_{j,k,m} (-1)^j
\e^{\pi \is r (2j+1+2m)/L}
t^{L(L+4)(2j^2+2j+k^2)+[2j+1+kL+2m(L+1)]^2}
\end{eq}
where we have used the fact that $r$ is even.

This expression is diagonalised by carrying out the sequence of
transformations $m\to m-j$, $k\to k+2j$, $k=3\ell+\alpha$,
$m=2n+\beta$, $j\to j-\ell$ and $\ell\to \ell-n$. This gives
\begin{eqnarray}
\lefteqn{
N_6=-2 \ib (L+1) \, t^{(L^2+4L+2)/2}
\sum_{\alpha=0}^2 \sum_{\beta=0}^1
t^{4\beta(\beta+1)
+2L(2\alpha^2+4\beta^2+2\alpha\beta+\alpha+2\beta)
+2L^2(\alpha^2+2\beta^2+2\alpha\beta)} } \nonumber \\
\lefteqn{\qquad \vphantom{\sum_a^a}
\qquad \qquad \times
\e^{\pi \is r(2\beta+1)/L}
E\!\left(\e^{4\pi \is r/L} \, t^{2(L+1)(L+4)(3+2\beta)},
t^{8(L+1)(L+4)}\right) } \\
\lefteqn{\qquad \vphantom{\sum_a^a}
\qquad \qquad \times
E\!\left(t^{4L(L+4)(2+\alpha)},t^{12L(L+4)}\right)
E\!\left(t^{2L(L+1)(5+4\alpha+6\beta)},t^{24L(L+1)}\right).}
\nonumber
\end{eqnarray}
Here the
terms with $\alpha=1$ vanish. The other terms may be combined to
\begin{eqnarray}
\lefteqn{
N_6=
2 \ib (L+1) \, \e^{-\pi \is r/L} \, t^{(L^2+4L+2)/2}
Q\left(t^{4L(L+4)}\right)
 \vphantom{\sum_a} } \nonumber \\
\lefteqn{ \qquad \qquad
\times \!
\left[ E\!\left(t^{10L(L+1)},t^{24L(L+1)}\right)+
t^{2L(L+1)} E\!\left(t^{2L(L+1)},t^{24L(L+1)}\right)
\right] \vphantom{\sum_a^a}}
\nonumber \\
\lefteqn{\qquad \qquad
\times \!
\left[ E\!\left(\e^{4\pi \is r/L} \,
t^{2(L+1)(L+4)},t^{8(L+1)(L+4)}\right)
\right.
\vphantom{\sum_a^a} } \\
\lefteqn{
\qquad \qquad \qquad \qquad \qquad \qquad \left.
-\e^{2\pi \is r/L} E\!\left(\e^{-4\pi \is r/L} \,
t^{2(L+1)(L+4)},t^{8(L+1)(L+4)}\right)
\right].
\vphantom{\sum^a} } \nonumber
\end{eqnarray}
If we again use the identities (\ref{eq:A.handy}),
and multiply the above expression by the correct
prefactors from the left-hand side of
(\ref{eq:A.denom4b}), we indeed obtain the right-hand side of
(\ref{eq:A.denom4b}).

%  #] denom4:

%  #] AppendixB:

%  #[ AppendixC:
\nsection{Theta functions}\label{app:A.elliptic}
In this appendix we list several definitions and relations
for the Jacobian $\vartheta$-functions, used in the
main text.
For a more complete introduction, we refer the reader
to {\em e.g.,} \cite{Whitwat}.

The four standard $\vartheta$-functions, of
{\em nome} $p$, $|p|<1$,  are defined as the
following infinite sums \cite{GR}:
\setlength{\mathindent}{0 cm}
\begin{eqnarray}
\lefteqn{
\te(u,p) = \te(u) =
-\ib \, \sum_{n=-\infty}^{\infty}
(-1)^n p^{(2n+1)^2/4} \: \e^{(2n+1)\is u}
=2 \, \sum_{n=0}^{\infty}
(-1)^n p^{(2n+1)^2/4} \sin (2n+1) u } \nonumber \\
\lefteqn{
\tt(u,p) = \tt(u) =
\sum_{n=-\infty}^{\infty}
p^{(2n+1)^2/4} \: \e^{(2n+1)\is u}
=2 \, \sum_{n=0}^{\infty}
p^{(2n+1)^2/4} \cos (2n+1) u } \nonumber \\
\lefteqn{
\td(u,p) = \td(u) =
\sum_{n=-\infty}^{\infty}
p^{n^2} \: \e^{2 n \is u }
= 1 + 2 \, \sum_{n=1}^{\infty}
p^{n^2} \cos 2 n u }
\label{eq:A.thetasums} \\
\lefteqn{
\tv(u,p) = \tv(u) = \sum_{n=-\infty}^{\infty}
(-1)^n p^{n^2} \e^{2 n \is u}
=1 + 2 \, \sum_{n=1}^{\infty}
(-1)^n p^{n^2} \cos 2 n u. } \nonumber
\end{eqnarray}
By virtue of Jacobi's triple product identity \cite{Andrews},
the theta functions admit a representation as infinite products
\setlength{\mathindent}{\mathin}
\begin{eqnarray}
\lefteqn{
\te(u) = 2 \, p^{1/4} \sin u \prod_{n=1}^{\infty}
(1-2 p^{2n} \cos 2u+p^{4n})(1-p^{2n}) } \nonumber \\
\lefteqn{
\tt(u) = 2 p^{1/4} \cos u \prod_{n=1}^{\infty}
(1+2 p^{2n} \cos 2u+p^{4n})(1-p^{2n})} \nonumber \\
\lefteqn{
\td(u) = \prod_{n=1}^{\infty}
(1+2p^{2n-1} \cos 2u+p^{4n-2})(1-p^{2n}) } \\
\lefteqn{
\tv(u) = \prod_{n=1}^{\infty}
(1-2p^{2n-1} \cos 2u+p^{4n-2})(1-p^{2n}).} \nonumber
\end{eqnarray}
Another function which proves to be useful is
\begin{equation}
E(x,p)=
\sum_{n=-\infty}^{\infty}(-1)^n p^{n(n-1)/2} x^n =
\prod_{n=1}^{\infty}
(1-p^{n-1}x)(1- p^n x^{-1})(1-p^{n}). \label{eq:A.Efunc}
\end{equation}
{}From its definition it follows immediately that
\begin{eqnarray}
E(x,p)&=&E(p x^{-1},p)=-xE(x^{-1},p) \nonumber \\
\vphantom{\sum_a^a}
E(x p^n,p)&=&(-x)^{-n} p^{-n(n-1)/2} E(x,p) \nonumber \\
E(p,p^3)&=& Q(p)  \vphantom{\sum_a}
\label{eq:A.limEfun} \\
\lim_{p\rightarrow 0} E(x p^a,p^b) &=& \left\{
\begin{array}{lc} 1 & \quad 0<a<b \\ 1-x & \quad a=0 \end{array} \right.
\nonumber
\end{eqnarray}
where the function $Q$ is defined as
\begin{equation}
Q(p) = \prod_{n=1}^{\infty} \left( 1-p^n \right).
\end{equation}
The $\vartheta$-functions can be expressed in terms of the $E$-function
as
\begin{eqnarray}
\lefteqn{
\te(u,p) = \ib p^{1/4} \, \e^{-\is u}
E\left(\e^{2\is u},p^2\right) } \nonumber \\
\lefteqn{
\tt(u,p) = p^{1/4} \, \e^{-\is u}
E\left(-\e^{2\is u},p^2\right) } \nonumber \\
\lefteqn{
\td(u,p) =
E\left(-p\,\e^{2\is u},p^2\right) } \label{eq:A.thetasasE} \\
\lefteqn{
\tv(u,p) =
E\left(p\,\e^{2\is u},p^2\right). } \nonumber
\end{eqnarray}
Likewise, the {\em conjugate modulus} transformation of the theta
functions, which relates $\vartheta$-functions of nome
$p=\exp(-\epsilon)$ to those of nome
\begin{equation}
p' = \e^{-\pi^2/\epsilon}
\end{equation}
can be written as
\begin{eqnarray}
\lefteqn{
\te(u,p) = \left(\frac{\pi}{\epsilon}\right)^{1/2}
\e^{-(u-\pi/2)^2/\epsilon} E(\e^{-2\pi u/\epsilon}, p'^2) }
\nonumber \\
\lefteqn{
\tt(u,p) = \left(\frac{\pi}{\epsilon}\right)^{1/2}
\e^{-u^2/\epsilon} E(p'\e^{-2\pi u/\epsilon}, p'^2) }
\nonumber \\
\lefteqn{
\td(u,p) = \left(\frac{\pi}{\epsilon}\right)^{1/2}
\e^{-u^2/\epsilon} E(-p'\e^{-2\pi u/\epsilon}, p'^2) }
\label{eq:A.congnome} \\
\lefteqn{
\tv(u,p) = \left(\frac{\pi}{\epsilon}\right)^{1/2}
\e^{-(u-\pi/2)^2/\epsilon} E(-\e^{-2\pi u/\epsilon}, p'^2). }
\nonumber
\end{eqnarray}
We note that these relations follow
directly from the Poisson summation formula.

%  #] AppendixC:

%  #] Appendix:

%  #[ Bibliography:

\end{document}